\documentclass[apj,iop,twocolappendix]{emulateapj}
\usepackage{natbib}
\usepackage{float}
\usepackage{graphicx}
\usepackage{epstopdf}
\usepackage{amsmath}
\begin{document}

\shorttitle{the \ion{He}{2} Proximity Effect}
\shortauthors{Khrykin et al.}

\title{the \ion{He}{2} Proximity effect and the lifetime of quasars}

\author{I.S.~Khrykin\altaffilmark{1,2}\footnotemark[*], J.F.~Hennawi\altaffilmark{1}, M.~McQuinn\altaffilmark{3}, G.~Worseck\altaffilmark{1}}

\footnotetext[*]{e-mail: khrykin@mpia-hd.mpg.de}
\altaffiltext{1}{ Max-Planck-Institut f{\"u}r Astronomie, K\"onigstuhl 17, D-69117 Heidelberg, Germany}
\altaffiltext{2}{Interntional Max Planck Research School for Astronomy \& Cosmic Physics at the University of Heidelberg, K\"onigstuhl 17, D-69117 Heidelberg, Germany}
\altaffiltext{3}{ University of Washington, Dep. of Astronomy, 3910 15th Ave NE, WA 98195-1580 Seattle, USA}

\begin{abstract}
  The lifetime of quasars is fundamental for understanding
  the growth of supermassive black holes, and is an important
  ingredient in models of the reionization of the intergalactic
  medium (IGM). However, despite various attempts to determine quasar
  lifetimes, current estimates from a variety of methods are 
  uncertain by orders of magnitude. This work combines cosmological
  hydrodynamical simulations and $1$D radiative transfer to
  investigate the structure and evolution of the \ion{He}{2}
  Ly$\alpha$ proximity zones around quasars at $z \simeq 3-4$. We show that the time evolution in the proximity zone can be described by a simple analytical model for the approach of the
  \ion{He}{2} fraction $x_{\rm HeII}\left( t \right)$ to ionization equilibrium,
  and use this picture to illustrate how the transmission
  profile depends on the quasar lifetime, quasar UV luminosity, and
  the ionization state of Helium in the ambient IGM (i.e. the
  average \ion{He}{2} fraction, or equivalently the
  metagalactic \ion{He}{2} ionizing background).  A significant
  degeneracy exists between the lifetime and the average \ion{He}{2}
  fraction, however the latter can be determined from measurements of
  the \ion{He}{2} Ly$\alpha$ optical depth far from quasars, allowing
  the lifetime to be measured.  We advocate stacking existing
  \ion{He}{2} quasar spectra at $z\sim 3$, and show that the shape of
  this average proximity zone profile is sensitive to lifetimes as
  long as $\sim 30$~Myr. At higher redshift $z\sim 4$ where the
  \ion{He}{2} fraction is poorly constrained, degeneracies will make
  it challenging to determine these parameters independently.
  Our analytical model for 
  \ion{He}{2} proximity zones should also provide a useful description
  of the properties of \ion{H}{1} proximity zones around quasars at $z \simeq 6-7$.

\end{abstract}


\section{Introduction}

Models of quasar and galaxy co-evolution \citep{Wyithe2003, Springel2005b, Hopkins2008, Conroy2013} posit that every massive galaxy underwent a luminous quasar phase, which is responsible for the growth of the supermassive black holes (BHs) that are found in the centers of all nearby bulge-dominated galaxies \citep{Soltan1982, KormRich95, Yu2002}. In many theories powerful feedback from this phase influences the properties of the galaxies themselves, potentially determining why some galaxies are red and dead while others remain blue \citep{Springel2005a, Hopkins2006}.

A holy grail of this research is the quasar lifetime and, relatedly, its \emph{duty cycle}, $t_{\rm dc}$, defined as the fraction of time that a galaxy hosts an active quasar.  This knowledge would shed light on the triggering mechanism for quasar activity (thought to be either major galaxy mergers or secular disk instabilities), on how gas funnels to the center of the galaxy from these mechanisms, and on the properties of the inner accretion disk \citep{Goodman2003, Hopkins2008,Hopkins2010}. 
It is well-known that the duty cycle
of a population of objects can be inferred by comparing its number
density and clustering strength \citep{CK1989,Martini2001,Haiman2001}. But to date this method has yielded only very weak constraints on the quasar duty cycle of $t_{\rm dc} \sim 10^6-10^9\,{\rm yr}$
\citep{AS05b,Croom2005,Shen2009,White2012,Conroy2013} because of uncertainties in the dark matter halo population of quasars \citep{White2012,Conroy2013}. Constraints on the duty cycle with comparable uncertainty come from comparing the time integral of the quasar luminosity function to the present day number density of black holes \citep{Yu2004}.\footnote{This inference suffers from the uncertainties related to the black hole demographics in local galaxy populations, and scaling relations \citep{KormendyHo2013}.}

Moreover, these methods that constrain $t_{\rm dc}$ do not shed light
on the duration of individual accretion episodes, i.e., the average
quasar lifetime ($t_{\rm lt}$). For instance, if quasars emit their
radiation in $\sim 1000$ bursts over the course of a Hubble time, with
each episode having duration of $t_{\rm lt}\sim 10^{5}~{\rm yr}$, this
would be indistinguishable from steady continuous emission for $t_{\rm
  lt}\sim 10^8~{\rm yr}$.  The former timescale of $t_{\rm lt}\sim
10^{5}~{\rm yr}$ is consistent with the picture of
\citet{Goodman2003}, who argues that the outer regions of quasar
accretion disks are unstable to gravitational fragmentation and cannot
be much larger than $\sim 1\,{\rm pc}$. Such small disks would need to
be replenished $\sim 100-1000$ times over to grow a SMBH, which could
generically result in episodic variability on timescales of
$10^5-10^6$\,yr.  However, the latter timescale $t_{\rm lt}\sim
10^8~{\rm yr}$ is roughly what is needed to grow the mass of a black
hole by one $e$-folding (the Salpeter time), and the timescale galaxy
merger simulations suggest for the duration of a quasar episode
\citep{Hopkins2005}.

\begin{figure}[!t]
\centering
 \includegraphics[width=1.0\linewidth]{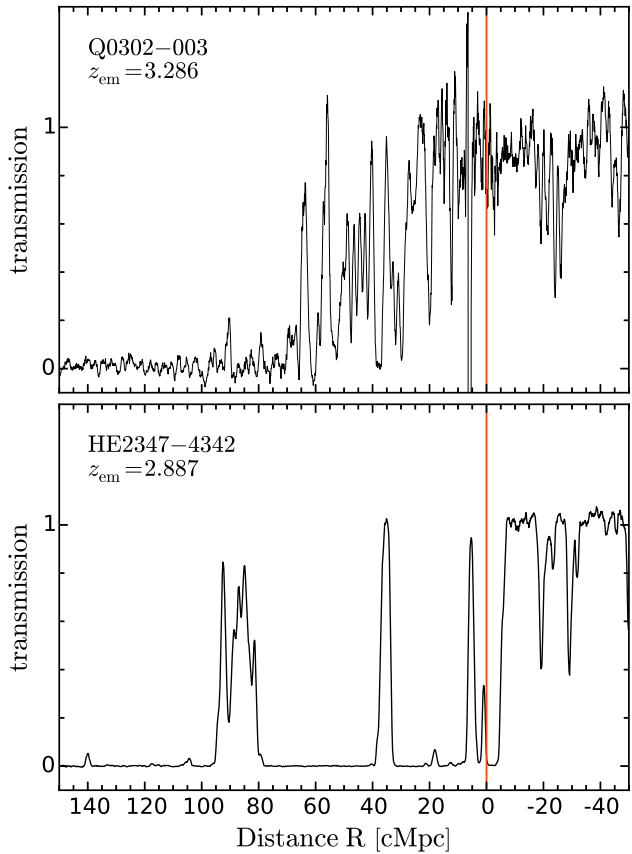}
 \caption{UV \ion{He}{2} Ly$\alpha$ absorption spectra ($R \sim 16,000$) of quasars Q~0302$-$003 (\emph{top}; $z_{\rm em} = 3.286$) and HE~2347$-$4342 (\emph{bottom}; $z_{\rm em} = 2.887$) as a function of comoving distance from the quasar. Both original spectra were rebinned by a factor of $3$ to yield the Nyquist sampling and then smoothed with a `boxcar' kernel of width $w = 11$. Red vertical lines indicate the location of the quasars.}
 \label{fig:Real}
\end{figure}

Current constraints on the quasar lifetime are weak ($t_{\rm lt}
\gtrsim 10^4$~yr; \citealp{Martini2004}), such that lifetimes
comparable to the Salpeter time are still plausible. This limit
derives from the line-of-sight \ion{H}{1} proximity effect -- the
enhancement in the ionization state of \ion{H}{1} in the quasar
environment as probed by the \ion{H}{1} Ly$\alpha$ forest. The
argument is that the presence of a line-of-sight \ion{H}{1} proximity
effect in $z\sim 2-4$ quasars implies that quasars have been emitting
continuously for an equilibration timescale, which corresponds to
about $10^4$\,yr in the $z\sim 2-4$ IGM.  \citet{Schawinski2010} and
\cite{Schawinski2015} argued for variability of several orders of
magnitude in quasar luminosity on short $\sim 10^5\,{\rm yr}$
timescales, based on the photoionization of quasar host galaxies and
light travel time arguments. However, these constraints are indirect
and plausible alternative scenarios related to AGN obscuration could
explain the observations without invoking short timescale quasar
variability. Furthermore, the discovery of quasar powered giant
Ly$\alpha$ nebulae at $z\sim 2$ \citep{Cantalupo2014, Hennawi2015} with
sizes of $\sim 500\,{\rm pkpc}$ implies quasar lifetimes of $\gtrsim
10^6{\rm yr}$, in conflict with the \citet{Schawinski2010} and
\citet{Schawinski2015} estimates. Recently, the presence of
high-equivalent width (EW$_{{\rm Ly}\alpha} \ge 100$\AA) Ly$\alpha$
emitters (LAEs) at large distances $\sim 3-20\,{\rm pMpc}$ from
hyper-luminous quasars has been used to argue for quasar lifetimes in
the range of $1\,{\rm Myr} \lesssim t_{\rm Q} \lesssim 30\,{\rm Myr}$,
based on the presumption that such LAEs result from quasar powered
Ly$\alpha$ fluoresence \citep{Trainor2013,Borisova2015}. However, at
such large distances $\sim 3-20\,{\rm pMpc}$ the fluorescent boost due
to the quasar is far fainter than the fluxes of the LAEs in the
\citet{Trainor2013} and \citet{Borisova2015} surveys, and hence some
other physical process intrinsic to the LAE and unrelated to quasar
radiation must be responsible for these sources\footnote{Using the
  expression for the fluorescent surface brightness in eqn.~(12) of
  \citet{Hennawi2013}, and assuming fluorescent LAEs have a diameter
  of $1.0^{\prime\prime}$, it can be shown that the expected
  fluorescent boost from a quasar at a distance of $R=15\,{\rm pMpc}$
  is a factor of $\gtrsim 400$ smaller than the fluxes of the
  \citet{Borisova2015} LAEs. The \citet{Trainor2013} survey probes
  deeper and considers smaller distances $R=3\,{\rm pMpc}$ where the
  fluorescent boost is larger, but a similar calculation shows the
  fluorescent Ly$\alpha$ emission is nevertheless a factor of $\gtrsim
  5$ smaller than the \citet{Trainor2013} LAEs.  Hence quasar powered
  Ly$\alpha$ fluorescence cannot be the mechanism powering high
  equivalent-width LAEs at such large distances.}.

There is an analogous proximity effect in the \ion{He}{2} Ly$\alpha$
forest that, as this paper shows, is much more sensitive to the quasar
lifetime than the line-of-sight \ion{H}{1} proximity effect.  The
\ion{He}{2} proximity effect has been detected at $2.7 < z < 3.9$
\citep{Hogan1997, Anderson1999, Heap2000, Syphers2014,
  Zheng2015}. Figure~\ref{fig:Real} shows the two well-studied
examples \citep{Shull2010, Syphers2014} that highlight the observed
variance in \ion{He}{2} proximity zone sizes and shapes
\citep{Zheng2015}. HE~2347$-$4342 may be either young ($t_{\rm
  Q}<1$~Myr, \citealp{Shull2010}) or peculiar due to an infalling
absorber \citep{Fechner2004}. Q~0302$-$003 shows a large proximity
zone of $60-100$~comoving Mpc depending on the local density field
\citep{Syphers2014}. Adopting a plausible range in the other relevant
parameters (quasar luminosity, IGM \ion{He}{2} fraction and IGM
clumpiness), \citet{Syphers2014} find that Q~0302$-$003 may have shone
for $0.2-31$~Myr. The simplifying assumptions of a homogeneous IGM
with a \ion{He}{2} fraction of unity only allow for rough estimates of
the quasar lifetime \citep{Hogan1997, Anderson1999, Zheng2015}.
  
In addition, constraints on the quasar lifetime have also been derived
from the so-called transverse proximity effect, i.e. the
enhancement of the UV radiation field around a foreground quasar which
gives rise to increased IGM transmission in a background
sightline. While several effects like anisotropic quasar emission,
episodic quasar lifetimes, and overdensities around quasars make a
statistical detection of the transverse proximity effect challenging
\citep[e.g.][]{Hennawi2006, Hennawi2007, Kirkman2008, Furlanetto2011,
  Prochaska2013}, it has been detected in a few cases, either as a
spike in the IGM transmission \citep{Jakobsen2003, Gallerani2008} or
as a locally harder UV radiation field in the background sightline
\citep{Worseck2006, Worseck2007, Goncalves2008, McQuinn2014}.  For
the handful of quasars for which such detections have been claimed, 
the transverse light crossing time between the foreground quasar and the background
sightline provides a lower limit to the quasar lifetime of $t_{\rm
  Q}=10-30$~Myr.

The goal of this paper is to understand whether the properties of
\ion{He}{2} Ly$\alpha$ line-of-sight proximity zones can constrain the
duration of the quasar phase. This work is motivated by the large number of \ion{He}{2} Ly$\alpha$ forest proximity zones that have been observed over the last five years with the Cosmic Origins Spectrograph (COS) onboard the \emph{Hubble Space Telescope} (HST; \citealp{Shull2010}, \citealp{Worseck2011}, \citealp{Syphers2012}). Another motivation is to generalize previous analyses from the restrictive assumption of $x_{\rm HeII}=1$. Even for ionized gas, \ion{He}{2} proximity
zones around $z \simeq 3$ quasars provide a much more powerful tool for constraining quasar lifetimes than \ion{H}{1} proximity zones. In order to produce a detectable proximity zone, the quasar must shine for a time comparable to or longer than the timescale for the IGM to attain equilibrium with the enhanced photoionization rate $\Gamma$, known as the equilibration timescale
$t_{\rm eq}\simeq \Gamma^{-1}$. Current measurements of the UV background in the $z\sim 3$ \ion{H}{1} Ly$\alpha$ forest yield an \ion{H}{1} photoionization rate $\Gamma_{\rm HI}^{\rm bkg} \simeq 10^{-12}\,{\rm s^{-1}}$ \citep{Becker2013} implying $t_{\rm eq} \simeq 3\times 10^4\,{\rm yr}$. On the other hand, as we argue in \S~\ref{sec:Conditions}, current optical depth measurements at $z\sim 3$ \citep{Worseck2011} imply a  \ion{He}{2} photoionization rate $\Gamma_{\rm HeII}^{\rm bkg} \simeq 10^{-15}\,{\rm s^{-1}}$, resulting in $t_{\rm eq} \simeq 3\times 10^7\,{\rm  yr}$. Thus, \ion{He}{2}
proximity zones can probe quasar lifetimes three orders of magnitude larger, closer to the Salpeter timescale. Moreover, the fact that the \ion{He}{2} background is $\sim 1000$ times lower than the
\ion{H}{1} background implies that the radius within which the quasar dominates over the background will be $\sim 30$ times larger for \ion{He}{2}, and given these much larger zones uncertainties due to density enhancements around the quasar will be much less significant \citep{FG2008}. 

This paper is organized as follows. We discuss the numerical
simulations we use and describe our radiative transfer algorithm in
\S~\ref{sec:code}. In \S~\ref{sec:Conditions} we explore the physical
conditions in the \ion{He}{2} proximity zones of $z \simeq 3$
quasars. We introduce the idea of computing stacked spectra and study
their dependence on the parameters governing quasars and the IGM in
\S~\ref{sec:xHeII}. We discuss our results in \S~\ref{sec:discussion}
and list our conclusions in \S~\ref{sec:Conclusions}.

Throughout this work we assume a flat $\Lambda$CDM cosmology with
Hubble constant $h=0.7$, $\Omega_{\rm m}=0.27$, $\Omega_{\rm
  b}=0.046$, $\sigma_8=0.8$ and $n_s = 0.96$ \citep{Larson2011}, and
helium mass fraction $Y_{\rm He}=0.24$. All distances are in unites of
\emph{comoving} Mpc, i.e., cMpc.

In order to avoid confusion, we distinguish between
several different timescales that govern the duration of quasar activity.
As noted above, the \emph{duty cycle} $t_{\rm dc}$ refers
to the total time that galaxies shine as active quasars integrated over the
age of the universe. On the other hand, one
of the goals of this paper is to understand the constraints that can be put on the
\emph{episodic lifetime} $t_{\rm episodic}$, which is the time spanned by a single episode
of accretion onto the SMBH. But in the context of proximity
effects in the IGM, one actually only constrains the quasar \emph{on-time},
which will denote as $t_{\rm Q}$. If we imagine that time $t=0$ corresponds to the
time when the quasar emitted light that is just now reaching our telescopes on Earth, then
the quasar on-time is defined such that the quasar turned on at time  $-t_{\rm Q}$ in the past.
This timescale is, in fact, a lower limit on the quasar episodic lifetime $t_{\rm episode}$, which
arises from the fact that we observe a proximity zone at $t=0$ when the quasar has
been shining for time $t_{\rm Q}$, whereas this quasar episode may indeed continue, 
which we can only record on Earth if we could conduct observations in the future.
For simplicity in the text, we will henceforth refer to the quasar on-time as the \emph{quasar lifetime} denoted by $t_{\rm Q}$, but the reader should always bear in mind that this is a lower limit on the quasar episodic lifetime. 

\section{Numerical Model}
\label{sec:code}

\subsection{Hydrodynamical Simulations}

To study the impact of quasar radiation on the intergalactic medium we
post-process the outputs of a smooth particle hydrodynamics (SPH)
simulation using a one-dimensional radiative transfer algorithm.  For
the cosmological hydrodynamical simulation, we use the Gadget-3 code
\citep{Springel2005c} with $2 \times 512^3$ particles and a box size of
$25h^{-1}$~cMpc. We extract 1000 sightlines, which we will refer to as
skewers, from the SPH simulation output at a snapshot corresponding to
$z=3.1$. In \S~\ref{sec:Gamma0} we also consider skewers at a higher
redshift of $z = 3.9$.  We identify quasars in the simulation as dark
matter halos with masses $M > 5 \times 10^{11} M_{\odot}$, by running
a friends-of-friends algorithm \citep{Davis1985} on the dark matter
particle distribution. While this mass threshold is $1$~dex lower than the halo mass inferred from quasar clustering measurements \citep{White2012,Conroy2013}, our simulation cube does not capture enough such massive
halos. However, because the \ion{He}{2} proximity effect extends many
correlation lengths ($r_0 \approx 10h^{-1}$~Mpc, \citealp{White2012}),
the use of the less clustered halos will not make a difference at most
radii we consider.

Starting from the location of the quasars, we create skewers by casting
rays through the simulation volume at random angles, and traversing
the box multiple times.  We use the periodic boundary conditions
to wrap a skewer through the box along the chosen direction. This procedure results in skewers from the quasar location through the intergalactic medium that are $318$~cMpc long,
significantly larger than the length of our simulation box and the \ion{He}{2} proximity region. Our
skewers have $27,150$ pixels which corresponds to a spatial interval of
${\rm d}R = 0.012$~cMpc or a velocity interval ${\rm d}v = 0.893\ {\rm
  km\ s^{-1}}$.  This is sufficient to resolve all the features in the \ion{He}{2} Ly$\alpha$ forest.

\subsection{Radiative Transfer}
\label{sec:RT}

We extract the one-dimensional density, velocity, and temperature
distributions along these skewers and use them as input to our
post-processing radiative transfer algorithm\footnote{Post-processing
  is a good approximation because the thermal state of the gas changes
  only when the quasar reionizes it, increasing its temperature by a
  factor of $\sim 2$ \citep{McQuinn2009}. Even in this case, the $\sim
  10$~Myr time since the quasar turned on is insufficient for
  intergalactic gas to respond dynamically, as the relaxation time is
  $(1+\delta)^{1/2}$ of the Hubble time, where $\delta$ is the gas
  overdensity.}, which is based on C$^2$-Ray algorithm
\citep{Mellema2006}. Because this algorithm is explicitly photon
conserving, it enables great freedom in the size of the grid cells and
the time step.  Our one-dimensional radiative transfer algorithm is
extremely fast, with one skewer calculation taking $\sim 5$ seconds on
a desktop machine, allowing us to run many realizations.
In this section, we describe the salient features of our radiative
transfer algorithm, and we refer the reader to the original
\citet{Mellema2006} paper on the C$^2$Ray code for additional details.

We put a single source of radiation, the quasar, at the beginning of each sightline, and trace the change in the ionization state and temperature of the IGM for a finite time, $t$, as the radiation is propagated\footnote{For simplicity we assume a `light bulb' model for a quasar in our simulations, which implies that the luminosity of the quasar does not change over time it is active.}.  The spectral energy distribution (SED) of the source is modeled
as a power-law, such that the quasar photon production rate
at any frequency $\nu \geq \nu_{\rm th}$ is 
\begin{equation}\label{eqn:1}
N_{\nu} = \frac{\alpha Q_{\rm 4Ry}}{\rm \nu_{th}} \left(\frac{\rm \nu}{\nu_{\rm th}}\right)^{-\alpha - 1}, 
\end{equation} 
where $Q_{\rm 4Ry}$ is the photon production rate above the \ion{He}{2} ionization threshold of
$4\,{\rm Ry}$ or a corresponding threshold frequency $\nu_{\rm th} = 1.316 \times 10^{16}\ {\rm Hz}$ corresponding to this threshold. We assume a spectral index of $\alpha = 1.5$, $f_{\nu} \sim \nu^{-\alpha}$, consistent with the measured values in stacked UV spectra of quasars \citep{Telfer2002,Shull2012,Lusso2015}.

The $r$-band magnitudes of \ion{He}{2} quasars in the HST/COS archive
range from $r=16.0-19.4$, with a median value of $r\simeq 18.25$. Following
the procedure described in \citet{Hennawi2006}, we
model the quasar spectral-energy distribution (SED) using a composite
quasar spectrum which has been corrected for IGM absorption
\citep{Lusso2015}. By integrating the \citet{Lusso2015} composite 
redshifted to $z = 3.1$ against the SDSS filter, we can relate
the $r$-band magnitude of a quasar to the photon production rate  
at the \ion{H}{1} Lyman limit $Q_{\rm 1Ry}$, which is then related to $Q_{\rm 4Ry}$ by
  $Q_{\rm 4Ry} = Q_{\rm 1Ry}\times 4^{-\alpha}$,
  assuming the same power law SED governs the quasar spectrum for
  frequencies blueward of $1$~Ry. This procedure implies that quasars in the
  range $r=16.0-19.4$ have $Q_{\rm 4Ry}=10^{56.0-57.4}$\ ${\rm
    s}^{-1}$, with median $r\simeq 18.25$ and $Q_{\rm 4Ry} = 10^{56.5}$\ ${\rm
    s}^{-1}$.
  
The quasar photoinization rate in every cell is given by
\begin{equation}\label{eqn:2}
\Gamma_{\rm QSO} = \int_{\nu_{\rm th}}^{\infty}\frac{N_{\rm
    \nu}e^{-\langle\tau_{\rm
      \nu}\rangle}}{h_{\rm P}\nu}\frac{1-e^{-\langle\delta\tau_{\rm
      \nu}\rangle}}{\langle n_{\rm HeII}\rangle V_{\rm
    cell}}d\nu,
\end{equation}
where $\langle\tau_{\nu}\rangle$ is the optical depth along the skewer
from the source to the current cell,
$\langle\delta\tau_{\nu}\rangle$ is the optical depth inside the cell, $\langle n_{\rm HeII} \rangle$ is the average number density of \ion{He}{2} in this cell, $V_{\rm cell}$ is the volume of the cell,
and $h_{\rm p}$ is Planck's constant. Here the angular brackets
indicate time averages over the discrete time step $\delta t$.

Combining with eqn.~(\ref{eqn:1}), we can then rewrite eqn.~(\ref{eqn:2}) as
\small
\begin{equation}\label{eqn:gammaqso}
\Gamma_{\rm QSO} = \frac{\alpha Q_{\rm 4Ry}}{n_{\rm HeII}V_{\rm cell}\nu_{\rm th}}\int_{\nu_{\rm th}}^{\infty}\left(\frac{\nu}{\nu_{\rm th}}\right)^{-(\alpha + 1)}e^{-\langle\tau_{\rm \nu}\rangle}\left(1-e^{-\langle\delta\tau_{\nu}\rangle}\right)d\nu.
\end{equation}
\normalsize 
Although this equation for the photoionization rate is an integral
over frequency, in practice the code is not tracking multiple
frequencies. For the power-law quasar SED that we have assumed, and
\ion{He}{2} ionizing photon absorption cross section
$\sigma_{\nu} \approx \sigma_{\rm th}(\nu\slash \nu_{\rm th})^{-3}$,
eqn.~(\ref{eqn:gammaqso}) has an analytic solution that depends on $\langle \tau_{\rm th}\rangle$ and $\langle \delta\tau_{\rm
  th}\rangle$, the time-averaged optical depth to the cell and inside the cell, 
respectively,
evaluated at the \ion{He}{2} ionization threshold. Thus, given the values for
these optical depths evaluated at the single edge frequency, 
we can compute the frequency-integrated photoionization rate.  

The foregoing has considered the case of a single source of radiation,
namely the quasar. This scenario is likely appropriate early on in
\ion{He}{2} reionization, when each quasar photoionizes its own
\ion{He}{3} zone. However, at later times, the intergalactic medium
is filled with \ion{He}{2} ionizing photons emitted by many
sources. In order to properly model the radiative transfer along our
skewer, we need to account for additional ionizations caused by this
intergalactic ionizing background. We approximate this background
$\Gamma_{\rm HeII}^{\rm bkg}$ as being a constant in space and time,
and simply add it into each pixel of our sightline to model the
presence of these other sources
\begin{equation}
\Gamma_{\rm tot} = \Gamma_{\rm QSO} + \Gamma_{\rm HeII}^{\rm
  bkg}.\label{eqn:gammatot}
\end{equation}
As such, this procedure does the full one-dimensional radiative
transfer to compute the photoionization rate $\Gamma_{\rm QSO}$
produced by the quasar, but it would treat the background as an
unattenuated and homogeneous radiation field present at every
location. However, the densest regions of the IGM will be capable of
self-shielding against ionizing radiation, thus giving rise to
\ion{He}{2} Lyman-limit systems (\ion{He}{2}-LLSs). Within these
systems our $1$D radiative transfer algorithm always properly
attenuates the quasar photoionization rate $\Gamma_{\rm QSO}$,
however, if we take the the \ion{He}{2} ionizing background
$\Gamma_{\rm HeII}^{\rm bkg}$ to be constant in every pixel along the
skewers, this would be effectively treating the \ion{He}{2}-LLSs as
optically thin, thus underestimating the \ion{He}{2} fraction in the
densest regions. In order to more accurately model the attenuation of
the \ion{He}{2} background by the \ion{He}{2}-LLSs, we implement an
algorithm described in \citet{McQuinn2010}. In
Appendix~\ref{ap:ImpactLLS} we summarize this approach and show that
the effect of self-shielding of \ion{He}{2}-LLSs to the \ion{He}{2}
background $\Gamma_{\rm HeII}^{\rm bkg}$ has a negligible effect on
the structure of the \ion{He}{2} proximity zones. We nevertheless
treat the \ion{He}{2} LLSs with this technique for all of the results
described in this paper. In \S~\ref{sec:Gamma0} we also discuss
the impact of spatial fluctuations in the \ion{He}{2} ionizing
background $\Gamma_{\rm HeII}^{\rm bkg}$ on our results (showing that
the average transmission profile is largely unaffected by such
fluctuations).

Having arrived at an expression for the photoionization rate in each
cell, the next step is to determine the evolution of the \ion{He}{2} fraction $x_{\rm HeII}$ in response to this time-dependent $\Gamma_{\rm tot}\left( t \right)$. The time evolution of the \ion{He}{2} fraction is given by
\begin{equation}\label{eqn:dxHeIIdt}
\frac{dx_{\rm HeII}}{dt} = -\Gamma_{\rm tot}(t) \, x_{\rm HeII} + \alpha_{\rm A} n_{\rm e} \left (1- x_{\rm HeII} \right),
\end{equation}
which if $n_{\rm e}$ is independent of $x_{\rm HeII}$ (which holds at the $6\%$ level) has the solution of
\begin{equation}\label{eqn:xHeIIfull}
x_{\rm HeII}\left(t\right) = x_{\rm HeII, 0} {\rm e}^{-\int_0^t dt' t_{\rm eq}(t')^{-1} } + \int_0^t dt' \alpha_{\rm A} n_{\rm e} {\rm e}^{-\int_0^{t'} dt'' t_{\rm eq}(t'')^{-1}}.
\end{equation}
where $x_{\rm HeII, 0}$ is the initial singly ionized fraction and $t_{\rm eq}$ is the equilibration timescale given by 
\begin{equation}\label{eqn:teq}
t_{\rm eq} = \left(\Gamma_{\rm tot} + n_{\rm e}\alpha_{\rm A}\right)^{-1}, 
\end{equation}
$n_{\rm e}$ is the electron density, and $\alpha_{\rm A}$ is the
Case A recombination coefficient. In the limit when $\Gamma_{\rm tot}(t)$ does not change over $t$ (and also the same for $\alpha_{\rm A} n_e$ which is less of an approximation) eqn.~(\ref{eqn:xHeIIfull}) simplifies considerably to
\begin{equation}\label{eqn:xHeII}
x_{\rm HeII}\left(t\right) = x_{\rm HeII, eq} + \left( x_{\rm HeII, 0} - x_{\rm HeII, eq} \right)e^{-t/t_{\rm eq}},
\end{equation}
where we have defined the equilibrium \ion{He}{2} fraction as
\begin{equation}\label{eqn:7}
  x_{\rm HeII,eq} = \alpha_{\rm A} n_{\rm e} t_{\rm eq}.
\end{equation}   
We discuss our treatment of \ion{He}{2} ionizing photons produced by
recombinations and the justification for adopting Case A in
\S~\ref{sec:recomb} below.  We have ignored collisional ionizations
which contribute negligibly in proximity zones, where gas is
relatively cool $\left( T \sim 10^4 {\rm K}\right)$, and are even more
highly suppressed for \ion{He}{2} relative to \ion{H}{1}.  Given that
the \ion{He}{2} fraction starts at an initial value $x_{\rm HeII,0}$,
eqn.~(\ref{eqn:xHeII}) gives the \ion{He}{2} fraction at a later time
$t$, provided that $\Gamma_{\rm tot}$, $n_{\rm e}$, and $\alpha_{\rm
  A}$ are constant over this interval.  As such, this equation is only
exact for infinitesimal time intervals.

Following \citet{Mellema2006}, we can compute the time averaged \ion{He}{2} fraction inside a
cell by averaging eqn.~(\ref{eqn:xHeII}) over the discrete time-step $\delta t$, yielding
\begin{equation}\label{eqn:xHeIIbar}
\langle x_{\rm HeII} \rangle = x_{\rm HeII,eq} + \left( x_{\rm HeII,t} - x_{\rm HeII,eq}\right)\left(1 - e^{-\frac{\delta t}{t_{\rm eq}}}\right)\frac{t_{\rm eq}}{\delta t}, 
\end{equation}
where $x_{\rm HeII,t}$ is the \ion{He}{2} fraction at the previous timestep $t$. We then 
use eqn.~(\ref{eqn:xHeIIbar}) to calculate the time-averaged optical depth in the cell
at the \ion{He}{2} edge
\begin{equation}\label{eqn:dtau}
\langle \delta \tau_{\rm th} \rangle = \langle x_{\rm HeII} \rangle n_{\rm He}\sigma_{\rm th}\Delta r, 
\end{equation}
where $\Delta r$ is the size of the cell. 
Likewise, the time-averaged optical depth to the cell $\langle \tau_{\rm th} \rangle$ is 
computed by adding, in causal order, all the $\langle \delta \tau_{\rm th} \rangle$
of the cells lying between the source and the cell under consideration. The electron density $\langle n_{\rm e}\rangle$ and number density of \ion{He}{2}
atoms $\langle n_{\rm HeII}\rangle$ are similarly computed using eqn.~(\ref{eqn:xHeIIbar}). 

The iterative process that we employ to find the new ionization state
in each cell (see the flow-chart in Fig~2 of \citealp{Mellema2006})
can be described as follows. Starting with the cell nearest the source
and moving outward, we:
\begin{enumerate}
  \item Set the mean \ion{He}{2} fraction $\langle x_{\rm HeII} \rangle$ to
    that given by the previous time-step (or the initial conditions). 
  \item Evaluate the optical depth between the source and the cell $\langle \tau_{\rm th} \rangle$ 
    by summing the $\delta \tau_{\rm th}$ from the previous time-step. 
  \item Iterate the following until convergence of the \ion{He}{2} fraction $\langle x_{\rm HeII} \rangle$ is achieved
    \begin{itemize}
    \item compute the time-averaged optical depth $\langle \delta \tau_{\rm th} \rangle$
within the cell (eqn.~\ref{eqn:dtau});
    \item compute the photoionization rate $\Gamma_{\rm tot}$ (eqns.~\ref{eqn:gammaqso} and \ref{eqn:gammatot});
    \item compute the mean electron number density based on the current mean ionization 
state;
    \item calculate the new \ion{He}{2} fraction $x_{\rm HeII} \left( t \right)$ at this time step (eqn.~\ref{eqn:xHeII}), as well as its time-step averaged value $\langle x_{\rm HeII}\rangle$ (eqn.~\ref{eqn:xHeIIbar})
    \item check for convergence
    \end{itemize}
    \item Once convergence is reached, advance to the next cell. 
\end{enumerate}
Following this procedure for every time-step, we thus integrate
the time-evolution over time interval $t_{\rm Q}$, 
which denotes the quasar lifetime, yielding the \ion{He}{2} fraction  $x_{\rm HeII}$
at each location in space in the proximity zone. 

\subsection{Heating and Cooling of the Gas}

We assume that the excess from photoionization of helium heats the surrounding gas.
In reality, instead of heating, the electron produced during the photoionization, can become a source of secondary (collisional) ionization, but they are unimportant for the highly ionized IGM, which will clearly be the case for \ion{He}{2} proximity zones. The amount of heat injected per time interval is given by \citep{Abel1999}

\begin{equation}
\frac{{\rm d} T}{{\rm d}t} \approx \frac{2 Y_{\rm He}}{3k_{\rm B}(8-5Y_{\rm He})} \langle E\rangle \frac{{\rm d}x_{\rm HeII}}{{\rm d}t} 
\end{equation}
where ${\rm d}x_{\rm HeII}/{\rm d}t$ is the change in the \ion{He}{2} fraction over the time step and $\langle E \rangle$ is the average excess energy of a photon above \ion{He}{2} ionization threshold, which is given by

\begin{equation}
\langle E \rangle = \left[\frac{\int_{\nu_{\rm th}}^{\infty} N_{\nu}\sigma_{\nu}e^{-\langle\tau_{\rm \nu}\rangle}\left(1-e^{-\langle\delta\tau_{\nu}\rangle}\right) \left( h_{\rm P}\nu - h_{\rm P}\nu_{\rm th}\right) {\rm d}\nu}{\int_{\nu_{\rm th}}^{\infty}N_{\nu}\sigma_{\nu} e^{-\langle\tau_{\rm \nu}\rangle}\left(1-e^{-\langle\delta\tau_{\nu}\rangle}\right) {\rm d}\nu}\right]
\end{equation}
For a power law SED, this frequency integral can be solved analytically, and yields a function that depends on the optical depth between the source and the current cell $\langle\tau_{\nu}\rangle$ and optical depth of the cell $\langle \delta \tau_{\nu} \rangle$. This function is evaluated at each time step and the resulting change in the temperature added to each cell. We ignore the heating caused by the \ion{He}{2} ionizing background $\Gamma_{\rm HeII}^{\rm bkg}$, since this is accounted for via the photoionization heating in the hydrodynamical simulations (and it does not depend on the normalization of the photoionizing background, which we freely adjust). We also do not include cooling in our simulations because the gas cooling time at mean density is comparable to the Hubble time which is much longer than the expected lifetime of the quasar (i.e., $t_{\rm cool} \gg t_{\rm Q}$) and thus can be safely ignored.

\subsection{Finite Speed of Light}
Our code works in the infinite speed of light limit, similar to
other one-dimensional radiative transfer codes which attempt to model
observations along the line-of-sight \citep{White2003, Shapiro2006, Bolton2007a,
  Lidz2007, Davies2014}. Thus, the results do not depend on the speed
of light, which may seem problematic because the light travel times
across the proximity region can be tens of Myr. However, in
Appendix~\ref{ap:lighttravel}, we show that because absorption observations also occur along the light-cone, the infinite speed of light assumption exactly describes the ionization state of the gas
probed by line-of-sight observations. 

\subsection{Recombinations}
\label{sec:recomb}

\begin{figure}[!t]
\centering
 \includegraphics[width=1.0\linewidth]{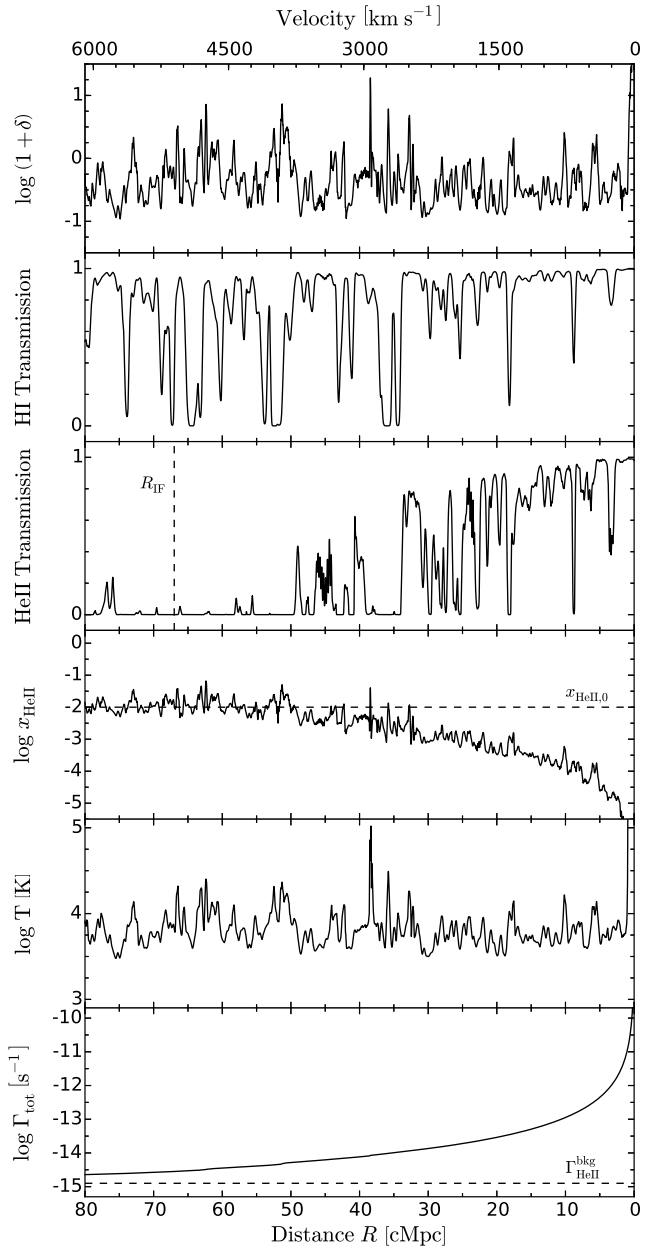}
 \caption{Example sightline at $z = 3.1$ from our radiative transfer code assuming a quasar has been on for $t_{\rm Q} = 10^7$yr at a photon production rate  $Q_{\rm 4Ry} = 10^{56.1}\ {\rm s^{-1}}$ with \ion{He}{2} ionizing background $\Gamma_{\rm HeII}^{\rm bkg} = 10^{-14.9}{\rm s^{-1}}$.
   The lower x-axis indicates distance $R$ from the quasar in units of comoving
   Mpc, whereas the upper x-axis is the corresponding velocity in 
   units of ${\rm km\ s^{-1}}$.
   Panels show (from top to bottom): the overdensity, transmitted flux in hydrogen, transmitted flux in helium, the \ion{He}{2} fraction $x_{\rm HeII}$, temperature $T$, photoionization rate $\Gamma_{\rm tot}$. The vertical dashed line in the panel with \ion{He}{2} transmission shows the extent of the quasar ionization front (see \S~\ref{sec:time_evol} for the description). We indicate the initial \ion{He}{2} fraction before the quasar is on by the dashed line in the panel with $x_{\rm HeII}$ evolution. The dashed line in the bottom panel indicates the value of the \ion{He}{2} ionizing background $\Gamma_{\rm HeII}^{\rm bkg}$.}
 \label{fig:ExSkewer}
\end{figure}

We use the Case A recombination coefficient throughout the paper.
Quasar proximity zones represent highly ionized media which 
typically have $x_{\rm HeII} \lesssim 10^{-2}$, making most cells optically
thin to ionizing photons produced by recombinations directly to the
ground state.  This recombination radiation is an additional source of
ionizations, but we do not include it in our computations. We show in Appendix~\ref{ap:recrad} that
neglecting secondary ionizations from recombination radiation is justified as it is negligible in
comparison to the quasar radiation.

\subsection{Radiative Transfer Outputs}

We calculate \ion{He}{2} spectra along each of the sightlines taken from the SPH simulations following the procedure described in \citet{Theuns1998}, which we outline in Appendix~\ref{ap:skewers}. Figure~\ref{fig:ExSkewer} shows various physical properties along an
example line-of-sight. The quasar is located on the right side of the
plot at $R=0$. These physical properties are drawn from a model that
has the quasar lifetime set to $t_{\rm Q} = 10$~Myr, the photon production
rate $Q_{\rm 4Ry} = 10^{56.1}\ {\rm s^{-1}}$, and the \ion{He}{2} ionizing background
$\Gamma_{\rm HeII}^{\rm bkg} = 10^{-14.9}{\rm s^{-1}}$, which is our preferred
value following the discussion in \S~\ref{sec:bkg}.

The uppermost panel of Figure~\ref{fig:ExSkewer} shows the gas density
along the skewer in units of the cosmic mean density, illustrating the level of density
fluctuations present in the $z = 3.1$ IGM. The second panel shows the
\ion{H}{1} transmitted flux, which exhibits the familiar absorption
signatures characteristic of the Ly$\alpha$ forest. A weak \ion{H}{1}
proximity effect \citep{Carswell1982, Bajtlik1988} is noticeable by
eye near the quasar for $R \lesssim 20$ cMpc. This weak \ion{H}{1}
proximity effect is expected: given the high value of the \ion{H}{1}
ionizing background, the region where the quasar radiation dominates
over the background is relatively small. Furthermore, the low
\ion{H}{1} Ly$\alpha$ optical depth at $z = 3.1$ reduces the contrast
between the proximity zone and regions far from the quasar. On the
contrary, the \ion{He}{2} transmission clearly indicates the large and
prominent ($\simeq 50$~cMpc) \ion{He}{2} proximity zone around the
quasar. At larger distances $R > 50$~cMpc the transmission drops to
nearly zero, giving rise to long troughs of Gunn-Peterson (GP; \citealp{GP1965}) absorption, as is commonly observed in the \ion{He}{2} transmission spectra of quasars observed with HST
\citep{Worseck2011,Syphers2014}. This GP absorption results from the
large \ion{He}{2} optical depth in the ambient IGM, which is in turn
set by our choice of $\Gamma_{\rm HeII}^{\rm bkg} = 10^{-14.9}{\rm
  s^{-1}}$. The \ion{He}{2} transmission follows the radial trend set
by the \ion{He}{2} fraction $x_{\rm HeII}$. As expected, close to the
quasar $R \lesssim 20$~cMpc, helium is highly ionized ($x_{\rm HeII} <
10^{-3}$) by the intense quasar radiation. At larger radii the quasar
photoionization rate weakens, dropping approximately as $R^{-2}$, as
indicated in the bottom panel. Eventually, at large distances $R \gtrsim 70$~cMpc the quasar radiation no
longer dominates over the the background $\Gamma_{\rm HeII}^{\rm bkg}$, and $x_{\rm HeII}$ gradually asymptotes to a value $x_{\rm HeII,0} \simeq 10^{-2}$, 
set by the chosen \ion{He}{2} ionizing background. 

\section{Physical Conditions in the Proximity Zone}
\label{sec:Conditions}
 
The transmitted flux in the proximity zone results from an interplay between different parameters. In order to constrain the quasar lifetime, $t_{\rm Q}$, independently from other parameters, such as the \ion{He}{2} ionizing background $\Gamma_{\rm HeII}^{\rm bkg}$ or the rate at which \ion{He}{2} ionizing photons are emitted by the quasar $Q_{\rm 4 Ry}$, we need to understand the impact of each one of them on the structure of the proximity zone. We begin first by considering the constraints on $\Gamma_{\rm HeII}^{\rm bkg}$ from observational data.

\subsection{Constraints on the \ion{He}{2} ionizing background $\Gamma_{\rm HeII}^{\rm bkg}$}
\label{sec:bkg}

In order to model the \ion{He}{2} proximity regions, we need to make
some assumptions about the background radiation field in the IGM. At
$z\gtrsim2.8$ this background may not be spatially uniform, as it has
been argued that \ion{He}{2} reionization is occurring
\citep{McQuinn2009b, Shull2010}.  \ion{He}{2} reionization is thought
to be driven by quasars turning on and emitting the hard photons
required to doubly ionize helium \citep{Madau1994, Miralda2000,
  McQuinn2009, Haardt2012, Compostella2013}.  At $z=3.1$, redshifts
characteristic of much of the COS data, the IGM is still likely to
consist of mostly reionized \ion{He}{3} regions, but there may be
some quasars that turn on in \ion{He}{2} regions.  The latter case
should become increasingly more likely with increasing redshift.  In this
paper we model the full range of possibilities, but let us first get
a sense for what currently available data implies about typical regions
of the IGM. 

\begin{figure}[!t]
\centering
 \includegraphics[width=1.0\linewidth]{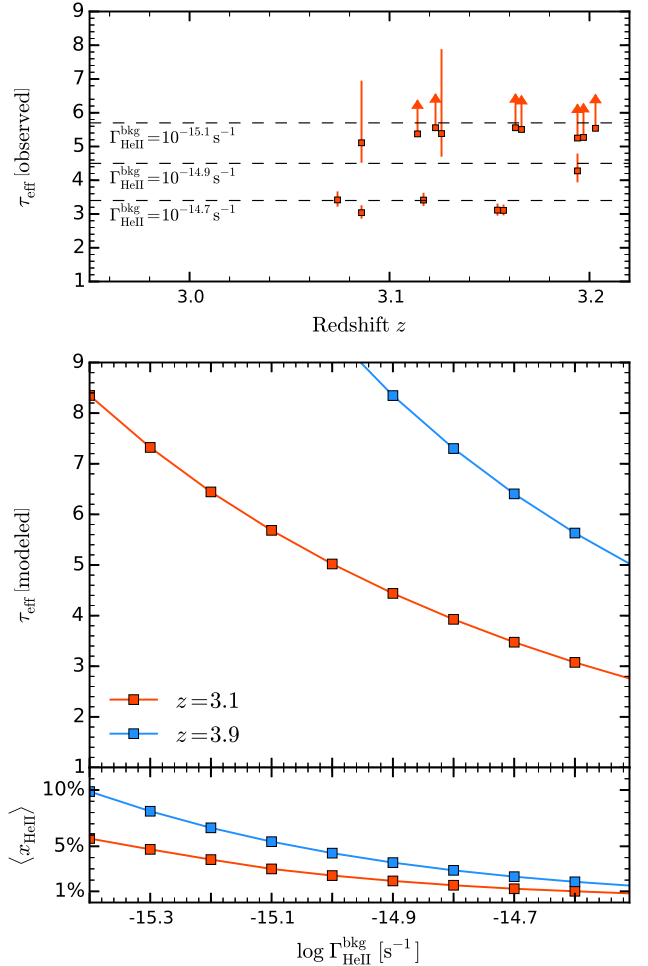}
\caption{Results of the radiative transfer simulations with just a uniform \ion{He}{2} ionizing background and no quasar source. The red curve corresponds to the calculations performed at redshift $z = 3.1$, while the blue curve is for $z = 3.9$. \emph{Top panel:} the observed distribution of the \ion{He}{2} effective optical depth at $z= 3.0-3.2$ from \citet{Worseck2014} \& Worseck et al. in prep. The dashed curves show the mean effective optical depths on the same spatial scale as data ($\Delta z = 0.04$) corresponding to the indicated $\Gamma_{\rm HeII}^{\rm bkg}$ value in the $1$D radiative transfer simulations. \emph{Middle panel}: \ion{He}{2} effective optical depth $\tau_{\rm eff}$ as a function of the \ion{He}{2} ionizing background field $\Gamma_{\rm HeII}^{\rm bkg}$ from our $1$D radiative transfer algorithm. \emph{Bottom panel}: \ion{He}{2} fraction $x_{\rm HeII}$ as a function of \ion{He}{2} background $\Gamma_{\rm HeII}^{\rm bkg}$.}
 \label{fig:F_tau}
\end{figure}

Recent measurements of the \ion{He}{2} effective optical depth
$\tau_{\rm eff}$ from \citet{Worseck2014} \& Worseck et al. in prep, albeit with large scatter,
imply $\langle \tau_{\rm eff}^{\rm HeII} \rangle \simeq 4-5$ on scales $\Delta z = 0.04$ ($\Delta R \simeq 40$~cMpc) at $z\sim 3.1$ (see the upper panel of Figure~\ref{fig:F_tau}). We use our $1$D radiative transfer algorithm to try to understand how these observational results constrain the \ion{He}{2}
ionizing background $\Gamma_{\rm HeII}^{\rm bkg}$. Similarly to
\citet{Worseck2014}, we exclude the proximity zone from our
calculations by turning the quasar off. The resulting transmission
through the IGM is solely due to the \ion{He}{2} background
$\Gamma_{\rm HeII}^{\rm bkg}$.
We then vary $\Gamma_{\rm HeII}^{\rm bkg}$, and calculate the effective optical depth
defined by $\tau_{\rm eff} \equiv - \ln \langle F\rangle$, where we take the average of the simulated transmission $F$ in $5$ regions along $100$ skewers, each region with size $\Delta z = 0.04$ equal to the size of the bin in the observations. We also calculate the average value of the \ion{He}{2} fraction $\langle x_{\rm HeII}\rangle$ in the same bins.

The results are shown in Figure~\ref{fig:F_tau}. The solid red line in
the middle panel shows the values of our modeled \ion{He}{2} effective
optical depth as a function of the \ion{He}{2} ionizing background
$\Gamma_{\rm HeII}^{\rm bkg}$. The corresponding \ion{He}{2} fraction $\langle x_{\rm HeII}\rangle $
is shown in the bottom panel of Figure~\ref{fig:F_tau}. We find that
the effective optical depth $\tau_{\rm eff} \simeq 4-5$ implies a characteristic 
\ion{He}{2} ionizing background of $\Gamma_{\rm HeII}^{\rm bkg} \simeq
10^{-14.9} {\rm s}^{-1}$, which will refer to henceforth as our
fiducial value for $z = 3.1$. This \ion{He}{2} background corresponds
to an average \ion{He}{2} fraction $\langle x_{\rm HeII}\rangle \simeq
0.02$\footnote{Note that our ability to constrain the average
  \ion{He}{2} fraction to be $x_{\rm HeII} \sim 10^{-2}$ contrasts
  sharply with the case of \ion{H}{1} GP absorption at
  $z\sim 6-7$. For hydrogen at these much higher redshifts the most
  sensitive measurements imply a lower limit on the \ion{H}{1}
  fraction $x_{\rm HI} \gtrsim 10^{-4}$ \citep{Fan2002,Fan2006}. This
  difference in sensitivity between \ion{He}{2} Ly$\alpha$ at $z \sim
  3$ and \ion{H}{1} Ly$\alpha$ at $z \sim 6$ results from several
  factors: 1) the \ion{He}{2} Ly$\alpha$ GP optical depth is $4$ times
  smaller than \ion{H}{1} Ly$\alpha$ due to the higher frequency of
  \ion{He}{2}; 
2) the abundance of helium is a factor of $\sim 12$ smaller than that of hydrogen 
3) the intergalactic medium is on average a factor of $5.4$ less dense at $z \sim 3$ compared to $z \sim 6$, but also the cosmological line element is $2.3$ times larger at $z \sim 6$ than at $z \sim 3$. 
All of these factors combined together therefore imply an increase of two orders of magnitude in the sensitivity of the \ion{He}{2} GP optical depth at $z\sim 3$ to the \ion{He}{2} fraction and hence the \ion{He}{2} ionizing background.}. 
While uniformity is likely not a good assumption
during \ion{He}{2}\ reionization, which makes the above number highly
approximate (and probably an underestimate for $x_{\rm HeII}$), it is a good assumption thereafter, i.e. $z\simeq 2.5$, where fluctuations of \ion{He}{2} ionizing background are on the order of unity \citep{McQuinn2014}.

We also preformed the same set of calculations at higher redshift of
$z = 3.9$ and plot the results as blue curves in Figure~\ref{fig:F_tau}. Note, that at higher redshifts the gas in the intergalactic medium is becoming more dense, and the steep redshift dependence
of the GP optical depth $\tau_{\rm GP}\propto (1+z)^{3/2}$ gives rise
to significantly higher optical depths at $z = 3.9$.
Thus at these high redshifts even \ion{He}{2} fractions of $x_{\rm HeII}\sim 0.01$
give rise to large effective optical depths $\tau_{\rm eff}\gtrsim 5$.
While we mostly consider $z = 3.1$, \S~\ref{sec:Gamma0} considers $z \simeq3.9$.

\subsection{Time Evolution of the Ionized Fraction}
\label{sec:time_evol}

\begin{figure*}[!t]
\centering \includegraphics[width=\textwidth]{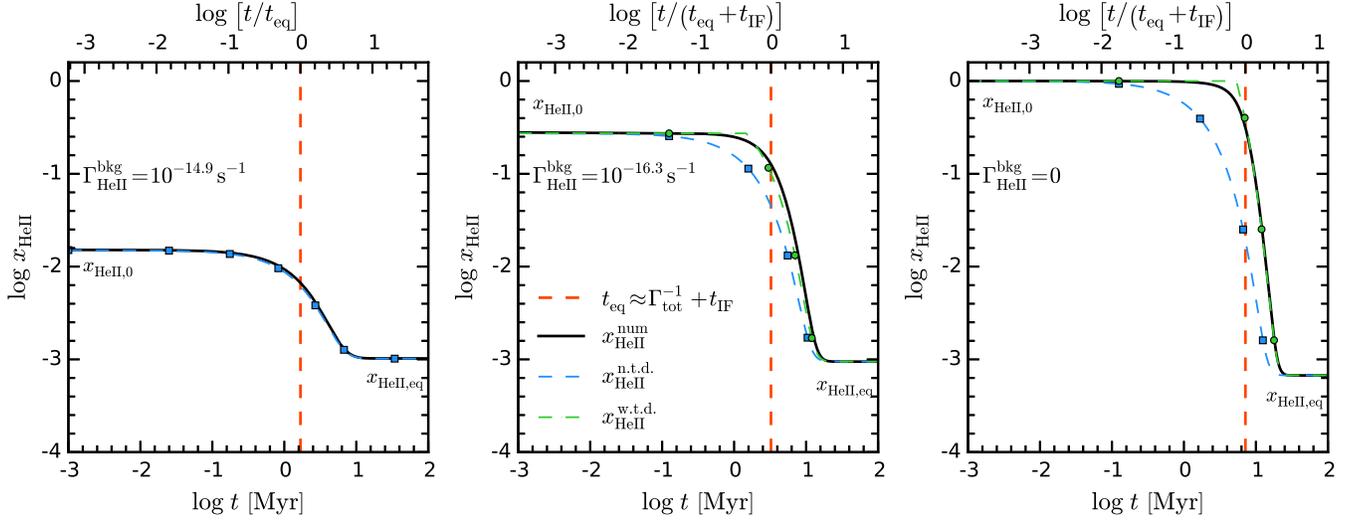}
 \caption{Time evolution of the \ion{He}{2} fraction, $x_{\rm
     HeII}\left(t, R\right)$, at a single location in the proximity
   zone at $R=25$\ cMpc, produced from our radiative transfer solution
   (solid black curve), compared to the analytical expression given by
   eqn.~(\ref{eqn:xHeIItqso}) (dashed blue curve with square
   markers). \emph{The left panel} shows the results for a model with
   $\Gamma_{\rm HeII}^{\rm bkg} = 10^{-14.9}{\rm s^{-1}}$ ($x_{\rm
     HeII,0} \simeq 0.02$). \emph{The middle panel} is the same, but
   for $\Gamma_{\rm HeII}^{\rm bkg} = 10^{-16.3}{\rm s^{-1}}$ ($x_{\rm
     HeII,0} \simeq 0.3$). The zero background model $\Gamma_{\rm
     HeII}^{\rm bkg} = 0$ (i.e., $x_{\rm HeII,0} = 1$) is shown in the
   \emph{right panel}. The dashed green curves correspond to the
   corrected analytical solution given by
   eqn.~(\ref{eqn:evol_sol_corr}). The \ion{He}{2} fraction $x_{\rm
     HeII,eq}$ and the equilibration time $t_{\rm eq}$ (red dashed
   vertical line in all panels) are evaluated using the quasar
   photoionization rate for a fully equilibrated IGM in the analytic
   expression.
   \label{fig:FxHeII}}
\end{figure*}

Consider the case of a quasar emitting radiation for time $t_{\rm Q}$
into a homogeneous IGM with \ion{He}{2} fraction $x_{\rm HeII}$.  The
time evolution of the \ion{He}{3} ionization front is governed by the
equation \citep{HaimanCen2001,Bolton2007a}
\begin{equation}
\frac{dR_{\rm IF}}{dt} = \frac{Q_{\rm 4Ry} - \frac{4}{3} R_{\rm IF}^3\alpha_{\rm HeIII}n_{\rm He}^2}{4\pi R_{\rm IF}^2 x_{\rm HeII}n_{\rm He}}, 
\end{equation}
which has the solution
\begin{equation}
  R_{\rm IF} = R_{S}\left[ 1 - {\rm exp}\left( -\frac{t_{\rm Q}}{x_{\rm HeII}t_{\rm rec}} \right) \right]^{1/3}\label{eqn:RIF}, 
\end{equation}
where $t_{\rm rec} = 1 \slash n_{\rm e}\alpha_{\rm A}$ is the recombination timescale and $R_{\rm S}$ is the classical
Str\"omgren radius $R_{\rm S} = \left( \frac{3Q_{\rm 4Ry}}{4\pi \alpha_{\rm A} n_{\rm HeIII} n_{\rm e}} \right)^{1\slash 3}$,
which is the radius of the sphere around a source of radiation,
within which ionizations are exactly balanced by recombinations.

Previous observational studies of the \ion{H}{1} Ly$\alpha$ proximity
zones around $z \simeq 6$ quasars have primarily focused on measuring
proximity zone sizes \citep{Cen2000, Madau2000, Mesinger2004, Fan2006, Carilli2010}, guided by the faulty intuition that, for assuming a highly neutral IGM $x_{\rm HI} \sim 1$, the location where the transmission profile goes to zero can be identified with the location of the ionization front $R_{\rm IF}$ in eqn.~(\ref{eqn:RIF}). However, the transmission profile for a \ion{H}{2} region expanding into a significantly neutral IGM can be difficult to
distinguish from that of a ``classical'' proximity zone embedded in an
already highly ionized IGM \citep{Bolton2007b,Maselli2007,Lidz2007}. This degenerate situation arises, because even very small residual neutral fractions in the
proximity zone $x_{\rm HI} \gtrsim 10^{-5}$ are sufficient to saturate
\ion{H}{1} Ly$\alpha$, which may occur well before the location of the
ionization front is reached \citep{Bolton2007a}. Thus naively
identifying the size of the proximity zone with the location of the
ionization front $R_{\rm IF}$, can lead to erroneous conclusions about
the parameters governing eqn.~(\ref{eqn:RIF}), e.g. the quasar
lifetime and the ionization state of the IGM. 

A similar degeneracy also exists for \ion{He}{2} proximity zones, which is exemplified in Figure~\ref{fig:ExSkewer} where the \ion{He}{2} transmission saturates at $R \simeq 50$~ cMpc, whereas the ionization front is located much further from the quasar $R_{\rm IF} \simeq 67$~cMpc (dashed vertical line). However, all previous work analyzing the structure of \ion{He}{2} proximity zones has been based on the assumption that the edge of the observed proximity zone can be identified with $R_{\rm IF}$ \citep{Hogan1997,Anderson1999,Syphers2014,Zheng2015}. In addition, the majority of studies have assumed that helium is completely singly ionized $x_{\rm HeII}=1$ for quasars at $3.2 < z < 3.5$ \citep{Hogan1997, Anderson1999, Zheng2015}, whereas our discussion in the previous section (see Figure~\ref{fig:F_tau}) indicates that at $z = 3.1$ observations of the effective optical depth suggest that $\Gamma_{\rm HeII}^{\rm bkg} = 10^{-14.9}{\rm s^{-1}}$ implying $x_{\rm HeII}\simeq 0.01$ (although the average $x_{\rm HeII}$ could be much larger if some regions are predominantly \ion{He}{2} at $z>2.8$, \citealt{Compostella2013, Worseck2014}). Hence, in many regions one is actually in the classical proximity zone regime, where radiation from the quasar increases the ionization level of nearby material which was already highly ionized to begin with and the location of the ionizaton front is irrelevant. Furthermore, as we describe below, the background level determines the
equilibration timescale $t_{\rm eq} \approx 1\slash \Gamma_{\rm
  HeII}^{\rm bkg}$ (see eqn.~\ref{eqn:teq}), which is the
characteristic time on which the \ion{He}{2} ionization state of IGM
gas responds to the changes in the radiation field. For our fiducial
value of the background $\Gamma_{\rm HeII}^{\rm bkg} = 10^{-14.9}{\rm
  s^{-1}}$, $t_{\rm eq} \simeq 2.5 \times 10^7\,{\rm yr}$ is
comparable to the Salpeter time. This suggests that the approach of
$x_{\rm HeII}$ to equilibrium is the most important physical effect in
\ion{He}{2} proximity zones, and in what follows we introduce a simple
analytical equation for understanding this time evolution.

The full solution to the time evolution of $x_{\rm HeII}$ is given by
eqn.~(\ref{eqn:xHeIIfull}), which involves a nontrivial integral
because $\Gamma_{\rm tot}$, $n_{\rm e}$ and $\alpha_{\rm A}$ are all
functions of time. Indeed, this is exactly the equation that is solved
at every grid cell in our $1$D radiative transfer calculation, by
integrating over infinitesimal timesteps (see
eqn.~\ref{eqn:xHeII}). However, in the limit of a highly ionized IGM
$x_{\rm HI} \ll 1$, $n_{\rm e} \propto (1+z)^3$ which is approximately
constant over the quasar lifetimes we consider. Furthermore, as we
will demonstrate later, the low singly ionized fraction $x_{\rm HeII}
\ll 1$ implies that the attenuation is small in most of the proximity
zone and, hence, the photoionization rate $\Gamma_{\rm tot}$ is also
approximately constant in time. Similarly, $\alpha_{\rm A}$ varies
only weakly with temperature (i.e., $\propto T^{-0.7}$), and given
that the temperature will also not vary significantly with time if the
HeII already has been reionized, $\alpha_{\rm A}$ can also be
approximated as constant. In this regime where $\Gamma_{\rm tot}$,
$n_{\rm e}$ and $\alpha_{\rm A}$ are constant in time,
eqn.~(\ref{eqn:xHeII}) reduces to the simpler expression evaluated at $t=t_{\rm Q}$, the quasar
lifetime: 
\begin{equation}\label{eqn:xHeIItqso}
x_{\rm HeII}\left(t_{\rm Q}\right) = x_{\rm HeII, eq} + \left( x_{\rm HeII, 0} - x_{\rm HeII, eq} \right)e^{-t_{\rm Q}/t_{\rm eq}},
\end{equation}
Given that the recombination timescale $t_{\rm rec} \equiv 1\slash \alpha_{\rm A}n_{\rm e}\simeq 10^9\,{\rm yr}$ is very long compared to the longest ionization timescales $1\slash
\Gamma^{\rm bkg}_{\rm HeiI }$, we can write $t_{\rm eq}\approx 1\slash
\Gamma_{\rm tot}$, $x_{\rm HeII,0}\approx \alpha_{\rm A}n_{\rm
  e}\slash \Gamma^{\rm bkg}_{\rm HeII}$, and $x_{\rm HeII,eq}\approx
\alpha_{\rm A}n_{\rm e}t_{\rm eq}$.

In the left panel of Figure~\ref{fig:FxHeII}, the solid black curve
shows the time evolution of $x_{\rm HeII}\left(t, R\right)$ at a
single location in the proximity zone $R=25$~cMpc, produced from our
radiative transfer solution for the quasar photon production rate
$Q_{\rm 4 Ry} = 10^{56.1}{\rm s^{-1}}$ and finite background case
$\Gamma_{\rm HeII}^{\rm bkg} = 10^{-14.9}{\rm s^{-1}}$ corresponding
to an initial \ion{He}{2} fraction $x_{\rm HeII,0} = 0.02$.  The
dashed blue curve is the analytical solution eqn.~(\ref{eqn:xHeIItqso}). 
where $x_{\rm HeII,eq}$ and $t_{\rm eq}$ have been evaluated from the
code outputs, using the total photoionization rate $\Gamma_{\rm tot}$
for a fully equilibrated IGM. In other words, we evaluate $\Gamma_{\rm
  tot}(t=\infty) = \Gamma_{\rm QSO}(t=\infty) + \Gamma_{\rm HeII}^{\rm
  bkg}$ where $t=\infty$ is taken to be our last output at
$t=100\,{\rm Myr}$. It is clear that for the case $\Gamma_{\rm
  HeII}^{\rm bkg} = 10^{-14.9}{\rm s^{-1}}$ and thus an initially
highly ionized IGM $x_{\rm HeII}\simeq 10^{-2}$, eqn.~(\ref{eqn:xHeIItqso}) provides an excellent match to the result of the full radiative transfer calculation.

Due to the patchy and inhomogeneous nature of \ion{He}{2} reionization
some regions on the IGM might, however, have a very high \ion{He}{2}
fraction. Therefore, it is important to check if our analytical
approximation also holds in this case. The middle and right panels of
Figure~\ref{fig:FxHeII} show the time evolution of the \ion{He}{2}
fraction at the same location and value of $Q_{\rm 4Ry}$, but now for
$\Gamma_{\rm HeII}^{\rm bkg} = 10^{-16.3}{\rm s^{-1}}$ and
$\Gamma_{\rm HeII}^{\rm bkg} = 0$, which correspond to $x_{\rm
  HeII,0}\simeq 0.3$ and $x_{\rm HeII,0}=1$ respectively.  The
analytical approximation (blue squares and curve) clearly fails to
reproduce the time evolution. Specifically, it predicts too rapid a
response to the quasar ionization relative to the true evolution, and
this discrepancy is largest for $\Gamma_{\rm HeII}^{\rm bkg}=0$ where
the true evolution to equilibrium is delayed by $\sim 5.5$~Myr.

Recall that, because we observe on the light cone, the speed of light
is effectively infinite in our code. Thus, at the location $R=25$~cMpc we
expect no delay in the response of the proximity zone to the quasar
radiation caused by finite light travel time effects. For the $x_{\rm
  HeII,0} \simeq 0.02$ ($\Gamma_{\rm HeII}^{\rm bkg} = 10^{-14.9}{\rm
  s^{-1}}$) case the ionization front travels at nearly the speed of
light and, because we observe on the light cone, there is thus no
noticeable delay between the evolution of the \ion{He}{2} fraction and
eqn.~(\ref{eqn:xHeIItqso}). However, if
the ionization front does not travel at the speed of light, which will
be the case for the lower backgrounds $\Gamma_{\rm HeII}^{\rm bkg} =
10^{-16.3}{\rm s^{-1}}$ and $\Gamma_{\rm HeII}^{\rm bkg} = 0$ and
correspondingly
higher \ion{He}{2} fractions ($x_{\rm HeII,0}\simeq 0.3$
and $x_{\rm HeII,0}=1$), then the time that it takes the ionization
front to propagate to the location $R = 25$~cMpc is no longer
negligible relative to the equilibration timescale, and the response
of the \ion{He}{2} fraction will be delayed.

We can estimate this time delay by noting that the location of the ionization front (see eqn.~\ref{eqn:RIF})
is given by
\begin{equation}
  R_{\rm IF} \approx  \left[\frac{3 Q_{\rm 4Ry} t_{\rm IF}}{4 \pi n_{\rm He} x_{\rm HeII}}\right]^{1/3}, 
  \label{eqn:RIF_sol}
\end{equation}
assuming that $t_{\rm Q} \ll x_{\rm HeII}t_{\rm rec} = (x_{\rm HeII}\slash 1.0)1.16 \times 10^9\,{\rm yr}$, valid for $x_{\rm HeII}\sim 0.3-1.0$ and the quasar lifetimes we consider. 
In this regime the ionization front is simply the radius of the ionized volume around the quasar.
Inverting this equation, we obtain that at a location $R$, the
time delay between the quasar turning on and the arrival of the first ionizing photons
is\footnote{The mean number density of helium $n_{\rm He}$ is calculated assuming an average overdensity $1+\delta \simeq 0.7$, similar to the value used in the radiative transfer solution for chosen skewer.}
\begin{equation}
  t_{\rm IF}\left(R \right) = 6.7 \ \left(\frac{x_{\rm HeII}}{1.0}\right)\left(\frac{{Q_{\rm 4Ry}}}{10^{56}\ {\rm s^{-1}}}\right)^{-1}\left( \frac{R}{25\ {\rm cMpc}} \right)^3 {\rm Myr}
\end{equation}
For $R = 25$~cMpc this delay is very nearly the delay seen in the middle and right panels of Figure~\ref{fig:FxHeII}, suggesting a simple physical interpretation for the behavior of the \ion{He}{2} fraction in the proximity zone for the $\Gamma_{\rm HeII}^{\rm bkg}=10^{-16.3}{\rm s^{-1}}$ and $\Gamma_{\rm HeII}^{\rm bkg}=0$ cases. Namely, eqn.~(\ref{eqn:xHeIItqso}) still describes the equilibration of the \ion{He}{2} fraction, but it must be modified to account for the delay in the arrival of the ionization front, only after which equilibration begins to occur. We thus write
\begin{equation}
  x_{\rm HeII}\left(t_Q, r\right) = x_{\rm HeII, eq} + \left(1 - x_{\rm HeII, eq}\right){\rm e}^{\frac{ -\left(t_{\rm Q} - t_{\rm IF}\right)} { t_{\rm eq} }} 
  \label{eqn:evol_sol_corr}
\end{equation}
The green curves in the middle and right panels of Figure~\ref{fig:FxHeII} illustrate that
the simple equilibration time picture, but now modified to account for a delay in the arrival
of the ionization front, provides a good description of the time evolution of $x_{\rm HeII}$ in the proximity zone for $\Gamma_{\rm HeII}^{\rm bkg}=10^{-16.3}{\rm s^{-1}}$ and $\Gamma_{\rm HeII}^{\rm bkg}=0$ cases.

To summarize, we have shown that the \ion{He}{2} fraction in quasar
proximity zones is governed by a simple analytical equation
(eqn.~\ref{eqn:xHeIItqso}),
which describes the exponential time evolution
from an initial pre-quasar ionization state $x_{\rm HeII,0}$ to an
equilibrium value $x_{\rm HeII,eq}$. The enhanced photoionization
rate near the quasar $\Gamma_{\rm tot}$ sets both the timescale of the
exponential evolution $t_{\rm eq} = 1\slash \Gamma_{\rm tot}$, and the
equilibrium value attained $x_{\rm HeII,eq}\approx \alpha_{\rm A}n_{\rm e}t_{\rm eq}$. For very high \ion{He}{2} fractions $x_{\rm HeII,0}\simeq 1$, this exponential equilibration is delayed by the
time it takes the sub-luminal ionization front to arrive to a given location.

\subsection{Degeneracy between Quasar Lifetime $t_{\rm Q}$ and \ion{He}{2} Ionizing Background $\Gamma_{\rm HeII}^{\rm bkg}$}
\label{sec:DEG}

Previous work on \ion{H}{1} proximity zones at $z\sim 6$ have
pointed out that the quasar lifetime and ionization state of
the IGM (or equivalently the ionizing background) are degenerate in
determining the location of the ionization front $R_{\rm IF}$
\citep{Bolton2007a,Bolton2007b,Lidz2007,Bolton2012}, which is readily
apparent from the exponent in eqn.~(\ref{eqn:RIF}). Although, many
studies simply assume a fixed value for the quasar lifetime of $t_{Q}=10^7\,{\rm yr}$ when making inferences about the ionization
state of the IGM \citep[but see][for a more careful treatment]{Bolton2012}. This degeneracy between
lifetime and ionizing background is complicated by the fact that, at
$z \sim 6$ only lower limits on the hydrogen neutral fraction $x_{\rm
  HI}$ (upper limits on the background photoionization rate
$\Gamma_{\rm HI}^{\rm bkg}$) can be obtained from lower limits on the
GP absorption optical depth. The situation is further exacerbated if
there are significant spatial fluctuations in the background caused 
by foreground galaxies that may have `pre-ionized' the IGM \citep{Lidz2007,Bolton2007b,
  Wyithe2008}.

\begin{figure}[!t]
\centering
 \includegraphics[width=1.\linewidth]{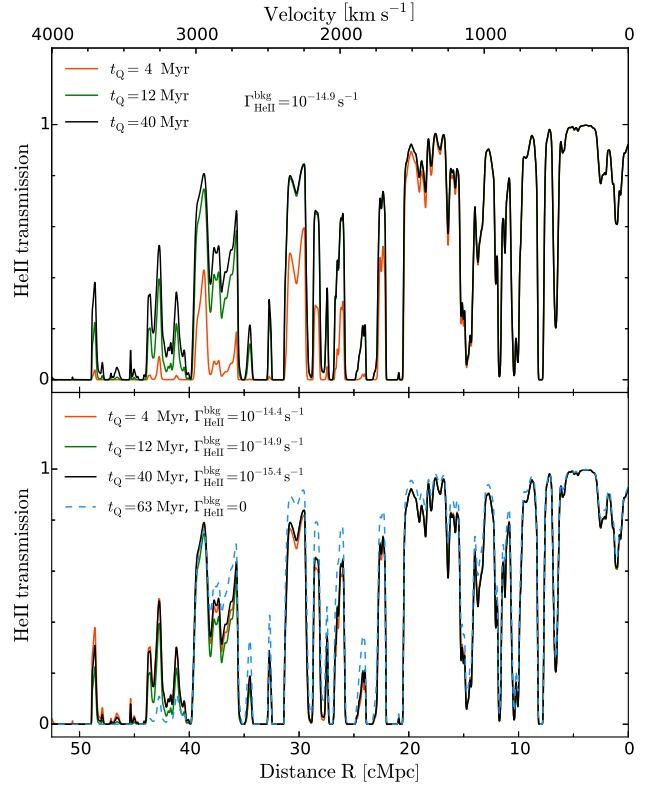}
 \caption{Illustration of the partial degeneracy between the \ion{He}{2} ionizing background and the quasar lifetime for the specified values of $\Gamma_{\rm HeII}^{\rm bkg}$ and $t_{\rm Q}$. \emph{The upper panel} shows three models with $t_{\rm Q}=4$~Myr (red), $t_{\rm Q}=12$~Myr (green) and $t_{\rm Q}=40$~Myr (black) with \ion{He}{2} background fixed to our fiducial value $\Gamma_{\rm HeII}^{\rm bkg} = 10^{-14.9}{\rm s^{-1}}$. \emph{The bottom panel} shows four models with the same values of quasar lifetime as before, but with different \ion{He}{2} backgrounds. The red curve corresponds to $t_{\rm Q}=4$~Myr and $\Gamma_{\rm HeII}^{\rm bkg} = 10^{-14.4} {\rm s^{-1}}$, whereas the solid black curve corresponds to $t_{\rm Q} = 40$~Myr and $\Gamma_{\rm HeII}^{\rm bkg} = 10^{-15.4} {\rm s^{-1}}$. The model with $t_{\rm Q} = 12$~Myr and $\Gamma_{\rm HeII}^{\rm bkg} = 10^{-14.9}{\rm s^{-1}}$ is shown by the green curve, and the dashed blue curve shows a $\Gamma_{\rm HeII}^{\rm bkg} = 0$ model with $t_{\rm Q} = 63$~Myr. The photon production rate is $Q_{\rm 4Ryd}=10^{56.1}{\rm s^{-1}}$ for all models.} 
 \label{fig:Degeneracy}
\end{figure}

An analogous degeneracy exists between $t_{\rm Q}$ and $x_{\rm HeII}$
for \ion{He}{2} proximity zones, as we illustrate in
Figure~\ref{fig:Degeneracy}. The upper panel shows three example
transmission spectra for the same skewer and value of \ion{He}{2}
background, but different values of quasar lifetime, which are
clearly distinguishable.
The situation changes if we
also allow the \ion{He}{2} background to vary, which is shown
in the bottom panel of Figure~\ref{fig:Degeneracy} where 
transmission spectra are plotted for the same skewer, but several distinct
combinations of lifetime and background. The nearly identical
resulting spectra indicate that the same degeneracy exists at $z
\simeq 3.1$ between the quasar lifetime $t_{\rm Q}$ and the value of
the \ion{He}{2} ionizing background $\Gamma_{\rm HeII}^{\rm bkg}$. In
what follows, we discuss this degeneracy in detail, aided by our
analytical model for the time evolution of the \ion{He}{2} fraction
from the previous section.

We can understand this degeneracy by rearranging eqn.~(\ref{eqn:xHeIItqso})
\begin{equation}\label{eqn:evol_sol_new}
  x_{\rm HeII}\left(t_{\rm Q}, R\right) = x_{\rm HeII, eq} \left[ 1 + \frac{\Gamma_{\rm QSO}}{\Gamma_{\rm HeII}^{\rm bkg}} {\rm e^{-\frac{t_{\rm Q}}{t_{\rm eq}\left(R\right)}}}  \right] 
\end{equation}
where for simplicity we have focused on the finite background case where the ionization
front time delay can be ignored. 
There are two regimes that are relevant to this degeneracy. First,
very near the quasar the attenuation of $\Gamma_{\rm QSO}$ is
negligible, implying that $t_{\rm eq}\propto \Gamma_{\rm QSO}^{-1} \propto R^2$ and given by
\begin{equation}
  t_{\rm eq} = 0.08\,\left(\frac{{Q_{\rm 4Ry}}}{10^{56}\ {\rm s^{-1}}}\right)^{-1}\left(\frac{R}{5\,{\rm cMpc}}\right)^2 \,{\rm Myr}. 
  \label{eqn:teq_lum_dist}
\end{equation}
At small distances ($R \lesssim 15$\ cMpc in Fig.~\ref{fig:Degeneracy}) in
the highly ionized `core' of the proximity zone, $t_{\rm eq} \ll t_{\rm Q}$
for the quasar lifetimes we consider, and eqn.~(\ref{eqn:evol_sol_new}) indicates that
the proximity zone structure depends only on the luminosity of the quasar, which determines
$x_{\rm HeII,eq}\approx \alpha_{\rm A} n_{\rm e}\slash \Gamma_{\rm QSO}$, but there is no sensitivity to either $t_{\rm Q}$ or $\Gamma_{\rm HeII}^{\rm bkg}$. 

Second, at larger distances the equilibration time grows as $t_{\rm
  eq}\propto R^2$ and will eventually be comparable to the quasar
lifetime. We define the equilbration distance $R_{\rm eq}$ to be the
location where $t_{\rm eq}(R_{\rm eq})\equiv t_{\rm Q}$, which gives
\begin{equation}
  R_{\rm eq} = 17\,\left(\frac{Q_{\rm 4Ry}}{10^{56}\ {\rm s^{-1}}}\right)^{1\slash 2} \left(\frac{t_{\rm Q}}{1\,{\rm Myr}}\right)^{1\slash 2}\,{\rm cMpc}, 
 \label{eqn:Req_lum_dist}
\end{equation}
where for simplicity we neglect the impact of attenuation of
$\Gamma_{\rm QSO}$. At distances comparable to the equilibration distance $R\sim R_{\rm eq}$, 
eqn.~(\ref{eqn:evol_sol_new})
indicates that the \ion{He}{2} fraction will be sensitive to
$t_{\rm Q}$. Note that at $R\sim R_{\rm eq}$ the quasar still dominates over
the background $\Gamma_{\rm QSO} \gg \Gamma_{\rm HeII}^{\rm bkg}$, such that
$x_{\rm HeII,eq}$ is still independent of $\Gamma_{\rm HeII}^{\rm bkg}$. One then
sees from eqn.~(\ref{eqn:evol_sol_new}) that for any change in
quasar lifetime $t_{\rm Q}$ one can always make a corresponding change to the
value of $\Gamma_{\rm HeII}^{\rm bkg}$ to yield the same value of $x_{\rm HeII}$. But note
that this degeneracy holds only at a single radius $R$, because
$t_{\rm eq} \left(R\right)$ is a function of $R$, whereas our
$\Gamma_{\rm HeII}^{\rm bkg}$ is assumed to be spatially
constant. Therefore, there is no way to choose a constant $\Gamma_{\rm
  HeII}^{\rm bkg}$ such that the $x_{\rm HeII} \left(t_{\rm Q},R
\right)$ matches different values of $t_{\rm Q}$ at all $R$. In
reality, $\Gamma_{\rm HeII}^{\rm bkg}$ will fluctuate spatially, but
it will not have the required dependence on $R$ to counteract the
lifetime dependence.

Similar arguments also apply when
$\Gamma_{\rm HeII}^{\rm bkg} = 0$, and the $x_{\rm HeII}(t,R)$
evolution is governed by eqn.~(\ref{eqn:evol_sol_corr}). In this case, the
time evolution depends only on the quasar lifetime $t_{\rm Q}$ and the
ionizing photon production rate of the quasar $Q_{\rm 4Ry}$.  Very
close to the quasar ($R \lesssim 15$~cMpc),
there is no sensitivity to quasar lifetime provided that $t_{\rm Q} - t_{\rm
  IF}(R) \gg t_{\rm eq}$, which is the case for the long quasar lifetime
model $t_{\rm Q}=63\,$~Myr shown in Figure~\ref{fig:Degeneracy} with
$\Gamma_{\rm HeII}^{\rm bkg}=0$ (dashed blue curve), which is indistinguishable
from the finite background models at small radii. Although note that
for much shorter quasar lifetimes $t_{\rm Q}\sim t_{\rm IF}\left(R\right)$ comparable to
the ionization front travel time, eqn.~(\ref{eqn:evol_sol_corr}) indicates one
may retain sensitivity to the quasar lifetime even in the core of the
zone. At larger distances $R \gtrsim 15$\,Mpc where $t_{\rm Q} -
t_{\rm IF}\left(R\right) \sim t_{\rm eq}$, the $\Gamma_{\rm HeII}^{\rm bkg}=0$ case becomes
sensitive to quasar lifetime according to eqn.~(\ref{eqn:evol_sol_corr}), but
Figure~\ref{fig:Degeneracy} still indicates that the transmission
profile is remarkably similar to the finite background case. In
principle $Q_{\rm 4 Ry}$ could be varied to produce a curve that
appears even more degenerate, but we do not explore that here (but see
the discussion in \S~\ref{sec:Gamma0}).

Finally, an obvious difference between the $\Gamma_{\rm HeII}^{\rm
  bkg}=0$ and finite background case is of course the transmission
level far from the quasar, which is zero for $\Gamma_{\rm HeII}^{\rm
  bkg}=0$, but corresponds to a finite value of $\tau_{\rm eff}$ for
$\Gamma_{\rm HeII}^{\rm bkg}\ne 0$ (see Figure~\ref{fig:F_tau}).  At
low redshifts $z\simeq 3$ where the effective optical depth can be
measured, this provides an independent constraint on $\Gamma_{\rm
  HeII}^{\rm bkg}$ which rules out a $\Gamma_{\rm HeII}^{\rm bkg}=0$
model. As discussed in \S~\ref{sec:bkg}, the enhanced sensitivity of
$\tau_{\rm eff}$ measurements to the \ion{He}{2} background level for
\ion{He}{2} GP absorption at $z\sim 3$, as compared to \ion{H}{1} GP
absorption at $z\sim 6-7$ (where only upper limits on the background
are available), constitutes an important difference between
\ion{He}{2} and \ion{H}{1} proximity zones, which can be leveraged to
break the degeneracy between the quasar lifetime $t_{\rm Q}$ and
\ion{He}{2} ionizing background $\Gamma_{\rm HeII}^{\rm bkg}$, as we
will elaborate on in the next section.

\section{Understanding the Structure of the Proximity Zone from Stacked Spectra}
\label{sec:xHeII}

\begin{figure}[!t]
\centering
 \includegraphics[width=1.0\linewidth]{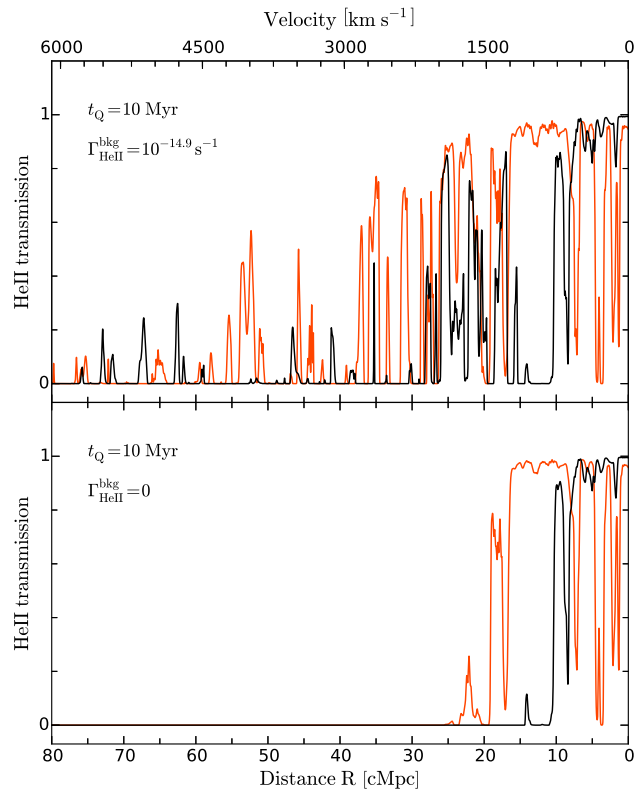}
\caption{\ion{He}{2} proximity region for two different density skewers and the same model parameters (\textit{top panel}: $t_{\rm Q} = 10$~Myr, $\Gamma_{\rm HeII}^{\rm bkg} = 10^{-14.9} {\rm s}^{-1}$, \textit{bottom panel}: $t_{\rm Q} = 10$~Myr, $\Gamma_{\rm HeII}^{\rm bkg} = 0$). Density fluctuations alter the proximity region profile even with other parameters fixed, acting as a source of uncertainty in our analysis.}
 \label{fig:Fluctuations}
\end{figure}

Density fluctuations in the intergalactic medium around the quasar
result in a significant variation in the proximity zone sizes for
individual sightlines, which complicates our ability to constrain any
parameters from quasar proximity regions. This effect is illustrated
in Figure~\ref{fig:Fluctuations}, where we show simulated transmission profiles
for the same model with values of quasar lifetime $t_{\rm Q} =
10$~Myr, \ion{He}{2} ionizing background of $\Gamma_{\rm HeII}^{\rm
  bkg}= 10^{-14.9}{\rm s^{-1}}$ and photon production rate of $Q_{\rm
  4 Ry} = 10^{56.1}{\rm s^{-1}}$, but for two skewers that have
different underlying density fields. The bottom panel shows the case
for the same quasar lifetime and photon production rate, but with the
\ion{He}{2} background set to zero. A nonzero background $\Gamma_{\rm HeII}^{\rm bkg}$ results in significantly more transmission far from the quasar, and concomitant
sightline-to-sightline scatter, increasing the ambiguity in
determining the edge of the \ion{He}{2} proximity zone \citep[see also discussion in][]{Bolton2007a,Bolton2007b,Lidz2007}.

\begin{figure*}[!t]
\centering
 \includegraphics[width=0.8\linewidth]{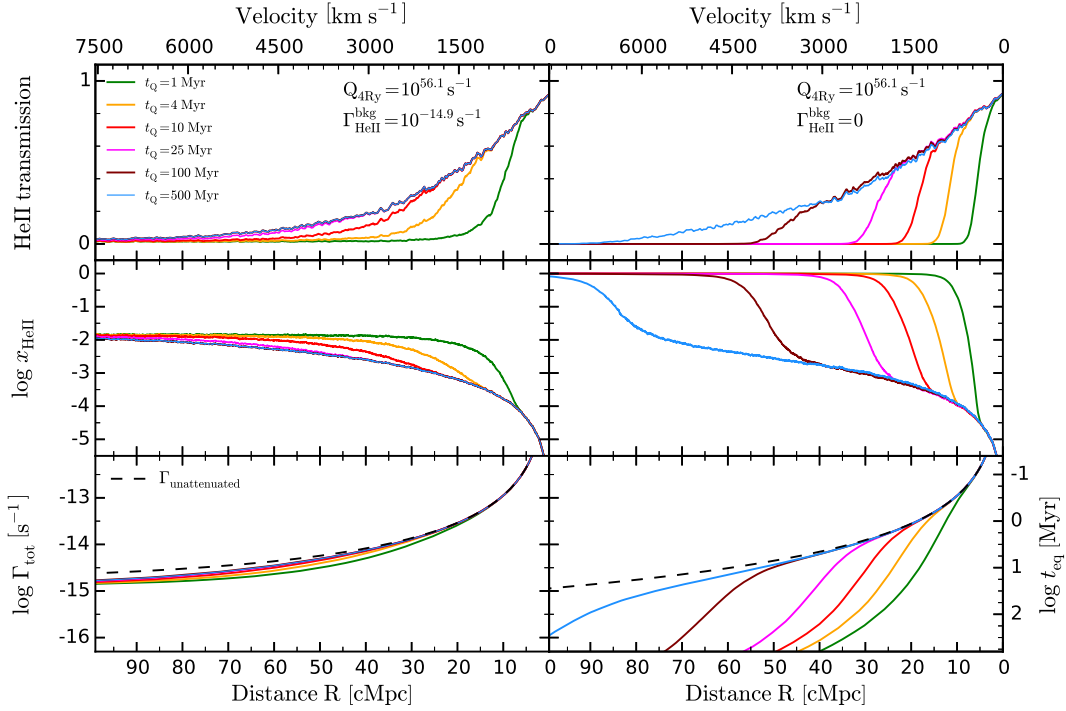}
\caption{Results of the radiative transfer simulations. Three panels show (from top to bottom): a) transmission profiles for different values of quasar lifetime and fixed values of $\Gamma_{\rm HeII}^{\rm bkg}=10^{-14.9}{\rm s^{-1}}$ and $Q_{\rm 4Ry}=10^{56.1}{\rm s^{-1}}$; b) evolution of the median \ion{He}{2} fraction $x_{\rm HeII}$ for each of the models; c) total photoionization rate $\Gamma_{\rm tot}$ (dashed line corresponds to the total unattenuated photoionization rate). Left side panels show models with \ion{He}{2} ionizing background $\Gamma_{\rm HeII}^{\rm bkg} = 10^{-14.9}{\rm s}^{-1}$, while the right side panels show models with zero background, i.e., $\Gamma_{\rm HeII}^{\rm bkg} = 0$. The equilibration timescale $t_{\rm eq}$ is plotted on the right $y$-axis in the bottom panel.}
\label{fig:StackLIFE}
\end{figure*}

One approach to mitigate the impact of these density fluctuations, is
to average them down by stacking different \ion{He}{2} proximity
regions, using potentially all $\sim 30$ \ion{He}{2} Ly$\alpha$ forest
sightlines observed to date \citep{Worseck2011, Syphers2012}. From the
perspective of the modeling, this also helps to isolate the salient
dependencies of the mean transmission profile on the model
parameters. In this section, we analyze stacked \ion{He}{2} Ly$\alpha$ profiles
and study their dependence on the three parameters that govern the structure
of the proximity zones: the quasar lifetime $t_{\rm Q}$, the \ion{He}{2} ionizing
background $\Gamma^{\rm bkg}_{\rm HeII}$, and the quasar photon production rate $Q_{\rm 4Ry}$.

\subsection{The Dependence on Quasar Lifetime $t_{\rm Q}$}
\label{sec:time}

In Figure~\ref{fig:StackLIFE} we show stacks of $1000$ skewers for a
sequence of proximity zone models with different quasar lifetimes in
the range $t_{\rm Q}=1-500$\,Myr, indicated by the colored curves, but
with other parameters ($\Gamma_{\rm HeII}^{\rm bkg}$ and $Q_{\rm
  4Ry}$) fixed. The panels show, from top to bottom, the stacked
transmission profiles in \ion{He}{2} Ly$\alpha$ region, the
\ion{He}{2} fraction $x_{\rm HeII}$, and the total photoionization
rate $\Gamma_{\rm tot} = \Gamma_{\rm QSO} + \Gamma_{\rm HeII}^{\rm
  bkg}$ together with the unattenuated photoionization rate
$\Gamma_{\rm unatt.}$.  The panels on the left show the models with
fixed values of $\Gamma_{\rm HeII}^{\rm bkg} = 10^{-14.9} {\rm
  s^{-1}}$, whereas the right side illustrates the $\Gamma_{\rm
  HeII}^{\rm bkg} = 0$ case.  The photon production rate has been set
to fiducial value $Q_{\rm 4 Ry}=10^{56.1}{\rm s^{-1}}$ throughout. 

Several qualitative trends are readily apparent from
Figure~\ref{fig:StackLIFE}. First, as was also mentioned in
\S~\ref{sec:DEG}, at the smallest radii $R \lesssim 5$~cMpc, there is
a `core' of the proximity zone, which is insensitive to the changes of the
quasar lifetime such that all transmission profiles overlap. Second, it
is clear that both with ($\Gamma_{\rm HeII}^{\rm bkg}=10^{-14.9}{\rm
  s}^{-1}$) and without ($\Gamma_{\rm HeII}^{\rm bkg}=0$) a
\ion{He}{2} ionizing background, increasing the quasar lifetime $t_{\rm
  Q}$ results in larger proximity zones, reflecting the longer time
that the nearby IGM has been exposed to the radiation from the
quasar. Third, in the presence of a \ion{He}{2} ionizing background
$\Gamma_{\rm HeII}^{\rm bkg}=10^{-14.9}{\rm s}^{-1}$, the transmission
profile shape loses sensitivity to quasar lifetime for models with
$t_{\rm Q} > 25$~Myr, whereas for $\Gamma_{\rm HeII}^{\rm bkg} =0$
case, the proximity zone continues to grow with increasing quasar
lifetime up to large values of $t_{\rm Q}=500$\,Myr.

We can gain a better physical understanding of the origin of these
trends from the equation that describes the time evolution
of the \ion{He}{2} fraction $x_{\rm HeII}\left( t \right)$ given by
eqn.~(\ref{eqn:xHeIItqso}) and discussed in \S~\ref{sec:time_evol}
and \S~\ref{sec:DEG}.  In
what follows we focus on the specific example $\Gamma_{\rm HeII}^{\rm
  bkg}=10^{-14.9}{\rm s}^{-1}$, but our discussion also applies to the
case of zero ionizing background, provided that the equation for the
time evolution is modified to account for the time-delay associated
with the arrival of the ionization front (see eqn.~(\ref{eqn:evol_sol_corr}) in
\S~\ref{sec:time_evol}).  At $z\sim
3$, observations of the effective optical depth strongly favor a finite
background (see \S~\ref{sec:bkg}) $\Gamma_{\rm HeII}^{\rm bkg}\simeq 10^{-14.9}{\rm s}^{-1}$
($x_{\rm HeII} \simeq 10^{-2}$), although most of the previous
interpretations of observed \ion{He}{2} proximity regions have
concentrated on the $\Gamma_{\rm HeII}^{\rm bkg}=0$ case
\citep{Shull2010,Syphers2014,Zheng2015}.

\begin{figure*}[!t]
\centering
 \includegraphics[width=\textwidth]{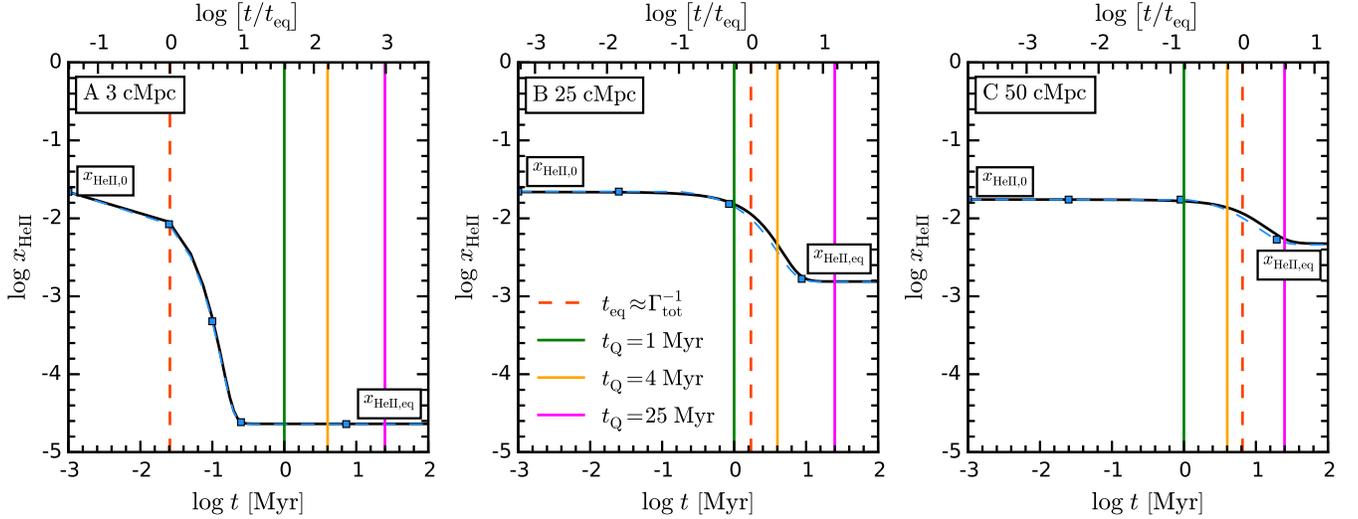}
 \caption{Time evolution of $x_{\rm HeII}\left(t,R\right)$ at
   different distances from the quasar: A) $3$~cMpc, B) $25$~cMpc, and C)
   $50$~cMpc. Solid black and dashed blue curves show $x_{\rm HeII}\left(t,R\right)$ evolution as output from the radiative
   transfer algorithm and analytical approximation based on
   eqn.~(\ref{eqn:xHeIItqso}), respectively. Red dashed vertical lines in
   each panel show the equilibration time $t_{\rm eq}$ at the distance
   $R$; green, orange and magenta vertical lines indicate three quasar
   lifetimes we consider here, i.e., $t_{\rm Q} = \left[1, 4, 25
     \right]$~Myr. These calculations use our fiducial value of the
   \ion{He}{2} ionizing background $\Gamma_{\rm HeII}^{\rm bkg} =
   10^{-14.9}{\rm s}^{-1}$, and have a photon production rate $Q_{\rm
     4Ry} = 10^{56.1}\, {\rm s^{-1}}$.}
 \label{fig:F_xHe1}
\end{figure*}

\begin{figure}[!b]
\centering
 \includegraphics[width=1.0\linewidth]{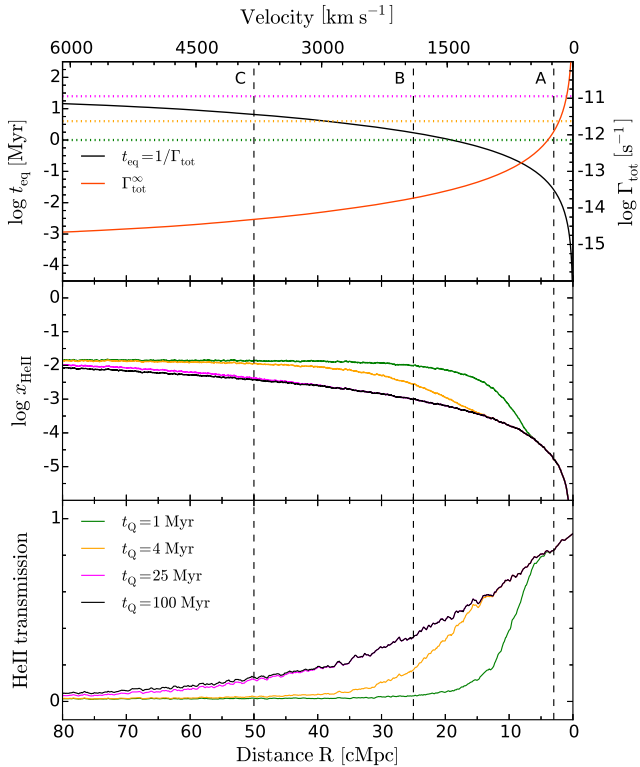}
\caption{ {\it Top panel}: total photoionization rate $\Gamma_{\rm tot}$ as a function of distance $R$ (red) and the equilibration time $t_{\rm eq} = 1 \slash \Gamma_{\rm tot}$ (black). Dashed vertical lines correspond to the locations A,B and C in Figure~\ref{fig:F_xHe1} where we traced the time-evolution of the \ion{He}{2} fraction. Three horizontal dotted lines in the upper panel correspond to the values of quasar lifetimes $t_{\rm Q}$ that we consider in Figure~\ref{fig:F_xHe1}. {\it Middle panel}: time evolution of the \ion{He}{2} fraction computed at $t = t_{\rm Q} = \left[1, 4, 25, 100 \right]\ {\rm Myr}$ as a function of distance $R$ from the quasar. {\it Bottom panel}: stacked transmission profiles for models with different quasar lifetimes (same as Figure~\ref{fig:StackLIFE}).}
 \label{fig:F_xHe2}
\end{figure}

The three panels in Figure~\ref{fig:F_xHe1} show the time evolution of
the \ion{He}{2} fraction at three different distances from the quasar
$R = \left[ 3, 25, 50 \right]$~cMpc, labeled A, B, and
C, respectively. The black curves show the average $x_{\rm HeII}\left( t \right)$ computed from
$100$ skewers, where the quasar was on for the entire $t_{\rm Q} =
100$~Myr that is shown. The dashed blue curves show the time
evolution from eqn.~(\ref{eqn:xHeIItqso}), where for the input parameters
we have averaged the outputs from the radiative transfer code.
Specifically, to compute the blue curves we take an average value of
photoionization rate $\langle \Gamma_{\rm tot} \rangle$ which is a
mean of 100 skewers and we do the same for the equilibrium $x_{\rm
  HeII,eq} = \langle \alpha_{\rm A}\left( {\rm T} \right) n_{\rm
  e}\rangle \slash \langle \Gamma_{\rm tot}\rangle$,  and initial
neutral fractions $x_{\rm HeII,0} = \langle \alpha_{\rm A}\left( {\rm
  T} \right) n_{\rm e}\rangle \slash \Gamma_{\rm HeII}^{\rm bkg}$
\footnote{Note, however, that taking the average values of $\langle
  \alpha_{\rm A} \left( {\rm T} \right) \rangle$ and $\langle n_{\rm
    e} \rangle$ separately does not reproduce the time integrated
  results from the radiative transfer because these two quantities are
  highly correlated due to the temperature-density relation and thus
  $\langle \alpha_{\rm A} \left( {\rm T} \right) n_{\rm e} \rangle
  \neq \langle \alpha_{\rm A} \left( {\rm T} \right) \rangle \langle
  n_{\rm e} \rangle$.}.

This procedure excellently reproduces the evolution given by the solid black curves, computed
from a full time integration of the radiative transfer.  Whereas we previously saw that this
analytical approximation provides a good fit to the time evolution of
the \ion{He}{2} fraction of a single skewer (see Figure~\ref{fig:FxHeII}),
it is somewhat surprising that it also works so well for the stacked
spectra using these averaged quantities.

As Figure~\ref{fig:F_xHe1} shows the full time evolution of
\ion{He}{2} fraction over $100$~Myr, observing a quasar with a given
quasar lifetime $t_{\rm Q}$ is equivalent to evaluating $x_{\rm He II}\left( t \right)$ at
the time $t=t_{\rm Q}$. The green, yellow, and magenta vertical lines
indicate three possible quasar lifetimes of $t_{\rm Q} = 1,\ 4,\ {\rm and}\ 25$~Myr, respectively.  The spatial profile of $x_{\rm HeII}$ and Ly$\alpha$ transmission are shown in the middle and
bottom panels of Figure~\ref{fig:F_xHe2} for the same three quasar
lifetime models (with the same line colors as in Figure~\ref{fig:F_xHe1}). The black curves in these two panels show the ``equilibrium'' $t=\infty$ profiles for $x_{\rm HeII}$ and the transmission, which we define to correspond to $t_{\rm Q}=100\,{\rm Myr}$, at which time $x_{\rm HeII}$ has fully equlibrated. The
vertical dashed lines labeled A, B, and C in Figure~\ref{fig:F_xHe2} indicate the three distances from the quasar $R = \left[ 3, 25, 50 \right]$~cMpc for which the time evolution is shown in Figure~\ref{fig:F_xHe1}.

First, consider location A in the inner `core' of the proximity zone
at a distance of $R = 3$~cMpc, at which the \ion{He}{2} fraction and
transmission profiles in Figure~\ref{fig:F_xHe2} are all identical for
the quasar lifetimes we consider. The uppermost panel of
Figure~\ref{fig:F_xHe2} shows the equilibration time as a function
of distance from the quasar, which indicates that at $R=3\,{\rm cMpc}$
$t_{\rm eq} = 0.025$\ Myr. The reason why all quasar lifetimes are
indistinguishable at this distance can be easily understood from the
$x_{\rm HeII}\left( t \right)$ evolution in the left panel of
Figure~\ref{fig:F_xHe1}. Irrespective of whether the quasar has been
emitting for $t_{\rm Q} = 1$~Myr (green), $t_{\rm Q} = 4$~Myr (yellow)
or $t_{\rm Q} = 16$~Myr (magenta), because at this distance the
equilibration time (red vertical dashed line) $t_{\rm eq} \ll t_{\rm
  Q}$, the IGM has already equilibrated $x_{\rm HeII,eq} \simeq 2.5\times
10^{-5}$, and there is thus no sensitivity to quasar lifetime (see
also the discussion in \S~\ref{sec:DEG} and eqn.~\ref{eqn:evol_sol_new}).

Note however that the $t_{\rm eq} \sim R^2$ dependence of equilibration time shown
in the top panel of Figure~\ref{fig:F_xHe2} indicates that at greater
distances, the equilibration time is larger and becomes comparable to
the lifetimes we consider. This manifests as significant differences in
the stacked $x_{\rm HeII}$ and transmission profiles for different
lifetimes in Figure~\ref{fig:F_xHe2}, which can again be understood
from the time evolution in Figure~\ref{fig:F_xHe1}. For example, consider
location B (middle panel of Figure~\ref{fig:F_xHe1}) at $R = 25$~cMpc,
at which the equilibration time is $t_{\rm eq} = 1.7$~Myr (red
vertical dashed line). For the shortest quasar lifetime of $1$~Myr
(green line), the IGM has not been illuminated long enough to
equilibrate, i.e., $t_{\rm Q} < t_{\rm eq}$, and thus still reflects the
\ion{He}{2} fraction $x_{\rm HeII,0}$ consistent with the $\Gamma_{\rm
  HeII}^{\rm bkg}$ that prevailed before the quasar turned on. This
$x_{\rm HeII,0} \simeq 2.2\times 10^{-2}$, much larger
than the equilibrium value $x_{\rm HeII,eq} \simeq 1.5\times 10^{-3}$,
explains why the corresponding transmission at location B in
Figure~\ref{fig:F_xHe2} (green curve) lies far below the fully
equilibrated model $t_{\rm Q}=100$~Myr (black curve).
Likewise, the $t_{\rm Q} = 4$~Myr model (yellow) is still in
the process of equilibriating, whereas the $t_{\rm Q} = 25$~Myr model
(magenta) has fully equilibrated, explaining the respective values
of the $x_{\rm HeII}$ and transmission for these models at location B in Figure~\ref{fig:F_xHe2}.

\begin{figure*}
\centering
 \includegraphics[width=0.8\linewidth]{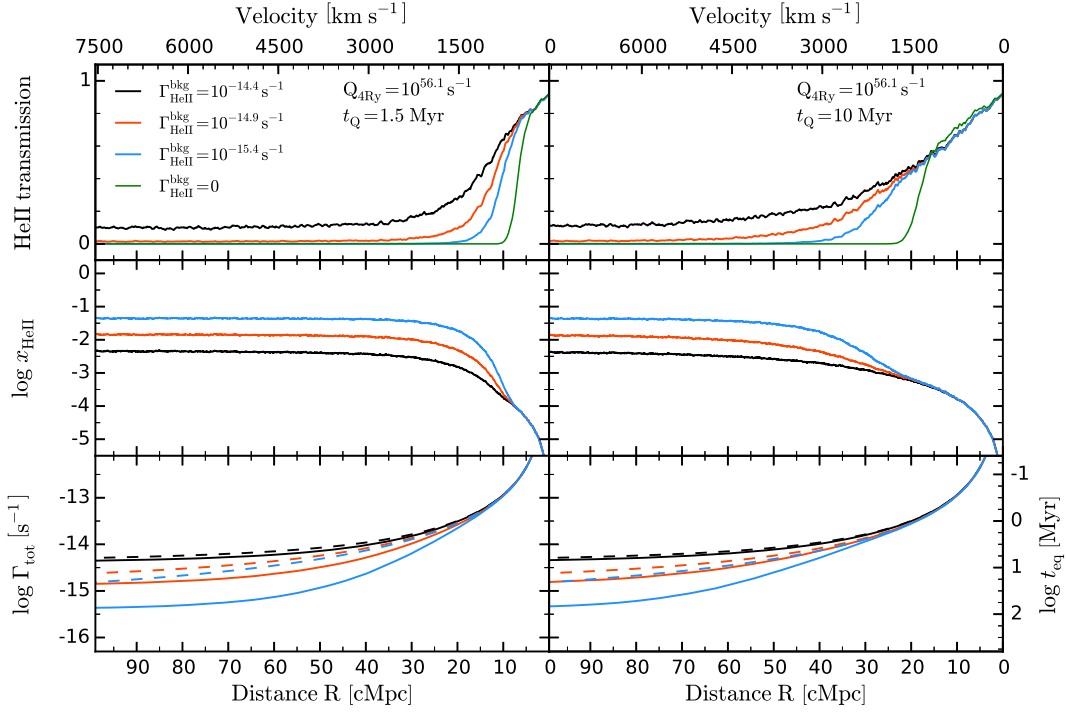}
\caption{Stacked proximity region trends when varying the \ion{He}{2} ionizing background, where each curve is calculated from $1000$ skewers. Lefthand panels show three models with $\Gamma_{\rm HeII}^{\rm bkg} = \left[ 10^{-14.4}, 10^{-14.9}, 10^{-15.4}, 0 \right] \ {\rm s}^{-1}$ for a quasar lifetime $t_{\rm Q} = 1.5$~Myr. Righthand panels show the same models of $\Gamma_{\rm HeII}^{\rm bkg}$ but for the quasar lifetime $t_{\rm Q} = 10$~Myr.  All panels take a photon production rate of $Q_{\rm 4 Ry} = 10^{56.1}{\rm s^{-1}}$. See Figure~\ref{fig:StackLIFE} for a description of the quantities plotted on the $y$-axis.}
 \label{fig:F_BKG}
\end{figure*}

\begin{figure*}
\centering
 \includegraphics[width=0.8\linewidth]{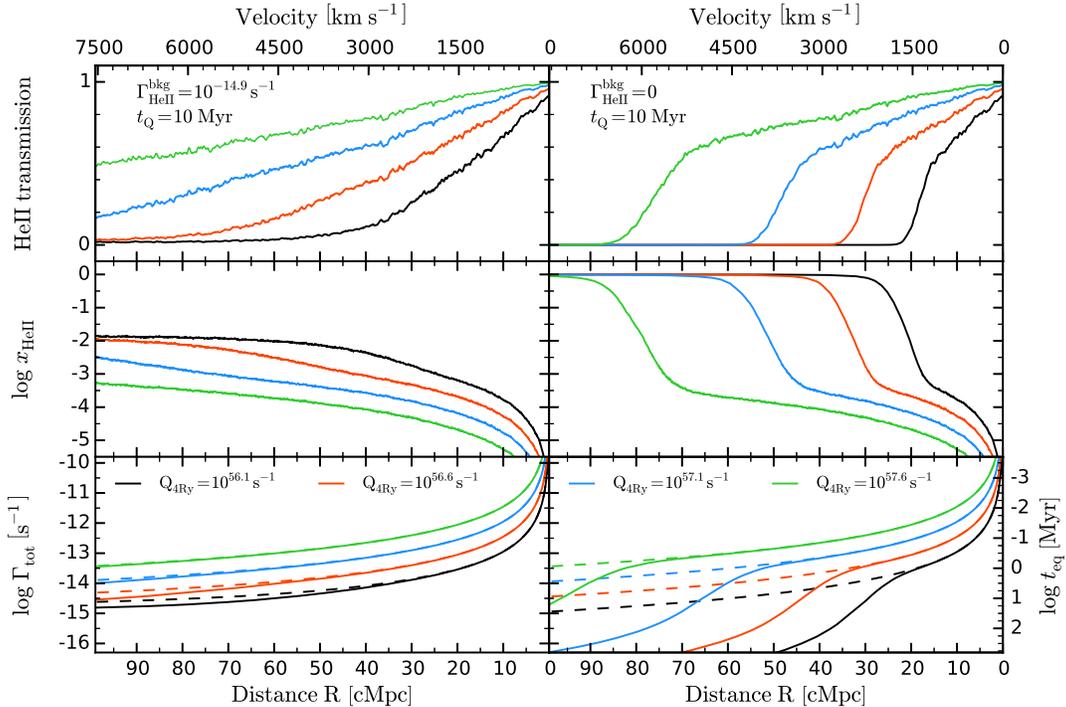}
\caption{Stacked proximity region trends with varying the quasar $4~$Ry photon production rate, ${\rm Q_{4Ry}}$, where each curve is calculated from $1000$ skewers. All curves assume a fixed quasar lifetime of $t_{\rm Q} = 10$~Myr and fixed \ion{He}{2} ionizing background $\Gamma_{\rm HeII}^{\rm bkg}$ where the lefthand panels take $\Gamma_{\rm HeII}^{\rm bkg} = 10^{-14.9}{\rm s}^{-1}$ and the righthand panels take $\Gamma_{\rm HeII}^{\rm bkg} = 0$ (i.e., $x_{\rm HeII}=1$). See Figure~\ref{fig:StackLIFE} for a description of the quantities plotted on the $y$-axis.} 
 \label{fig:StackLUMI}
\end{figure*}

Because equilibration time increases with distance from the quasar,
the stacked transmission profile becomes sensitive to larger quasar
lifetimes at larger radii. This is evident from the transmission
profiles in Figure~\ref{fig:StackLIFE} and Figure~\ref{fig:F_xHe2},
where models with progressively larger lifetimes peel off from the
equilibrium model ($t_{\rm Q}=100$ Myr) at progressively larger
radii. However, eventually far from the quasar, this sensitivity
saturates, as $t_{\rm eq}$ approaches its asymptotic value $t_{\rm eq}
\sim 1 \slash \Gamma_{\rm HeII}^{\rm bkg}$ (see upper panel of
Figure~\ref{fig:F_xHe2}), which for our fiducial $\Gamma_{\rm
  HeII}^{\rm bkg} = 10^{-14.9}{\rm s}^{-1}$ corresponds to $t_{\rm
  eq}=25$\ Myr. This saturation effect is illustrated by the time
evolution at location C, $R = 50$~cMpc from the quasar, in the right
panel of Figure~\ref{fig:F_xHe1}.  If the quasar has been illuminating
the IGM for $t_{\rm Q} = 25$~Myr (magenta curve), the \ion{He}{2} fraction has nearly
reached equilibrium $x_{\rm HeII,eq}$, and thus at location C
Figure~\ref{fig:F_xHe2} exhibits only a small but still noticeable
difference between the $x_{\rm HeII}$ at $t_{\rm Q}=25$~Myr and the
equilibrium model (black curve), and consequently the transmission
profiles (lower panel) hardly differ at all. It would clearly be
extremely challenging to distinguish between different models with
$t_{\rm Q} > 25$~Myr. Finally, at the largest radii $R>R_{\rm bkg} =
70$~cMpc, where $R_{\rm bkg}$ is defined to be the location where
$\Gamma_{\rm HeII}^{\rm bkg} = \Gamma_{\rm QSO}$, the quasar no longer
dominates over the background, and all the transmission profiles
converge to the mean transmission set by the \ion{He}{2}
background. Finally, we again note that if $\Gamma_{\rm HeII}^{\rm
  bkg} =0$, there is no such asymptote in the equilibration time, and
the transmission profile continues to be sensitive to values of quasar
lifetime as large as $t_{\rm Q}=500$\,Myr as shown in
Figure~\ref{fig:StackLIFE}.  However, in practice for very long quasar
lifetimes and therefore very large proximity zones, one might
eventually encounter locations in the universe where the background is
no longer zero.

\subsection{The Dependence on the \ion{He}{2} background $\Gamma_{\rm HeII}^{\rm bkg}$}

In Figure~\ref{fig:F_BKG} we illustrate the impact of varying the
\ion{He}{2} ionizing background $\Gamma_{\rm HeII}^{\rm bkg}$ on the
stacked transmission profile, with $t_{\rm Q}$ and $Q_{\rm 4Ry}$ held
fixed. The left panels show a quasar lifetime of $t_{\rm Q} =
1.5$~Myr, whereas the right show $t_{\rm Q} = 10$~Myr. Four different
values for $\Gamma_{\rm HeII}^{\rm bkg}$ are plotted, including the
$\Gamma_{\rm HeII}^{\rm bkg}=0$ case.

From Figure~\ref{fig:F_BKG}, we see that in the inner core $R <
5\,{\rm cMpc}$ of the proximity zone, the transmission profile is
independent of $\Gamma_{\rm HeII}^{\rm bkg}$, analogous to the
behavior in Figure~\ref{fig:StackLIFE}, where we saw that the core is
also independent of $t_{\rm Q}$.  As discussed in \S~\ref{sec:DEG}
(see also the previous section), this insensitivity to $\Gamma_{\rm
  HeII}^{\rm bkg}$ and $t_{\rm Q}$ can be understood from
eqn.~(\ref{eqn:evol_sol_new}) governing the time evolution of $x_{\rm
  HeII}$.  For $t_{\rm eq} \ll t_{\rm Q}$, the IGM has already
equilibrated and $x_{\rm HeII}(t_{\rm Q}) \approx x_{\rm HeII,eq}$. At
small distances $R \ll R_{\rm bkg}$ the quasar dominates over the
background $\Gamma_{\rm QSO} \gg \Gamma_{\rm HeII}^{\rm bkg}$ and attenuation is negligible, 
hence the equilibrium \ion{He}{2} fraction $x_{\rm HeII,eq}\propto \Gamma_{\rm QSO}^{-1}$ is determined by
the quasar photon production rate alone, and is independent of the background
and quasar lifetime.

In the previous section we argued that the equilibration time picture
explains why the transmission profiles for progressively larger
lifetimes peel off from the equilibrium model ($t_{\rm Q}=100$~Myr) at
progressively larger radii (see Figure~\ref{fig:StackLIFE}). The
curves in Figure~\ref{fig:F_BKG} illustrate that varying the ionizing
background has a different effect, namely to change the slope of the
transmission profile about this peel off point, as well as to
determine the transmission level far from quasar. The different
response of the transmission profile to these two parameters, $t_{\rm
  Q}$ and $\Gamma_{\rm HeII}^{\rm bkg}$, results from the functional form of
eqn.~(\ref{eqn:evol_sol_new}). The stronger peel off behavior with $t_{\rm
  Q}$ is due to the exponential dependence on the quasar lifetime
$t_{\rm Q}$ in eqn.~(\ref{eqn:evol_sol_new}), whereas the milder variation
of the slope with $\Gamma_{\rm HeII}^{\rm bkg}$ arises because of the
inverse proportionality of $x_{\rm HeII,eq}$ on $\Gamma_{\rm
  HeII}^{\rm bkg}$. Furthermore, at large distances from the quasar
where $\Gamma_{\rm tot}\simeq \Gamma_{\rm HeII}^{\rm bkg}$, $x_{\rm HeII,eq}$
approaches $x_{\rm HeII,0}\propto 1\slash \Gamma_{\rm HeII}^{\rm bkg}$ and the
background sets the absorption level as expected.  As we also noted
in \S~\ref{sec:DEG}, the different dependence of the
transmission profile on $t_{\rm Q}$ and $\Gamma_{\rm HeII}^{\rm bkg}$ suggests that the
degeneracy between these parameters could be broken by the
shape of the transmission profile, which we discuss further in \S~\ref{sec:Gamma0}.

\subsection{The Dependence on the Photon Production Rate $Q_{\rm 4Ry}$}
\label{sec:Q_ped}

Figure~\ref{fig:StackLUMI} shows the effect of the photon production
rate on the structure of the proximity zone.  The quasar lifetime has
been set to the value $t_{\rm Q} = 10$~Myr, and the background is
$\Gamma_{\rm HeII}^{\rm bkg} = 10^{-14.9}{\rm s}^{-1}$ on the left and
$\Gamma_{\rm HeII}^{\rm bkg} = 0$ on the right. One can see that the
impact of the photon production rate $Q_{\rm 4Ry}$ on the resulting
transmission profile is twofold. First, as expected, more luminous
quasars produce larger proximity zones, i.e., increasing the photon
production rate by $0.5$~dex expands the characteristic size of the
proximity zone by a factor of $\sim 2$. Second, besides increasing
the overall size, the slope of the stacked transmission profile
becomes shallower when $Q_{\rm 4Ry}$ is increased.

The increase in proximity zone size with $Q_{\rm 4Ry}$ can be easily
understood in the equilibration time picture. From
eqn.~(\ref{eqn:Req_lum_dist}), we see that the equilibration distance
$R_{\rm eq}$, which sets the location where the transmission profile
becomes sensitive to $t_{\rm Q}$ and approaches the mean transmission
level of the IGM (see Figure~\ref{fig:StackLIFE}), scales as $R_{\rm eq}\propto Q_{\rm 4Ry}^{1/2}$.
Hence an increase (decrease) in photon production rate $Q_{\rm 4Ry}$ results in a
larger (smaller) characteristic size of the proximity zone.

To understand the change in transmission profile slope with $Q_{\rm
  4Ry}$, consider that in the core of the proximity zone where $t_{\rm
  eq}\ll t_{\rm Q}$, the \ion{He}{2} fraction has equilibrated and is
given by $x_{\rm HeII}\left( t \right)=x_{\rm HeII,eq}$, where $x_{\rm HeII,eq}
\propto Q_{\rm 4Ry}^{-1}$. As the transmitted flux is just the exponential
of a constant times $x_{\rm HeII,eq}$, an increase (decrease) in $Q_{\rm 4Ry}$ makes this exponent smaller (larger) and thus the slope of the transmission profile becomes shallower (steeper).   

\subsection{The Dependence on the spectral slope $\alpha$}

\begin{figure}[!b]
\centering
 \includegraphics[width=1\linewidth]{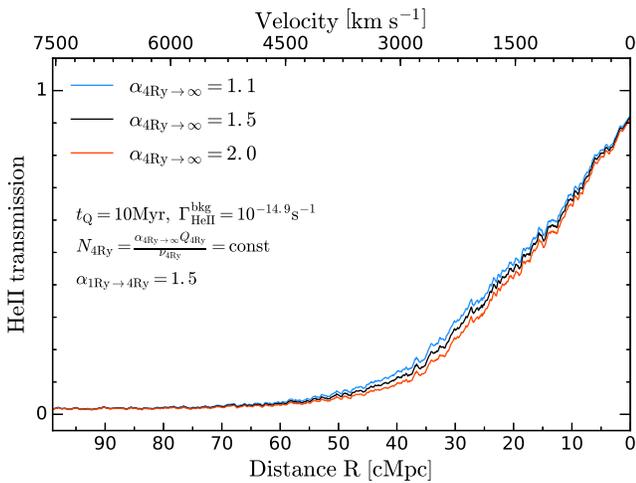}
  \caption{Stacked proximity region trends with varying the slope of quasar SED above $4$~Ry, $\alpha_{\rm 4Ry \rightarrow \infty}$, where each curve is calculated from $1000$ skewers. All curves assume a fixed quasar lifetime of $t_{\rm Q} = 10$~Myr and fixed \ion{He}{2} ionizing background $\Gamma_{\rm HeII}^{\rm bkg} = 10^{-14.9}{\rm s}^{-1}$.}
 \label{fig:F_alpha}
\end{figure}

As we illustrated in \S~\ref{sec:RT}, the total quasar photon production rate $Q_{\rm 4Ry}$ at frequencies above the \ion{He}{2} ionization threshold depends on the assumed slope of the quasar SED $\alpha$ blueward of $4$~Ry. Throughout the paper we have assumed that $\alpha = 1.5$ is constant at all frequencies blueward of $1$~Ry and simply varied the value of $Q_{\rm 4Ry}$ to illustrate its effect on the stacked transmission profiles (see \S~\ref{sec:Q_ped}). However, the spectral slope regulates the number of hard photons with long mean free path which can affect the transmission profiles at large distances from the quasar. Therefore, it is not obvious whether $Q_{\rm 4Ry}$ and $\alpha$ impact the structure of the proximity zone in the same way.

We thus run a set of radiative transfer simulations to further explore this question. In order to disentangle the effects of $Q_{\rm 4Ry}$ and $\alpha$ on the transmission profiles we chose to fix the quasar specific luminosity $N_{\rm 4Ry}$ at $4$~Ry (see eqn.~\ref{eqn:1}), which can be deduced from the observable quasar specific luminosity at the \ion{H}{1} ionization threshold of $1$~Ry by scaling it down to \ion{He}{2} ionization threshold with constant spectral slope $\alpha_{\rm 1Ry \rightarrow 4Ry} = 1.5$. We then freely vary the spectral slop $\alpha_{\rm 4Ry \rightarrow \infty}$ within the range $\alpha_{\rm 4Ry \rightarrow \infty} = 1.1 - 2.0$. Although, the actual slope of quasar SED at $\lambda \leq 912 {\rm \AA}$ is not currently well known, the chosen range of $\alpha_{\rm 4Ry \rightarrow \infty}$ values is motivated by the constraints on the power-law index from \citet{Lusso2015}, who found $\alpha_{\nu} = 1.70 \pm 0.61$ at $\lambda \leq 912 {\rm \AA}$.
  
Figure~\ref{fig:F_alpha} illustrates the effect of the spectral slope on the structure of the proximity zone. The quasar lifetime and \ion{He}{2} background are fixed at $t_{\rm Q} = 10$~Myr and $\Gamma_{\rm HeII}^{\rm bkg} = 10^{-14.9}{\rm s^{-1}}$, respectively. Three curves show stacked transmission profiles for our fiducial value $\alpha_{\rm 4Ry \rightarrow \infty} = 1.5$ (black), harder slope $\alpha_{\rm 4Ry \rightarrow \infty} = 1.1$ (blue), and softer slope $\alpha_{\rm 4Ry \rightarrow \infty} = 2.0$ (red). It is apparent from Figure~\ref{fig:F_alpha} that the effect of the spectral slope on the structure of the proximity zone is negligible in comparison to the effect of the overall photon production rate $Q_{\rm 4Ry}$ (see Figure~\ref{fig:StackLUMI}).

Consider that when specific luminosity $N_{\rm 4Ry}$ is fixed
  at $4 {\rm Ry}$, the quasar photon production rate $Q_{\rm 4Ry}$
  scales with $\alpha_{\rm 4Ry \rightarrow \infty}$ as $Q_{\rm 4Ry}
  \propto \alpha_{\rm 4Ry \rightarrow \infty}^{-1}$ (see
  eqn.~\ref{eqn:1}). Thus, varying the value of $\alpha_{\rm 4Ry \rightarrow \infty}$, results
  in $25-45\%$ changes in the photon production rate $Q_{\rm 4Ry}$
  (which is the ratio of assumed spectral slopes $\alpha_{\rm 4Ry
    \rightarrow \infty}$ in different models).  Thus, one would expect
  a significant difference in the resulting transmission according to
  Figure~\ref{fig:StackLUMI}. However, the more important quantity here is
  the cross-section weighted quasar \ion{He}{2} photoionization rate
  $\Gamma_{\rm QSO}$ given by eqn.~(\ref{eqn:gammaqso}), which
  regulates the time-evolution of the \ion{He}{2} fraction (and,
  consequently, the transmission profile). Since the ionization
  cross-section scales as $\sigma_{\nu} \propto \left( \nu \slash
  \nu_{\rm th} \right)^{-3}$, the resulting dependence of the
  \ion{He}{2} photoionization rate in the unattenuated limit on the
  spectral slope is $\Gamma_{\rm QSO} \propto \left( \alpha_{\rm 4Ry
    \rightarrow \infty}+3 \right)^{-1}$. Hence, the difference between
  the photoionization rates of the fiducial model $\alpha_{\rm 4Ry \rightarrow \infty} = 1.5$
  and the two models in Figure~\ref{fig:F_alpha} with $\alpha_{\rm 4Ry \rightarrow \infty} =
  1.1$ and $\alpha_{\rm 4Ry \rightarrow \infty} = 2.0$ is only $\Delta
  \Gamma \approx 10\%$. Therefore, there is no significant effect in
  the resulting transmission in \ion{He}{2} proximity zone even for
  relatively large variations in the assumed spectral slope of quasar
  SED blueward of $4$~Ry.

\section{Discussion} 
\label{sec:discussion}

In the previous section we assumed only single values of quasar lifetime and \ion{He}{2} background when exploring their impact on the stacked transmission profiles. However, this approach is probably too simplistic because in reality these parameters are expected to vary. The \ion{He}{2} background will fluctuate from one line-of-sight to another due to the density fluctuations.  Analogously, quasars will also have a distribution of the lifetimes. Moreover, the complete reionization of intergalactic helium is a temporally extended process.  Thus, the conditions in the IGM will evolve and might affect the sensitivity of our models to the quasar and IGM parameters. Therefore, in this section we want to: $\left(1 \right)$ check 
if the results we obtained in previous sections still hold if we 
consider more realistic models with a distribution of
quasar lifetimes and \ion{He}{2} ionizing backgrounds as one
expects to encounter in the universe, and $\left(2 \right)$ check if our results are valid at higher redshifts where the \ion{He}{2} ionizing background cannot be determined from effective optical depth measurements. 

In what follows we consider two diagnostic methods. First, we use the same stacked transmission profiles that we discussed in previous sections. Second, we study the distribution of the \ion{He}{2} proximity zone sizes, the statistics that has been previously used in the literature to characterize high-redshift \ion{H}{1} and \ion{He}{2} proximity zones. 

\subsection{The Effect of the Distribution of Quasar Lifetimes and \ion{He}{2} Backgrounds on the stacked Transmission Profiles}

In what follows we study the impact of the distribution of quasar
lifetimes and \ion{He}{2} backgrounds on the shape of the stacked
transmission profiles in \ion{He}{2} Ly$\alpha$ regions at $z = 3.1$
using the same stacking technique described in \S~\ref{sec:xHeII}.

\begin{figure}[!t]
\centering
 \includegraphics[width=1\linewidth]{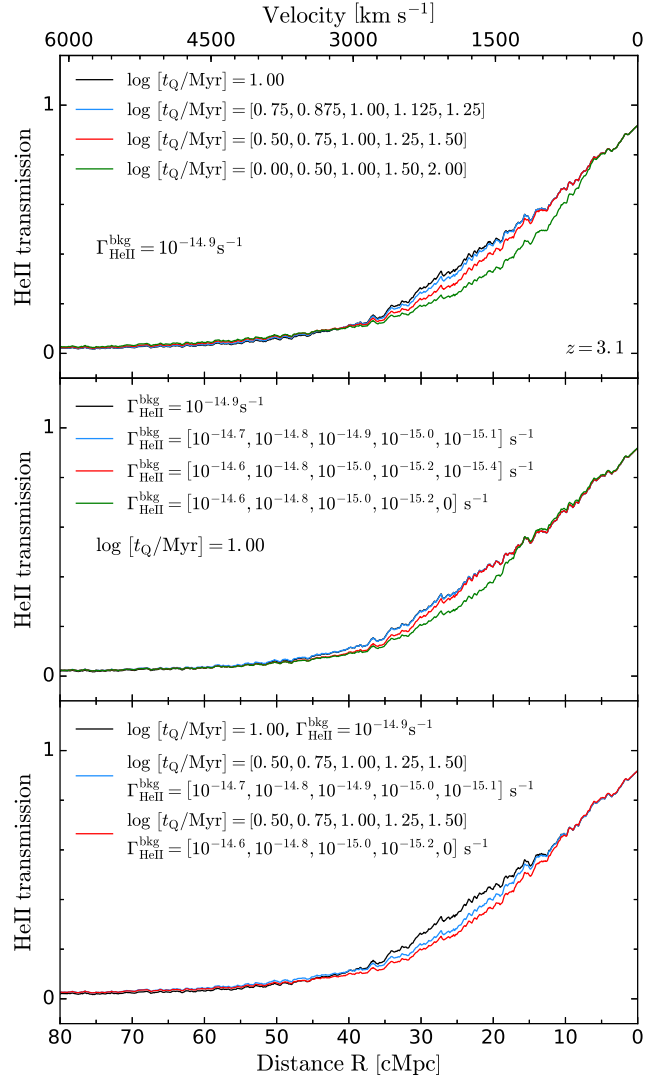}
  \caption{{\it Top panel:} effect of the quasar lifetime distribution on the stacked transmission profile. The \ion{He}{2} background is fixed for all models to $\Gamma_{\rm HeII}^{\rm bkg} = 10^{-14.9}{\rm s^{-1}}$. Each of the curves represents a stack of $1000$ skewers. The black curve shows the stacked profile of the model with single value of quasar lifetime $t_{\rm Q} = 10$~Myr, while other curves correspond to the stacked profiles of the models that take distribution of $t_{\rm Q}$ with different widths: ${\rm log}\ ( t_{\rm Q}  \slash {\rm Myr} ) = \left[ 0.75, 0.875, 1.00, 1.125, 1.25\right]$ (blue), ${\rm log}\ ( t_{\rm Q}  \slash {\rm Myr} ) = \left[ 0.50, 0.75, 1.00, 1.25, 1.50\right]$ (red), ${\rm log}\ (t_{\rm Q} \slash {\rm Myr}) = \left[ 0.00, 0.50, 1.00, 1.50, 2.00 \right]$ (green). {\it Middle panel:} The same, but for distribution of \ion{He}{2} backgrounds. Quasar lifetime is fixed to ${\rm log}\ (t_{\rm Q} \slash {\rm Myr}) = 1.00$. The black curve has a fiducial value of $\Gamma_{\rm HeII}^{\rm bkg} = 10^{-14.9}{\rm s^{-1}}$, and other curves take the distribution of backgrounds, i.e., $\Gamma_{\rm HeII}^{\rm bkg}= \left[ 10^{-14.7}, 10^{-14.8}, 10^{-14.9}, 10^{-15.0}, 10^{-15.1}\right]{\rm s^{-1}}$ (blue), $\Gamma_{\rm HeII}^{\rm bkg} = \left[ 10^{-14.6}, 10^{-14.8}, 10^{-15.0}, 10^{-15.2}, 10^{-15.4}\right]{\rm s^{-1}}$ (red), and $\Gamma_{\rm HeII}^{\rm bkg} = \left[ 10^{-14.6}, 10^{-14.8}, 10^{-15.0}, 10^{-15.2}, 0\right]{\rm s^{-1}}$ (green). {\it Bottom panel:} The combined effect of the distributions of quasar lifetimes and \ion{He}{2} backgrounds.}
 \label{fig:F_eff}
\end{figure}

First, we consider the distribution of quasar lifetimes only, while
keeping \ion{He}{2} background and quasar photon production rate fixed
to our fiducial values $\Gamma_{\rm HeII}^{\rm bkg}=10^{-14.9}{\rm
  s^{-1}}$ and $Q_{\rm 4 Ry} = 10^{56.1}{\rm s^{-1}}$,
respectively. The upper panel of Figure~\ref{fig:F_eff} shows the
comparison between the stacked transmission profiles of our fiducial model with single quasar lifetime $t_{\rm Q} = 10$~Myr
and three models with different distributions of $t_{\rm Q}$. We model the distribution of quasar lifetimes as a uniform sampling from $5$ discrete $t_{\rm Q}$ values centered on ${\rm log}\ ( t_{\rm Q} \slash {\rm Myr} ) = 1.00$ spanning a total range of $0 \leq {\rm
  log}\ ( t_{\rm Q} \slash {\rm Myr} ) \leq 2$, but with different widths $\Delta {\rm log}\ (t_{\rm Q}/{\rm Myr})=\left[0.5,1.0,2.0\right]$. This is done by constructing stacks of $1000$ skewers, where
each skewer is randomly chosen from one of the single lifetime models
over the range specified by the width. The blue curve in the upper panel of Figure~\ref{fig:F_eff} represents the stack of skewers taken from models with ${\rm log}\ \left( t_{\rm Q} \slash
{\rm Myr} \right)= \left[0.75,0.875,1.00,1.125,1.25\right]$, red is a stack of skewers
with ${\rm log}\ \left( t_{\rm Q} \slash {\rm Myr} \right) =
\left[0.50,0.75,1.00,1.25,1.50\right]$ and green is ${\rm log}\ \left( t_{\rm Q}
\slash {\rm Myr} \right) = \left[0.00,0.50,1.00,1.50,2.00\right]$. The numbers in square brackets represent the values of $t_{\rm Q}$ or $\Gamma_{\rm HeII}^{\rm bkg}$ (see below) used in the models that contributed to the stacks of the distribution models. 

The upper panel of Figure~\ref{fig:F_eff} clearly shows that in
comparison to the single lifetime model, there appears to be a
reduction in the transmission in the range $R \simeq 10 - 40$~cMpc in
the stacked spectra of models with the distribution of quasar
lifetimes. This is due to the skewers with lower values of $t_{\rm
  Q}$. In these models the IGM did not have enough time to respond to
the changes in the radiation field caused by the quasar and still
reflects a higher \ion{He}{2} fraction set by the \ion{He}{2} ionizing
background, resulting in decreased transmission. Furthermore, the
transmission becomes more depressed as the width of the quasar
lifetime distribution increases, indicating that stacked transmission
profiles are also sensitive to the width of the quasar lifetime distribution. 

Similarly, we now investigate how the distribution of \ion{He}{2}
backgrounds affects the stacked transmission profiles. As we discussed
in \S~\ref{sec:bkg}, our fiducial value of \ion{He}{2} background
$\Gamma_{\rm HeII}^{\rm bkg} = 10^{-14.9}{\rm s^{-1}}$ is derived from 
measurements of the \ion{He}{2} effective optical depth $\tau_{\rm
  eff}$ which give $\tau_{\rm eff} \simeq 4.5$ at $z = 3.1$ (see
Figure~\ref{fig:F_tau}). Therefore, the distribution of \ion{He}{2}
ionizing backgrounds should also correspond to the same observed mean
effective optical depth of $\tau_{\rm eff} \simeq 4.5$. Analogously to
\S~\ref{sec:bkg}, we run our radiative transfer calculation
with the quasar turned off for $100$ skewers chosen from several models with different values of
the \ion{He}{2} background. We focus on two uniform distributions for
$\Gamma_{\rm HeII}^{\rm bkg}$ which yield the same effective optical depth $\tau_{\rm eff} \simeq 4.5$: 
$\Gamma_{\rm HeII}^{\rm bkg} = \left[10^{-14.7}, 10^{-14.8}, 10^{-14.9},
  10^{-15.0}, 10^{-15.1} \right]{\rm s^{-1}}$ and $\Gamma_{\rm HeII}^{\rm bkg} =
\left[ 10^{-14.6},10^{-14.8}, 10^{-15.0}, 10^{-15.2},10^{-15.4} \right]{\rm s^{-1}}$. We
then run our $1$D radiative transfer algorithm with the quasar on for
$t_{\rm Q} = 10$~Myr and calculate $10$ different models with the above
mentioned values of \ion{He}{2} background. Similar to the
distribution of quasar lifetimes, we calculate two stacked
transmission profiles using $1000$ skewers randomly chosen from these
models.

The results of this exercise are shown in the middle panel of
Figure~\ref{fig:F_eff}, where the black curve corresponds to the model
with a single value of \ion{He}{2} background fixed at our fiducial
value of $\Gamma_{\rm HeII}^{\rm bkg} = 10^{-14.9}{\rm s^{-1}}$, the
blue curve shows the model with a background distribution that spans
a range of $0.4$~dex $\Gamma_{\rm HeII}^{\rm bkg} = \left[10^{-14.7}-10^{-15.1} \right]{\rm s^{-1}}$, 
and the red curve is for the model spanning $0.8$~dex $\Gamma_{\rm HeII}^{\rm bkg} = \left[
  10^{-14.6} - 10^{-15.4} \right]{\rm s^{-1}}$. We also include a model in which $20\%$ of the IGM at $z \simeq 3$ is still represented by the regions with $x_{\rm HeII} = 1$, or, equivalently, $\Gamma_{\rm HeII}^{\rm bkg} = 0$, which is shown in the middle panel of Figure~\ref{fig:F_eff} by the green curve. It is apparent that for the distributions we consider here, varying the
\ion{He}{2} background such that the mean effective optical depth is fixed, has only a small effect on the resulting
transmission profile. However, if measurements of the \ion{He}{2} effective optical depth cannot rule out the existence of the regions of IGM with high fractions of singly ionized helium, including them into the distribution of backgrounds can produce a stronger effect on the stacked transmission profile. Nevertheless, comparing to the upper panel of Figure~\ref{fig:F_eff} we can conclude that the distribution of quasar lifetimes has a dominant effect on the stacked transmission profile and future attempts to model stacked proximity zone should take this into account. This is illustrated in the bottom panel of Figure~\ref{fig:F_eff} where we combine these effects and model both the distributions of quasar lifetimes $t_{\rm Q}$ and \ion{He}{2} backgrounds $\Gamma_{\rm HeII}^{\rm bkg}$ simultaneously, by combining skewers drawn from different models in both parameters. Note, that even in the model where $20\%$ of IGM has $\Gamma_{\rm HeII}^{\rm bkg} = 0$ ($x_{\rm HeII} = 1$), when convolved with a broad distribution of quasar lifetimes, the impact of these regions with $x_{\rm HeII} = 1$ is not very significant.

In \S~\ref{sec:xHeII} we showed that variations in the quasar lifetime
and \ion{He}{2} background impact stacked transmission profiles in
distinct ways. Our analysis here shows that the width of
the distribution of quasar lifetimes, can significantly change the
shape of the stacked transmission profile, and should be included in
any attempt to model real observations. Nevertheless, we argue that
that at $z \simeq 3.1$ one should be able to put interesting
constraints on the quasar lifetime given that the 
average \ion{He}{2} background can be determined from the level of
transmission in the IGM as quantified by effective optical depth measurements (see Figure~\ref{fig:F_tau} in \S~\ref{sec:bkg}), and due to the fact that the stacked transmission profile is
relatively insensitive to fluctuations in the \ion{He}{2} background.

\subsection{Sensitivity to the Quasar Lifetime and \ion{He}{2} Background at Higher Redshifts}
\label{sec:Gamma0} 

Recall Figure~\ref{fig:F_tau}, where the blue curve shows the
evolution of the mean \ion{He}{2} effective optical depth $\tau_{\rm
  eff}$ and the mean \ion{He}{2} fraction at $z = 3.9$ in our
simulations. At this higher redshift, the $\tau_{\rm GP}\propto
(1+z)^{3/2}$ dependence of the optical depth, implies that even
backgrounds which correspond to a IGM with on average relatively low 
\ion{He}{2} fraction of only $\simeq 0.03-0.04$, correspond to
very large effective optical depth is $\tau_{\rm eff} \simeq
8-9$. Thus, unlike the situation at $z \simeq 3$, it would be
extremely challenging to measure an optical depth that high with
HST/COS, making it virtually impossible to measure the \ion{He}{2}
background and thus distinguish between IGM with low \ion{He}{2} fraction ($x_{\rm
  HeII} < 0.05$) and the high \ion{He}{2} fraction ($x_{\rm HeII} = 1$) at $z \simeq
4$. It is, therefore, interesting to explore whether the shape of
the transmission profile of \ion{He}{2} proximity zones can be used
to independently probe both the quasar lifetime and the \ion{He}{2}
background at $z\sim 4$, where the background cannot be independently constrained.

\begin{figure}[!t]
\centering
 \includegraphics[width=1\linewidth]{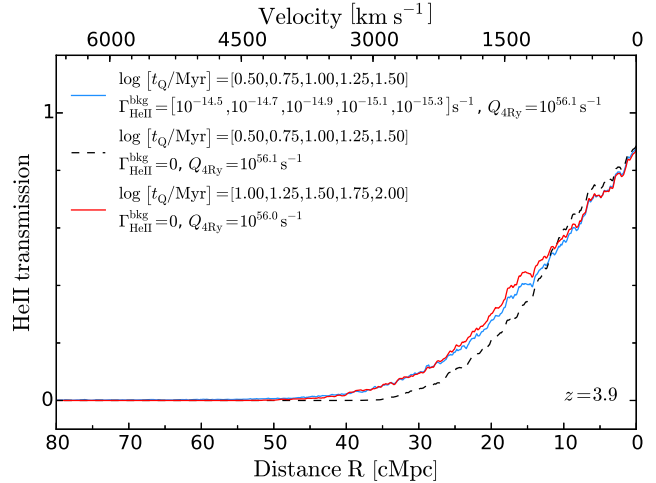}
  \caption{Comparison between the stacked finite \ion{He}{2} background model with $\Gamma_{\rm HeII}^{\rm bkg} = \left[ 10^{-14.5}, 10^{-14.7}, 10^{-14.9}, 10^{-15.1}, 10^{-15.3} \right]\ {\rm s^{-1}}$, ${\rm log}\ (t_{\rm Q} \slash {\rm Myr})= \left[0.50, 0.75,1.00,1.25,1.50 \right]$ (solid blue curve) and zero \ion{He}{2} background models ($x_{\rm HeII} = 1$ initially) with the same quasar lifetime distribution (dashed black curve). Both models have a fiducial value of photon production rate $Q_{\rm 4Ry} = 10^{56.1}{\rm s^{-1}}$. The solid red curve shows the model with ${\rm log}\ (t_{\rm Q} \slash {\rm Myr})= \left[1.00, 1.25,1.50,1.75,2.00 \right]$ and reduced photon production rate $Q_{\rm 4Ry} = 10^{56.0}{\rm s^{-1}}$. All calculations are performed at quasar redshift $z = 3.9$.}
 \label{fig:F_39}
\end{figure}

We begin by applying our method to skewers drawn from the same
hydrodynamical simulation at redshift $z = 3.9$. Following our
approach in the previous subsection, we choose a finite background
model with uniform distribution $\Gamma_{\rm HeII}^{\rm bkg} = \left[
  10^{-14.5}, 10^{-14.7}, 10^{-14.9}, 10^{-15.1}, 10^{-15.3} \right]{\rm s^{-1}}$.
For the distribution of lifetimes we adopt ${\rm log}\ (t_{\rm
  Q}\slash {\rm Myr}) = \left[0.50, 0.75, 1.00, 1.25, 1.50 \right]$, and fix the
photon production rate to $Q_{\rm 4Ry} = 10^{56.1}{\rm s^{-1}}$.  The
resulting stacked transmission profile is shown by the blue curve in
Figure~\ref{fig:F_39}. Despite the finite background and relatively
small average \ion{He}{2} fraction $\langle x_{\rm HeII}\rangle \simeq 0.05$,
far from the quasar $R > 50$~cMpc, the average transmission is very
nearly zero reflecting the large effective optical depth $\tau_{\rm
  eff} \simeq 8-9$ for this model.
The dashed black curve in Figure~\ref{fig:F_39} shows the stacked
transmission profile for a model with the same distribution of quasar
lifetimes, but with $\Gamma_{\rm HeII}^{\rm bkg} = 0$ and hence an IGM
with $\langle x_{\rm HeII}\rangle =1.0$, and $Q_{\rm 4Ry}$ fixed to
the same value. In the absence of the \ion{He}{2}
background, one sees that the proximity zone is smaller, and the
transmission approaches zero at smaller distance from the quasar than
for the finite background model. This significant difference
in the transmission profile naively suggests that proximity zones can be used
to determine the value of the background.

However, clearly one way of compensating for this difference between
the transmission profiles is to consider: $1$) a distribution
with longer quasar lifetimes and, $2$) a change in the quasar
photon production rate $Q_{\rm 4Ry}$, for the zero background model.
Both of these parameter variations change the size of the proximity zone,
which could make the two models look more similar. In principle the
photon production rate $Q_{\rm 4Ry}$ should be determined by our knowledge
of the quasar magnitudes, however in practice the average
quasar SED is not well constrained at energies above $4$~Ry, giving rise
to at least $\sim 50\%$ relative uncertainty in $Q_{\rm 4Ry}$ \citep[see
  e.g.][]{Lusso2015}. To illustrate these parameter degeneracies, 
we increase the values of $t_{\rm Q}$ in
our distribution for the zero background model by $0.50$~dex
to  ${\rm log}\ (t_{\rm Q}\slash {\rm Myr}) = \left[1.00, 1.25, 1.50, 1.75, 2.00 \right]$, and simultaneously
slightly reduce the value of the photon production rate by $0.1$~dex to 
$Q_{\rm 4 Ry} = 10^{56.0}{\rm s^{-1}}$. The result of this exercise is
shown by the solid red curve in Figure~\ref{fig:F_39}, which shows
that the the transmission profiles for a highly doubly ionized helium
($\langle x_{\rm HeII}\rangle = 0.05$) and a completely singly ionized
helium ($\langle x_{\rm HeII}\rangle = 1.0$) are essentially indistinguishable.

In conclusion we see, that at $z \sim 4$ the
limited sensitivity of HST/COS combined with the steep rise of
effective optical depth with redshift (see Figure~\ref{fig:F_tau}),
implies that one will likely only be able to place lower limits on the
the average effective optical depth $\tau_{\rm eff}$, and hence upper
limits on the value of the \ion{He}{2} background. The two very
different models that we considered for the \ion{He}{2} background in
Figure~\ref{fig:F_39} are expected to be consistent with such limits.
Without an independent measurement on the \ion{He}{2} background, and
given our poor knowledge of the SEDs of quasars above 4 Ryd, the
similarity of the models in Figure~\ref{fig:F_39} (blue and red
curves) illustrate that it will be extremely challenging to break the
degeneracies between quasar lifetime, ionizing background, and photon
production rate, given existing data or even data that might be
collected in the future with HST/COS. Thus, in contrast with $z\sim
3$, where an independent determination of the the \ion{He}{2}
background from effective optical depth measurements allows one to
infer the quasar lifetime from proximity zones, the proximity zones of
$z\sim 4$ quasars alone cannot independently constrain the quasar
lifetime and ionization state of the IGM. Nevertheless, a degenerate combination
of these parameters would still be extremely informative, and could be combined
with other measurements to yield tighter constraints.

\begin{figure}[!t]
\centering
\includegraphics[width=1.\linewidth]{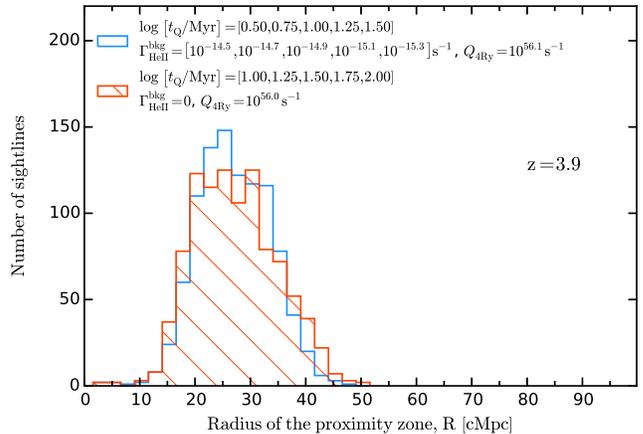}
\caption{Distribution of the \ion{He}{2} proximity zone sizes measured from radiative transfer calculations on $1000$ skewers. The models are the same as in Figure~\ref{fig:F_39}. The size of the proximity zone is defined as the location where the smoothed spectra crosses the threshold $F = 0.1$ (see text for details). The overlap of the histograms makes it impossible to precisely distinguish one model from another.}
\label{fig:Histogram}
\end{figure}

\subsection{Distribution of the Proximity Zone Sizes}
\label{sec:FLUC}

Previous studies of \ion{H}{1} proximity zones at $z \simeq 6$ have concentrated on the location of the `edge' of the ionized regions in order to infer the unknown parameters governing proximity zones.
In this section we adopt a similar technique to investigate if it is a better diagnostic tool than stacking, considered in the previous section, for constraining the properties of the IGM and quasars at $z = 3.9$.

We follow previous conventions \citep{Fan2006} and define the size of
the \ion{He}{2} proximity zone to be the location where the
appropriately smoothed transmission profile crosses the threshold
value $F = 0.1$ for the first time. For the choice of smoothing we
follow conventions that have been adopted in the study of \ion{H}{1}
proximity zones at $z\sim 6$
\citep{Fan2006,Carilli2010,Bolton2007a,Lidz2007}. Specifically,
following the work of \citet{Fan2006}, these studies smooth the
spectra by a Gaussian filter with ${\rm FWHM = 20\AA}$ in the observed
frame, which corresponds to a velocity interval $\Delta v \sim
700\ {\rm km\ s^{-1}}$\ or proper distance $R_{\rm prop} \simeq
0.97$\ Mpc. We adopt the same value of the smoothing scale in proper
units $R_{\rm prop}^{z = 3.9} = 0.97$\ Mpc, which corresponds to a
comoving scale of $R_{\rm com} = 4.75$\ cMpc at $z = 3.9$, or a
velocity interval $\Delta v \sim 410\,{\rm km\,s^{-1}}$. This is
approximately twice the FWHM of HST/COS (for G140L grating).

Figure~\ref{fig:Histogram} shows the distribution of the proximity
zone sizes determined in this way measured from a set of $1000$
skewers for the two models shown in Figure~\ref{fig:F_39} whose
stacked spectra were degenerate. Given the large degree of overlap
between the histograms for these two models, and the
relatively small number $\simeq 8$ of $z \gtrsim 3.5$ quasars with HST/COS
spectra it is clear that it will be extremely challenging to measure the value of quasar lifetime or \ion{He}{2} background using this definition of the proximity zone size.  In \S~\ref{sec:xHeII} we discussed how density fluctuations also introduce scatter in the distribution of the proximity zone sizes and thus complicate our ability to infer parameters (see Figure~\ref{fig:Fluctuations}). We conclude that this statistical approach of measuring the sizes of \ion{He}{2} proximity zones does not result in higher sensitivity to the properties of quasars or the IGM parameters at high redshift $z\sim 4$.

\section{Mock Observations}

\begin{figure}[!t]
\centering
\includegraphics[width=1.\linewidth]{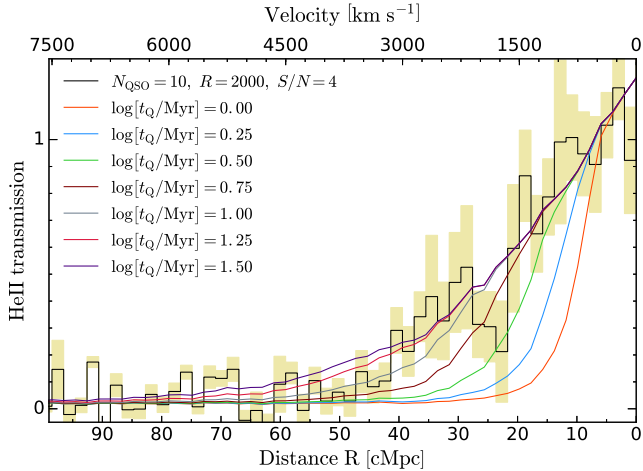}
\caption{Comparison of the mock observed \ion{He}{2} proximity zones to the results of radiative transfer simulations. The histogram shows the stack of $10$ simulated skewers from our fiducial model with $t_{\rm Q} = 10$~Myr, $\Gamma_{\rm HeII}^{\rm bkg} = 10^{-14.9}{\rm s^{-1}}$ and $Q_{\rm 4Ry}=10^{56.1}{\rm s^{-1}}$. The resolution of each skewer has been reduced to $R=2000$ to match low-resolution HST/COS data, the noise has been added with resulting $S/N = 4$. The yellow shaded area shows 1-$\sigma$ errorbars calculated via bootstrapping. Models with different values of quasar lifetime $t_{\rm Q}$ consist of $1000$ skewers also reduced to the same resolution.}
\label{fig:real2}
\end{figure}

In order to get a feeling for the the constraints on quasar
  lifetime $t_{\rm Q}$ that can be obtained with existing samples of
  \ion{He}{2} proximity zone spectra \citep{Worseck2011, Worseck2014,
    Syphers2012} we perform a simple comparison of our model stacked
  transmission profiles to a mock observational dataset. Specifically,
  we randomly select $10$ skewers\footnote{This is approximately the
    number of \ion{He}{2} Ly$\alpha$ spectra in HST/COS archive at $z
    \simeq 3.1\pm 0.2$.} from our fiducial model with $t_{\rm Q} =
  10$~Myr and mock up the properties of real HST/COS data by
  convolving these spectra with a Gaussian line-spread function
  matched to HST/COS moderate resolution spectra $R = \lambda \slash
  \Delta \lambda = 2000$ or FWHM=$150~{\rm km~s^{-1}}$ (comparable to the G140L
  grating). We also add Gaussian distributed noise to each spectrum assuming 
  a constant signal-to-noise ratio $S/N = 4$. The black
  histogram in Figure~\ref{fig:real2} shows the resulting stacked
  transmission profile in the \ion{He}{2} proximity zone region. We
  also calculate 1-$\sigma$ errorbars for this mock dataset using the
  bootstrap technique, which is illustrated by the yellow shaded
  area. We then compare the stack with several simulated models with
  different values of $t_{\rm Q}$ ($1000$ skewers per model with the
  same resolution of $R=2000$). It is apparent from the example 
  shown in Figure~\ref{fig:real2} that we should be able to measure the
  quasar lifetime within a factor of $\sim 2-3$ uncertainty.
  
We plan to use the method described in this paper for the follow-up analysis of real observational data. However, there is a range of uncertainties that can complicate the estimation of average quasar lifetime from
  the stacked transmission profiles. First, in order to stack quasar
  spectra one needs to take care of the flux normalization, which
  depends on the fitted quasar continuum. Statistical errors on the
  power-law fit to the quasar continuum in the \ion{He}{2} Ly$\alpha$
  forest are typically few per cent for the low $S/N$ spectra
  \citep{Worseck2011, Worseck2014}. However, given the large variance
  indicated by the yellow shaded region in Figure~\ref{fig:real2},
  continuum errors will not significantly impact our results since
  flucutations in the IGM dominate the scatter.  Similarly, the
  spectral resolution does not affect our constraints on quasar
  lifetime because we choose to stack individual spectra. Thus, we are
  more interested in the information contained in the shape and the
  slope of the stacked transmission profile, but not the individual
  absorption lines in the \ion{He}{2} Ly$\alpha$ forest. In addition,
  large errors in the quasar systemic redshift that result from using
  rest-frame UV emission lines (typical uncertainties are $500 -
  1000\ {\rm km\ s^{-1}}$; \citealp{Shen2011}) can result in a $5-10$~cMpc uncertainty in the
  quasar location. 
  While these redshift errors can be easily modeled, they will nevertheless degrade our constraints on the parameters governing the proximity zone. However, these uncertainities can be easily mitigated by
  obtaining near-infrared spectra of the narrow forbidden [\ion{O}{3}] line, which is an excellent tracer of the systemic redshift \citep{Boroson2005}, or by obtaining spectra of the low-ionization \ion{Mg}{2} line which traces
  systemic redshift to $\sim \delta v \simeq 200~{\rm km~s^{-1}}$\citep{Shen2011}. The detection of
  molecular emission lines with ALMA can also be used to establish the systemic redshift \citep{Carilli2013}.

\section{Summary \& Conclusions}
\label{sec:Conclusions}

We have used a combination of numerical hydrodynamical simulations and
a $1$D radiative transfer algorithm to study the evolution of the
\ion{He}{2} proximity zones around quasars at $z \simeq 3$ and $z\simeq
4$.  We have analyzed the effects of the quasar lifetime ($t_{\rm
  Q}$), the average \ion{He}{2} ionizing background ($\Gamma_{\rm
  HeII}^{\rm bkg}$), and the photon production rate of the quasar ($Q_{\rm
  4Ry}$) on the transmission profiles of \ion{He}{2} proximity zones.

Previous work analyzing the structure of \ion{He}{2} proximity zones
have assumed that the edge of the observed proximity zone can be
identified with the radius of the quasar ionization front $R_{\rm IF}$
\citep{Hogan1997, Anderson1999, Zheng2015}. We showed that the
\ion{He}{2} proximity zones of quasars in an IGM with $x_{\rm
  HeII}\simeq 1$ and quasar lifetimes $t_{\rm Q}\simeq 30\,{\rm Myr}$,
can look identical to those in an IGM with primarily doubly ionized
helium $x_{\rm HeII}\simeq 0.03-0.05$ and shorter lifetimes $t_{\rm
  Q}\simeq 10\,{\rm Myr}$. Thus, analogous to the situation of
\ion{H}{1} proximity zones around $z\sim 6$ quasars
\citep{Bolton2007a, Lidz2007, Maselli2007}, naively identifying the
size of the proximity zone with the location of the ionization front
can lead to erroneous conclusions about the parameters which govern
them. Furthermore, whereas the majority of previous work
\citep{Hogan1997, Anderson1999, Zheng2015} has assumed that helium is
completely singly ionized $x_{\rm HeII}\simeq 1$, we argued that
observations of the effective optical depth at $z = 3.1$ suggest the characteristic
$\Gamma_{\rm HeII}^{\rm bkg} = 10^{-14.9}{\rm s^{-1}}$ and hence
Helium is highly doubly ionized in \ion{He}{3} regions $x_{\rm HeII}\simeq 0.02$. Thus \ion{He}{2}
proximity zones are more likely to in a regime where radiation from the
quasar increases the ionization level of nearby material which was
already highly ionized to begin with, making the location of the
ionizaton front irrelevant.

We introduced a new and more appropriate way of thinking about
proximity zones in terms of the time-evolution of the \ion{He}{2}
fraction and its approach to equilibrium, which governs both the case
of a quasar turning on in an IGM for which Helium is singly ionized,
as well as case where the IGM is already highly doubly ionized.  We
presented a simple analytical formula describing this time evolution,
and showed that it agrees with the results of detailed radiative
transfer calculations. This model was used to understand how proximity
zone properties reflect the quasar lifetime, \ion{He}{2}-ionizing
background, and the photon production rate, and the degeneracies
between these parameters. The approach to ionization equilibrium in
the proximity zone is set by the equilibration timescale, which is the
inverse of the local photoionization rate $t_{\rm eq}\simeq
\Gamma_{\rm tot}^{-1}$. It follows that for an IGM in which Helium is
already highly doubly ionized, which is likely to be the case at $z =
3.1$, the maximum value of the quasar lifetime that one can probe is
$t_{\rm Q}^{\rm max} \simeq 1/\Gamma_{\rm HeII}^{\rm bkg} \simeq 2.5
\times 10^7\,{\rm yr}$, which is comparable to the Salpeter time.

IGM density fluctuations result in significant dispersion in the sizes
and properties of the \ion{He}{2} proximity zone, making it difficult
to isolate parameter dependencies. In order to eliminate this source
of variation, we investigated stacking proximity zones for ensembles
of quasars.  At $z \simeq 3.1$ we find that because the value of
\ion{He}{2} ionizing background is determined independently from the
measurements of the \ion{He}{2} effective optical depth, the degeneracy
which exists between quasar lifetime and \ion{He}{2} background can be
broken. Therefore the resulting stacked transmission profiles at $z
\simeq 3.1$ can be used to determine the quasar
lifetime. Unfortunately, in contrast to the case of $z \simeq 3.1$,
the \ion{He}{2} ionizing background is poorly constrained at higher redshifts
$z \simeq 4$,  because the increase in the \ion{He}{2} effective optical
depth, implies the transmission is beyond the sensitivity limit of HST/COS. Combined
with additional uncertainties in the quasar SED above $4$~Ry, it will
be extremely challenging to determine quasar lifetime or \ion{He}{2}
background independently using \ion{He}{2} proximity zones of $z
\simeq 4$ quasars.  We also showed that the shape of the stacked
transmission profiles is highly sensitive to the distribution of
quasar lifetimes, which should be taken into account in any attempt of
modeling the observational data. On the other hand, a broad distribution
of \ion{He}{2} backgrounds does not impact the stacked profiles
significantly.

In the future we plan to stack the existing $\sim 30$ spectra of \ion{He}{2} proximity zones to measure the average quasar
lifetime. In the context of the next generation of UV space missions, it
would also be extremely interesting to acquire \ion{He}{2} Ly$\alpha$ transmission
spectra covering the proximity zones of $z \simeq 2-4$ quasars, which would
allow one to constrain the quasar lifetime as a function of luminosity
and redshift, as well as characterize the ionization state of the IGM
as a function of cosmic time. In addition, we note that the
general equilibration time picture explored in this work in the
context of \ion{He}{2} proximity zones is also applicable to study the
properties of the \ion{H}{1} proximity zones around $z \simeq 6-7$
quasars. These studies would also benefit from stacking of multiple
sightlines, which would yield interesting insights into the evolution of
quasars and the intergalactic medium.

Finally, if helium in the IGM was significantly singly
  ionized before a quasar turns on, photoelectric heating by the
  expanding quasar ionization front can boost the temperature of the
  surrounding gas producing a so-called ``thermal proximity effect"
  \citep{Meiksin2010}, which might be detectable via the thermal
  broadening of absorption lines in the \ion{H}{1} Ly$\alpha$
  forest \citep{Bolton2012}. In the future work we will use the radiative transfer code developed
  here to explore the thermal proximity effect in detail, and study its potential
  for constraining the quasar lifetime and the \ion{He}{2} fraction in the IGM.
\section{Acknowledgements}

We would like to thank members of the ENIGMA \footnote{http://www.mpia.de/ENIGMA/} group at the Max-Planck-Institut f{\"u}r Astronomie (MPIA) for useful discussions and comments on the paper. We are grateful to the anonymous referee for comments and suggestions, which
greatly improved the text. J.F.H. acknowledges generous support from the Alexander von Humboldt foundation in the context of the Sofja Kovalevskaja Award. The Humboldt foundation is funded by the German Federal Ministry for Education and Research. M.M. acknowledges support from NASA through a grant from the Space Telescope Institute, HST-AR-13903.00. G.W. has been supported by the Deutsches Zentrum f\"ur Luft- und Raumfahrt (DLR) under contract 50 OR 1317. Partly based on observations made with the NASA/ESA Hubble Space Telescope (Programs 11528 and 12033) obtained from the Mikulski Archive for Space Telescopes (MAST). The Space Telescope Science Institute is operated by the Association of Universities for Research in Astronomy, Inc., under NASA contract NAS5-26555.

\bibliography{ref}

\begin{thebibliography}{}
\expandafter\ifx\csname natexlab\endcsname\relax\def\natexlab#1{#1}\fi

\bibitem[{{Abel} \& {Haehnelt}(1999)}]{Abel1999}
{Abel}, T., \& {Haehnelt}, M.~G. 1999, \apjl, 520, L13

\bibitem[{{Adelberger} \& {Steidel}(2005)}]{AS05b}
{Adelberger}, K.~L., \& {Steidel}, C.~C. 2005, \apj, 630, 50

\bibitem[{{Anderson} {et~al.}(1999){Anderson}, {Hogan}, {Williams}, \&
  {Carswell}}]{Anderson1999}
{Anderson}, S.~F., {Hogan}, C.~J., {Williams}, B.~F., \& {Carswell}, R.~F.
  1999, \aj, 117, 56

\bibitem[{{Bajtlik} {et~al.}(1988){Bajtlik}, {Duncan}, \&
  {Ostriker}}]{Bajtlik1988}
{Bajtlik}, S., {Duncan}, R.~C., \& {Ostriker}, J.~P. 1988, \apj, 327, 570

\bibitem[{{Becker} {et~al.}(2013){Becker}, {Hewett}, {Worseck}, \&
  {Prochaska}}]{Becker2013}
{Becker}, G.~D., {Hewett}, P.~C., {Worseck}, G., \& {Prochaska}, J.~X. 2013,
  \mnras, 430, 2067

\bibitem[{{Bolton} {et~al.}(2012){Bolton}, {Becker}, {Raskutti}, {Wyithe},
  {Haehnelt}, \& {Sargent}}]{Bolton2012}
{Bolton}, J.~S., {Becker}, G.~D., {Raskutti}, S., {et~al.} 2012, \mnras, 419,
  2880

\bibitem[{{Bolton} \& {Haehnelt}(2007a)}]{Bolton2007a}
{Bolton}, J.~S., \& {Haehnelt}, M.~G. 2007a, \mnras, 374, 493

\bibitem[{{Bolton} \& {Haehnelt}(2007b)}]{Bolton2007b}
---. 2007b, \mnras, 381, L35

\bibitem[{{Borisova} {et~al.}(2015){Borisova}, {Lilly}, {Cantalupo},
  {Prochaska}, {Rakic}, \& {Worseck}}]{Borisova2015}
{Borisova}, E., {Lilly}, S.~J., {Cantalupo}, S., {et~al.} 2015, ArXiv e-prints,
  arXiv:1510.00029

\bibitem[{{Boroson}(2005)}]{Boroson2005}
{Boroson}, T. 2005, \aj, 130, 381

\bibitem[{{Cantalupo} {et~al.}(2014){Cantalupo}, {Arrigoni-Battaia},
  {Prochaska}, {Hennawi}, \& {Madau}}]{Cantalupo2014}
{Cantalupo}, S., {Arrigoni-Battaia}, F., {Prochaska}, J.~X., {Hennawi}, J.~F.,
  \& {Madau}, P. 2014, \nat, 506, 63

\bibitem[{{Carilli} \& {Walter}(2013)}]{Carilli2013}
{Carilli}, C.~L., \& {Walter}, F. 2013, \araa, 51, 105

\bibitem[{{Carilli} {et~al.}(2010){Carilli}, {Wang}, {Fan}, {Walter}, {Kurk},
  {Riechers}, {Wagg}, {Hennawi}, {Jiang}, {Menten}, {Bertoldi}, {Strauss}, \&
  {Cox}}]{Carilli2010}
{Carilli}, C.~L., {Wang}, R., {Fan}, X., {et~al.} 2010, \apj, 714, 834

\bibitem[{{Carswell} {et~al.}(1982){Carswell}, {Whelan}, {Smith}, {Boksenberg},
  \& {Tytler}}]{Carswell1982}
{Carswell}, R.~F., {Whelan}, J.~A.~J., {Smith}, M.~G., {Boksenberg}, A., \&
  {Tytler}, D. 1982, \mnras, 198, 91

\bibitem[{{Cen} \& {Haiman}(2000)}]{Cen2000}
{Cen}, R., \& {Haiman}, Z. 2000, \apjl, 542, L75

\bibitem[{{Cole} \& {Kaiser}(1989)}]{CK1989}
{Cole}, S., \& {Kaiser}, N. 1989, \mnras, 237, 1127

\bibitem[{{Compostella} {et~al.}(2013){Compostella}, {Cantalupo}, \&
  {Porciani}}]{Compostella2013}
{Compostella}, M., {Cantalupo}, S., \& {Porciani}, C. 2013, \mnras, 435, 3169

\bibitem[{{Conroy} \& {White}(2013)}]{Conroy2013}
{Conroy}, C., \& {White}, M. 2013, \apj, 762, 70

\bibitem[{{Croom} {et~al.}(2005){Croom}, {Boyle}, {Shanks}, {Smith}, {Miller},
  {Outram}, {Loaring}, {Hoyle}, \& {da {\^A}ngela}}]{Croom2005}
{Croom}, S.~M., {Boyle}, B.~J., {Shanks}, T., {et~al.} 2005, \mnras, 356, 415

\bibitem[{{Davies} {et~al.}(2014){Davies}, {Furlanetto}, \&
  {McQuinn}}]{Davies2014}
{Davies}, F.~B., {Furlanetto}, S.~R., \& {McQuinn}, M. 2014, ArXiv e-prints,
  arXiv:1409.0855

\bibitem[{{Davis} {et~al.}(1985){Davis}, {Efstathiou}, {Frenk}, \&
  {White}}]{Davis1985}
{Davis}, M., {Efstathiou}, G., {Frenk}, C.~S., \& {White}, S.~D.~M. 1985, \apj,
  292, 371

\bibitem[{{Fan} {et~al.}(2002){Fan}, {Narayanan}, {Strauss}, {Lupton},
  {Becker}, {White}, {Pentericci}, \& {Rix}}]{Fan2002}
{Fan}, X., {Narayanan}, V.~K., {Strauss}, M.~A., {et~al.} 2002, in Lighthouses
  of the Universe: The Most Luminous Celestial Objects and Their Use for
  Cosmology, ed. M.~{Gilfanov}, R.~{Sunyeav}, \& E.~{Churazov}, 309

\bibitem[{{Fan} {et~al.}(2006){Fan}, {Strauss}, {Becker}, {White}, {Gunn},
  {Knapp}, {Richards}, {Schneider}, {Brinkmann}, \& {Fukugita}}]{Fan2006}
{Fan}, X., {Strauss}, M.~A., {Becker}, R.~H., {et~al.} 2006, \aj, 132, 117

\bibitem[{{Faucher-Gigu{\`e}re} {et~al.}(2008){Faucher-Gigu{\`e}re}, {Lidz},
  {Zaldarriaga}, \& {Hernquist}}]{FG2008}
{Faucher-Gigu{\`e}re}, C.-A., {Lidz}, A., {Zaldarriaga}, M., \& {Hernquist}, L.
  2008, \apj, 673, 39

\bibitem[{{Fechner} {et~al.}(2004){Fechner}, {Baade}, \&
  {Reimers}}]{Fechner2004}
{Fechner}, C., {Baade}, R., \& {Reimers}, D. 2004, \aap, 418, 857

\bibitem[{{Furlanetto} \& {Lidz}(2011)}]{Furlanetto2011}
{Furlanetto}, S.~R., \& {Lidz}, A. 2011, \apj, 735, 117

\bibitem[{{Gallerani} {et~al.}(2008){Gallerani}, {Ferrara}, {Fan}, \&
  {Choudhury}}]{Gallerani2008}
{Gallerani}, S., {Ferrara}, A., {Fan}, X., \& {Choudhury}, T.~R. 2008, \mnras,
  386, 359

\bibitem[{{Gon{\c c}alves} {et~al.}(2008){Gon{\c c}alves}, {Steidel}, \&
  {Pettini}}]{Goncalves2008}
{Gon{\c c}alves}, T.~S., {Steidel}, C.~C., \& {Pettini}, M. 2008, \apj, 676,
  816

\bibitem[{{Goodman}(2003)}]{Goodman2003}
{Goodman}, J. 2003, \mnras, 339, 937

\bibitem[{{Gunn} \& {Peterson}(1965)}]{GP1965}
{Gunn}, J.~E., \& {Peterson}, B.~A. 1965, \apj, 142, 1633

\bibitem[{{Haardt} \& {Madau}(2012)}]{Haardt2012}
{Haardt}, F., \& {Madau}, P. 2012, \apj, 746, 125

\bibitem[{{Haiman} \& {Cen}(2001)}]{HaimanCen2001}
{Haiman}, Z., \& {Cen}, R. 2001, in Astronomical Society of the Pacific
  Conference Series, Vol. 222, The Physics of Galaxy Formation, ed.
  M.~{Umemura} \& H.~{Susa}, 101

\bibitem[{{Haiman} \& {Hui}(2001)}]{Haiman2001}
{Haiman}, Z., \& {Hui}, L. 2001, \apj, 547, 27

\bibitem[{{Heap} {et~al.}(2000){Heap}, {Williger}, {Smette}, {Hubeny}, {Sahu},
  {Jenkins}, {Tripp}, \& {Winkler}}]{Heap2000}
{Heap}, S.~R., {Williger}, G.~M., {Smette}, A., {et~al.} 2000, \apj, 534, 69

\bibitem[{{Hennawi} \& {Prochaska}(2007)}]{Hennawi2007}
{Hennawi}, J.~F., \& {Prochaska}, J.~X. 2007, \apj, 655, 735

\bibitem[{{Hennawi} \& {Prochaska}(2013)}]{Hennawi2013}
---. 2013, \apj, 766, 58

\bibitem[{{Hennawi} {et~al.}(2015){Hennawi}, {Prochaska}, {Cantalupo}, \&
  {Arrigoni-Battaia}}]{Hennawi2015}
{Hennawi}, J.~F., {Prochaska}, J.~X., {Cantalupo}, S., \& {Arrigoni-Battaia},
  F. 2015, Science, 348, 779

\bibitem[{{Hennawi} {et~al.}(2006){Hennawi}, {Prochaska}, {Burles}, {Strauss},
  {Richards}, {Schlegel}, {Fan}, {Schneider}, {Zakamska}, {Oguri}, {Gunn},
  {Lupton}, \& {Brinkmann}}]{Hennawi2006}
{Hennawi}, J.~F., {Prochaska}, J.~X., {Burles}, S., {et~al.} 2006, \apj, 651,
  61

\bibitem[{{Hogan} {et~al.}(1997){Hogan}, {Anderson}, \& {Rugers}}]{Hogan1997}
{Hogan}, C.~J., {Anderson}, S.~F., \& {Rugers}, M.~H. 1997, \aj, 113, 1495

\bibitem[{{Hopkins} {et~al.}(2008){Hopkins}, {Hernquist}, {Cox}, \& {Kere{\v
  s}}}]{Hopkins2008}
{Hopkins}, P.~F., {Hernquist}, L., {Cox}, T.~J., \& {Kere{\v s}}, D. 2008,
  \apjs, 175, 356

\bibitem[{{Hopkins} {et~al.}(2006){Hopkins}, {Hernquist}, {Cox}, {Robertson},
  \& {Springel}}]{Hopkins2006}
{Hopkins}, P.~F., {Hernquist}, L., {Cox}, T.~J., {Robertson}, B., \&
  {Springel}, V. 2006, \apjs, 163, 50

\bibitem[{{Hopkins} {et~al.}(2005){Hopkins}, {Hernquist}, {Martini}, {Cox},
  {Robertson}, {Di Matteo}, \& {Springel}}]{Hopkins2005}
{Hopkins}, P.~F., {Hernquist}, L., {Martini}, P., {et~al.} 2005, \apjl, 625,
  L71

\bibitem[{{Hopkins} \& {Quataert}(2010)}]{Hopkins2010}
{Hopkins}, P.~F., \& {Quataert}, E. 2010, \mnras, 407, 1529

\bibitem[{{Jakobsen} {et~al.}(2003){Jakobsen}, {Jansen}, {Wagner}, \&
  {Reimers}}]{Jakobsen2003}
{Jakobsen}, P., {Jansen}, R.~A., {Wagner}, S., \& {Reimers}, D. 2003, \aap,
  397, 891

\bibitem[{{Kirkman} \& {Tytler}(2008)}]{Kirkman2008}
{Kirkman}, D., \& {Tytler}, D. 2008, \mnras, 391, 1457

\bibitem[{{Kormendy} \& {Ho}(2013)}]{KormendyHo2013}
{Kormendy}, J., \& {Ho}, L.~C. 2013, \araa, 51, 511

\bibitem[{{Kormendy} \& {Richstone}(1995)}]{KormRich95}
{Kormendy}, J., \& {Richstone}, D. 1995, \araa, 33, 581

\bibitem[{{Larson} {et~al.}(2011){Larson}, {Dunkley}, {Hinshaw}, {Komatsu},
  {Nolta}, {Bennett}, {Gold}, {Halpern}, {Hill}, {Jarosik}, {Kogut}, {Limon},
  {Meyer}, {Odegard}, {Page}, {Smith}, {Spergel}, {Tucker}, {Weiland},
  {Wollack}, \& {Wright}}]{Larson2011}
{Larson}, D., {Dunkley}, J., {Hinshaw}, G., {et~al.} 2011, \apjs, 192, 16

\bibitem[{{Lidz} {et~al.}(2007){Lidz}, {McQuinn}, {Zaldarriaga}, {Hernquist},
  \& {Dutta}}]{Lidz2007}
{Lidz}, A., {McQuinn}, M., {Zaldarriaga}, M., {Hernquist}, L., \& {Dutta}, S.
  2007, \apj, 670, 39

\bibitem[{{Lusso} {et~al.}(2015){Lusso}, {Worseck}, {Hennawi}, {Prochaska},
  {Vignali}, {Stern}, \& {O'Meara}}]{Lusso2015}
{Lusso}, E., {Worseck}, G., {Hennawi}, J.~F., {et~al.} 2015, ArXiv e-prints,
  arXiv:1503.02075

\bibitem[{{Madau} \& {Meiksin}(1994)}]{Madau1994}
{Madau}, P., \& {Meiksin}, A. 1994, \apjl, 433, L53

\bibitem[{{Madau} \& {Rees}(2000)}]{Madau2000}
{Madau}, P., \& {Rees}, M.~J. 2000, \apjl, 542, L69

\bibitem[{{Martini}(2004)}]{Martini2004}
{Martini}, P. 2004, Coevolution of Black Holes and Galaxies, 169

\bibitem[{{Martini} \& {Weinberg}(2001)}]{Martini2001}
{Martini}, P., \& {Weinberg}, D.~H. 2001, \apj, 547, 12

\bibitem[{{Maselli} {et~al.}(2007){Maselli}, {Gallerani}, {Ferrara}, \&
  {Choudhury}}]{Maselli2007}
{Maselli}, A., {Gallerani}, S., {Ferrara}, A., \& {Choudhury}, T.~R. 2007,
  \mnras, 376, L34

\bibitem[{{McQuinn}(2009b)}]{McQuinn2009b}
{McQuinn}, M. 2009b, \apjl, 704, L89

\bibitem[{{McQuinn} {et~al.}(2009a){McQuinn}, {Lidz}, {Zaldarriaga},
  {Hernquist}, {Hopkins}, {Dutta}, \& {Faucher-Gigu{\`e}re}}]{McQuinn2009}
{McQuinn}, M., {Lidz}, A., {Zaldarriaga}, M., {et~al.} 2009a, \apj, 694, 842

\bibitem[{{McQuinn} \& {Switzer}(2010)}]{McQuinn2010}
{McQuinn}, M., \& {Switzer}, E.~R. 2010, \mnras, 408, 1945

\bibitem[{{McQuinn} \& {Worseck}(2014)}]{McQuinn2014}
{McQuinn}, M., \& {Worseck}, G. 2014, \mnras, 440, 2406

\bibitem[{{Meiksin} {et~al.}(2010){Meiksin}, {Tittley}, \&
  {Brown}}]{Meiksin2010}
{Meiksin}, A., {Tittley}, E.~R., \& {Brown}, C.~K. 2010, \mnras, 401, 77

\bibitem[{{Mellema} {et~al.}(2006){Mellema}, {Iliev}, {Alvarez}, \&
  {Shapiro}}]{Mellema2006}
{Mellema}, G., {Iliev}, I.~T., {Alvarez}, M.~A., \& {Shapiro}, P.~R. 2006, NA,
  11, 374

\bibitem[{{Mesinger} \& {Haiman}(2004)}]{Mesinger2004}
{Mesinger}, A., \& {Haiman}, Z. 2004, \apjl, 611, L69

\bibitem[{{Miralda-Escud{\'e}} {et~al.}(2000){Miralda-Escud{\'e}}, {Haehnelt},
  \& {Rees}}]{Miralda2000}
{Miralda-Escud{\'e}}, J., {Haehnelt}, M., \& {Rees}, M.~J. 2000, \apj, 530, 1

\bibitem[{{Prochaska} {et~al.}(2013{\natexlab{a}}){Prochaska}, {Hennawi}, \&
  {Simcoe}}]{Prochaska2013a}
{Prochaska}, J.~X., {Hennawi}, J.~F., \& {Simcoe}, R.~A. 2013{\natexlab{a}},
  \apjl, 762, L19

\bibitem[{{Prochaska} {et~al.}(2015){Prochaska}, {O'Meara}, {Fumagalli},
  {Bernstein}, \& {Burles}}]{Prochaska2015}
{Prochaska}, J.~X., {O'Meara}, J.~M., {Fumagalli}, M., {Bernstein}, R.~A., \&
  {Burles}, S.~M. 2015, \apjs, 221, 2

\bibitem[{{Prochaska} {et~al.}(2013{\natexlab{b}}){Prochaska}, {Hennawi},
  {Lee}, {Cantalupo}, {Bovy}, {Djorgovski}, {Ellison}, {Lau}, {Martin},
  {Myers}, {Rubin}, \& {Simcoe}}]{Prochaska2013}
{Prochaska}, J.~X., {Hennawi}, J.~F., {Lee}, K.-G., {et~al.}
  2013{\natexlab{b}}, \apj, 776, 136

\bibitem[{{Schawinski} {et~al.}(2015){Schawinski}, {Koss}, {Berney}, \&
  {Sartori}}]{Schawinski2015}
{Schawinski}, K., {Koss}, M., {Berney}, S., \& {Sartori}, L.~F. 2015, \mnras,
  451, 2517

\bibitem[{{Schawinski} {et~al.}(2010){Schawinski}, {Evans}, {Virani}, {Urry},
  {Keel}, {Natarajan}, {Lintott}, {Manning}, {Coppi}, {Kaviraj}, {Bamford},
  {J{\'o}zsa}, {Garrett}, {van Arkel}, {Gay}, \& {Fortson}}]{Schawinski2010}
{Schawinski}, K., {Evans}, D.~A., {Virani}, S., {et~al.} 2010, \apjl, 724, L30

\bibitem[{{Shapiro} {et~al.}(2006){Shapiro}, {Iliev}, {Alvarez}, \&
  {Scannapieco}}]{Shapiro2006}
{Shapiro}, P.~R., {Iliev}, I.~T., {Alvarez}, M.~A., \& {Scannapieco}, E. 2006,
  \apj, 648, 922

\bibitem[{{Shen} {et~al.}(2009){Shen}, {Strauss}, {Ross}, {Hall}, {Lin},
  {Richards}, {Schneider}, {Weinberg}, {Connolly}, {Fan}, {Hennawi}, {Shankar},
  {Vanden Berk}, {Bahcall}, \& {Brunner}}]{Shen2009}
{Shen}, Y., {Strauss}, M.~A., {Ross}, N.~P., {et~al.} 2009, \apj, 697, 1656

\bibitem[{{Shen} {et~al.}(2011){Shen}, {Richards}, {Strauss}, {Hall},
  {Schneider}, {Snedden}, {Bizyaev}, {Brewington}, {Malanushenko},
  {Malanushenko}, {Oravetz}, {Pan}, \& {Simmons}}]{Shen2011}
{Shen}, Y., {Richards}, G.~T., {Strauss}, M.~A., {et~al.} 2011, \apjs, 194, 45

\bibitem[{{Shull} {et~al.}(2010){Shull}, {France}, {Danforth}, {Smith}, \&
  {Tumlinson}}]{Shull2010}
{Shull}, J.~M., {France}, K., {Danforth}, C.~W., {Smith}, B., \& {Tumlinson},
  J. 2010, \apj, 722, 1312

\bibitem[{{Shull} {et~al.}(2012){Shull}, {Stevans}, \& {Danforth}}]{Shull2012}
{Shull}, J.~M., {Stevans}, M., \& {Danforth}, C.~W. 2012, \apj, 752, 162

\bibitem[{{Soltan}(1982)}]{Soltan1982}
{Soltan}, A. 1982, \mnras, 200, 115

\bibitem[{{Springel}(2005c)}]{Springel2005c}
{Springel}, V. 2005c, \mnras, 364, 1105

\bibitem[{{Springel} {et~al.}(2005a){Springel}, {Di Matteo}, \&
  {Hernquist}}]{Springel2005a}
{Springel}, V., {Di Matteo}, T., \& {Hernquist}, L. 2005a, \apjl, 620, L79

\bibitem[{{Springel} {et~al.}(2005b){Springel}, {Di Matteo}, \&
  {Hernquist}}]{Springel2005b}
---. 2005b, \mnras, 361, 776

\bibitem[{{Syphers} {et~al.}(2012){Syphers}, {Anderson}, {Zheng}, {Meiksin},
  {Schneider}, \& {York}}]{Syphers2012}
{Syphers}, D., {Anderson}, S.~F., {Zheng}, W., {et~al.} 2012, \aj, 143, 100

\bibitem[{{Syphers} \& {Shull}(2014)}]{Syphers2014}
{Syphers}, D., \& {Shull}, J.~M. 2014, \apj, 784, 42

\bibitem[{{Telfer} {et~al.}(2002){Telfer}, {Zheng}, {Kriss}, \&
  {Davidsen}}]{Telfer2002}
{Telfer}, R.~C., {Zheng}, W., {Kriss}, G.~A., \& {Davidsen}, A.~F. 2002, \apj,
  565, 773

\bibitem[{{Theuns} {et~al.}(1998){Theuns}, {Leonard}, {Efstathiou}, {Pearce},
  \& {Thomas}}]{Theuns1998}
{Theuns}, T., {Leonard}, A., {Efstathiou}, G., {Pearce}, F.~R., \& {Thomas},
  P.~A. 1998, \mnras, 301, 478

\bibitem[{{Trainor} \& {Steidel}(2013)}]{Trainor2013}
{Trainor}, R., \& {Steidel}, C.~C. 2013, \apjl, 775, L3

\bibitem[{{White} {et~al.}(2012){White}, {Myers}, {Ross}, {Schlegel},
  {Hennawi}, {Shen}, {McGreer}, {Strauss}, {Bolton}, {Bovy}, {Fan},
  {Miralda-Escude}, {Palanque-Delabrouille}, {Paris}, {Petitjean}, {Schneider},
  {Viel}, {Weinberg}, {Yeche}, {Zehavi}, {Pan}, {Snedden}, {Bizyaev},
  {Brewington}, {Brinkmann}, {Malanushenko}, {Malanushenko}, {Oravetz},
  {Simmons}, {Sheldon}, \& {Weaver}}]{White2012}
{White}, M., {Myers}, A.~D., {Ross}, N.~P., {et~al.} 2012, \mnras, 424, 933

\bibitem[{{White} {et~al.}(2003){White}, {Becker}, {Fan}, \&
  {Strauss}}]{White2003}
{White}, R.~L., {Becker}, R.~H., {Fan}, X., \& {Strauss}, M.~A. 2003, \aj, 126,
  1

\bibitem[{{Worseck} {et~al.}(2007){Worseck}, {Fechner}, {Wisotzki}, \&
  {Dall'Aglio}}]{Worseck2007}
{Worseck}, G., {Fechner}, C., {Wisotzki}, L., \& {Dall'Aglio}, A. 2007, \aap,
  473, 805

\bibitem[{{Worseck} {et~al.}(2014){Worseck}, {Prochaska}, {Hennawi}, \&
  {McQuinn}}]{Worseck2014}
{Worseck}, G., {Prochaska}, J.~X., {Hennawi}, J.~F., \& {McQuinn}, M. 2014,
  ArXiv e-prints, arXiv:1405.7405

\bibitem[{{Worseck} \& {Wisotzki}(2006)}]{Worseck2006}
{Worseck}, G., \& {Wisotzki}, L. 2006, \aap, 450, 495

\bibitem[{{Worseck} {et~al.}(2011){Worseck}, {Prochaska}, {McQuinn},
  {Dall'Aglio}, {Fechner}, {Hennawi}, {Reimers}, {Richter}, \&
  {Wisotzki}}]{Worseck2011}
{Worseck}, G., {Prochaska}, J.~X., {McQuinn}, M., {et~al.} 2011, \apjl, 733,
  L24

\bibitem[{{Wyithe} {et~al.}(2008){Wyithe}, {Bolton}, \&
  {Haehnelt}}]{Wyithe2008}
{Wyithe}, J.~S.~B., {Bolton}, J.~S., \& {Haehnelt}, M.~G. 2008, \mnras, 383,
  691

\bibitem[{{Wyithe} \& {Loeb}(2003)}]{Wyithe2003}
{Wyithe}, J.~S.~B., \& {Loeb}, A. 2003, \apj, 595, 614

\bibitem[{{Yu} \& {Lu}(2004)}]{Yu2004}
{Yu}, Q., \& {Lu}, Y. 2004, \apj, 602, 603

\bibitem[{{Yu} \& {Tremaine}(2002)}]{Yu2002}
{Yu}, Q., \& {Tremaine}, S. 2002, \mnras, 335, 965

\bibitem[{{Zheng} {et~al.}(2015){Zheng}, {Syphers}, {Meiksin}, {Kriss},
  {Schneider}, {York}, \& {Anderson}}]{Zheng2015}
{Zheng}, W., {Syphers}, D., {Meiksin}, A., {et~al.} 2015, \apj, 806, 142

\end{thebibliography}

\begin{appendix}

\section{Appendix A: Light Travel Effects}
\label{ap:lighttravel}  

The radiative transfer algorithm we used in this study works under the assumption of the infinite speed of light. This assumption can result in the nonphysical ionization fronts traveling with speed greater than a speed of light. However, previous works \citep{White2003, Shapiro2006, Bolton2007a,Lidz2007, Davies2014}
 have found that the infinite speed of light assumption describes exactly the ionization state of the gas along the light cone. Here we will show that this is indeed correct. 

Let $X\left(r,t\right)$ be the gas property (ionization state of the gas influenced by the quasar radiation) at position $r$ from the quasar and time $t$. It is influenced by the properties at position $r' > r$ if $r'$ is located on the backward light cone. Time evolution of the property $X\left(r,t\right)$ is, thus, given by

\begin{equation}
\begin{split}
X\left(r,t\right) = X\left(r',t - dt\right) + \\ 
\int_{r}^{\infty} dr' f \left( r', t - \left[ r' - r \right] / c \right) dt
\end{split}
\end{equation} 
where $f$ is the function describing the ionization state of the gas at the position $r'$ and the luminosity of the quasar at time $ \left(r_{QSO} -r\right)/c$ in the past. The differential form of this equation is described by

\begin{equation}
\frac{dX\left(r,t - r/c\right)}{dt} = \int_{r}^{\infty} dr' f \left( r', t - r' / c \right) 
\end{equation}

Evaluating the property $X\left(r,t\right) $ on the light cone gives

\begin{equation}
\frac{dX\left(r,t_{LC}\right)}{dt_{LC}} = \int_{r}^{\infty} dr' f \left( r', t_{LC}\right) 
\end{equation}
where time on the light cone $t_{LC} = t - r/c$

This solution is identical to the one given by the infinite speed of light approximation

\begin{equation}
\frac{dX\left(r,t\right)}{dt} = \int_{r}^{\infty} dr' f \left( r', t\right) 
\end{equation}
As long as the boundary conditions for the $X\left(r,t\right)$ are the same as the light cone time boundary conditions the infinite speed of light code provides the same solution for the state of gas as solving for this state on the light cone. The light cone solution is the solution relevant for the line-of-sight proximity region.

\section{Appendix B: The Impact of Lyman-Limit Systems}
\label{ap:ImpactLLS}

For the expected values of the \ion{He}{2} ionizing background (see
\S~\ref{sec:Conditions}), clouds which have \ion{H}{1} Ly$\alpha$ forest column
densities of $N_{\rm HI} \gtrsim 10^{15}\,{\rm cm^{-2}}$ \citep{McQuinn2009}
will be dense enough to self-shield against \ion{He}{2} ionizing photons, making them
much more abundant than the higher column density \ion{H}{1} Lyman-limit
systems ($N_{\rm HI} > 10^{17.3}\,{\rm cm^{-2}}$). For such self-shielding absorbers, the quasar photoionization rate $\Gamma_{\rm QSO}$ is always properly attenuated within the \ion{He}{2}-LLS,
experiencing the correct $e^{-\tau}$ attenuation in every pixel along
the sightline, as determined by our radiative transfer
algorithm. However, the photoionization rate due to the \ion{He}{2} ionizing
background, $\Gamma_{\rm HeII}^{\rm bkg}$, was introduced as a
constant homogeneous quantity in every pixel, and this effectively
treats these overdense self-shielded regions as being optically thin. Ignoring this attenuation will give 
\ion{He}{2} fractions in the overdense \ion{He}{2}-LLSs which are systematically too low, which then
further manifests as errors in our radiative transfer from the quasar. An
exact treatment of this problem is clearly impossible with a
one-dimensional radiative transfer implementation, since the full
solution requires attenuating the UV background coming from rays in
all directions. In order to model the self-shielding of
\ion{He}{2}-LLSs from the UV background more accurately, we adapt an algorithm first
introduced in \citet{McQuinn2010}.

First, the algorithm identifies the locations of LLS along the skewer,
defined as the pixels with overdensities above a threshold value of
$1+\delta > 5$. Regions that have overdensities $1+\delta < 5$
experience an initial unattenuated background photoionization rate
$\Gamma_{\rm HeII}^{\rm bkg}$.  Whereas, the \ion{He}{2} ionizing
background is attenuated inside of each segment with $1 + \delta > 5$
based on the optical depth to neighboring cells. Specifically, the
algorithm assumes half of $\Gamma_{\rm HeII}^{\rm bkg}$ comes to the
overdense region from each side.  It then calculates the optical depth
in one direction $\tau_{< i} = \sigma_{\rm HeII}\times N_{{\rm
    HeII},<i}$ for all pixels inside the overdense segment, where $i$
is the pixel number, $\sigma_{\rm HeII}$ is the \ion{He}{2}
cross-section at the ionizing edge, and $N_{{\rm HeII},<i}$ is the \ion{He}{2} column density to the pixel $i$ from pixels $ < i$ above the
overdensity threshold. Similarly, the code computes the contribution
to the optical depth $\tau_{>i}$ arising in the other direction from
pixels $> i$. The \ion{He}{2} ionizing background coming from each
direction, $\Gamma_{\rm HeII}^{\rm bkg}\slash 2$, is then attenuated
inside the overdense segment by $e^{-\tau_{<i}}$ and $e^{-\tau_{>i}}$,
respectively, and these two contributions are summed to give the final
attenuated background photoionization rate in the cell.  This
procedure is performed at each time step of the radiative transfer
evolution, and is iterated until convergence is achieved.

Without self-shielding in \ion{He}{2}-LLSs, our procedure for
initializing the $x_{\rm HeII}$ in the code is straightforward. With
the quasar off, the only source of radiation is the background, and
since this background is assumed to be present in each pixel, the
initial \ion{He}{2} fractions are set by simply imposing optically thin
photoionization equilibrium. With self-shielding however, the value of
the background is attenuated in the LLS pixels, and these pixels must
be initialized with their correct equilibrium \ion{He}{2} fraction. But
because we compute the attenuation by integrating out to determine the
optical depth in two directions, the level of attenuation depends on
the \ion{He}{2} fraction in other neighboring pixels.  To ensure that the
quasar radiative transfer begins with the correct initial \ion{He}{2}
fractions at $t=0$, we adopt the following approach.  Similar to the
case without self-shielding, we first initialize each pixel assuming
optically thin ionization equilibrium with the unattenuated UV background, which
underestimates the \ion{He}{2} fractions in LLS pixels. We start the
calculation at time $t=-5 \times t_{\rm eq}$ with the quasar turned
off, where $t_{\rm eq} = 1\slash \Gamma_{\rm HeII}^{\rm bkg}$ is the
characteristic timescale for the \ion{He}{2} fraction to reach its
equilibrium value (see eqn.~\ref{eqn:teq}). This procedure ensures
that photoionization equilibrium with the background is achieved at
all pixels, including those self-shielding pixels for which the
background is attenuated. The quasar is then turned on at time $t=0$
and the photoionization rate becomes $\Gamma_{\rm tot} = \Gamma_{\rm
  QSO} + \Gamma_{\rm HeII}^{\rm bkg}$, and thus the radiative transfer
is evolved with the appropriate initial conditions everywhere.
 
\ion{He}{2} Lyman-limit systems are dense self-shielding absorbers that can act to stop the ionization front in its tracks.  They will add variance to the profile of \ion{He}{2} proximity zones. 
These systems are captured in simulations, but as we have mentioned our default radiative transfer algorithm makes their helium overly doubly ionized by the uniform ionizing background.

\begin{figure}[t]
\centering
 \includegraphics[width=0.9\linewidth]{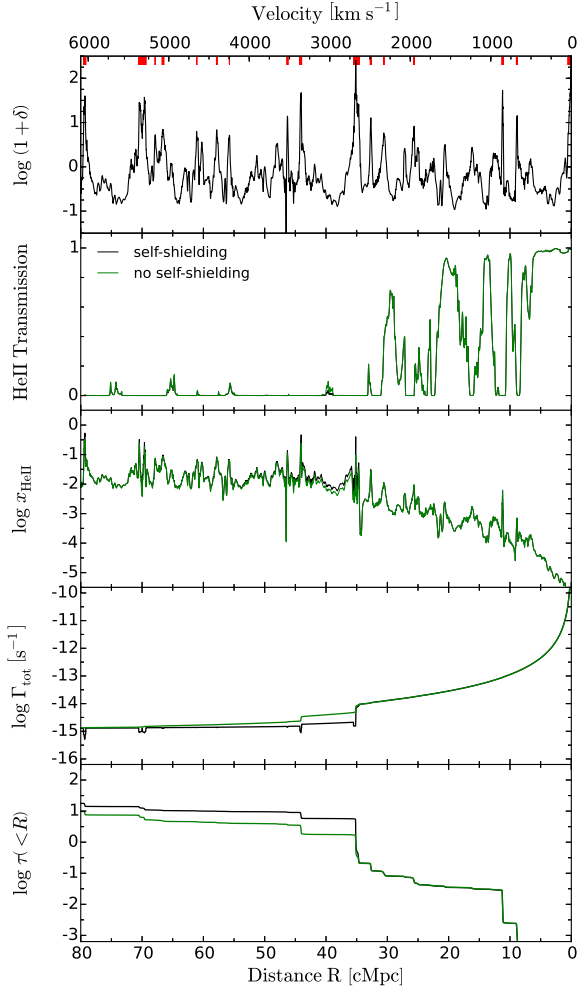}
  \caption{Impact of Lyman Limit Systems on the proximity zone. From top to bottom panels show: distribution of the overdensities (all locations of \ion{He}{2}-LLS are marked by the red ticks on top), transmission in \ion{He}{2}, evolution of the \ion{He}{2} fraction $x_{\rm HeII}$, total photoionization rate $\Gamma_{\rm tot}$ and the Lyman continuum optical depth $\tau \left( < R \right)$.}
 \label{fig:F_LLS}
\end{figure}

Figure~\ref{fig:F_LLS} shows an example skewer from our radiative transfer algorithm. We ran our algorithm for $10$~Myr and use our fiducial value of \ion{He}{2} ionizing background (see \S~\ref{sec:Conditions}) of $\Gamma_{\rm HeII}^{\rm bkg} = 10^{-14.9}{\rm s^{-1}}$ and $Q_{\rm 4Ry} = 10^{56.1}{\rm s^{-1}}$. 

Note that ignoring LLSs can lead to inaccurate results for three
different reasons: 1) the initial \ion{He}{2}
fractions are not correct in the LLS pixels because the pixels are
initialized under the assumption that they are optically thin, and the
initial \ion{He}{2} fractions in these pixels are thus set too low.  These
self-shielding pixels should have very high \ion{He}{2} fractions, but instead they start
out highly doubly ionized and thus reach their equilibrium ionization
state too soon; 
2) the total photoionization rate $\Gamma_{\rm tot}$ in
the \ion{He}{2}-LLS pixels is incorrect because the UVB is not attenuated,
hence their $t_{\rm eq}$ are too short, and they thus evolve away from
their (incorrect) initial state too fast; 3) these two effects result in 
incorrect \ion{He}{2} fractions at various \ion{He}{2}-LLS
locations along the skewer, which then implies that the amount of attenuation
between any location and the quasar is incorrectly
modeled, the photoionization rate $\Gamma_{\rm tot}$ is overestimated everywhere, and thus evolution of the \ion{He}{2} fractions in the
proximity zone will proceed too quickly.

Now first consider locations very close to the quasar, i.e., $R
\ll R_{\rm bkg}$, such that the quasar dominates over the
background. For these locations the presence of \ion{He}{2}-LLSs make no
difference. Referring to the three reasons described above, we can see
that this is true because: 1) very near the quasar
$t_{\rm eq}(R) \sim 10^4 - 10^5\,{\rm yr}$ and thus for the typical quasar lifetimes 
we consider $t_{\rm eq} \ll t_{\rm Q}$ and the \ion{He}{2} fraction has
already reached its equilibrium value and has no memory of the
incorrect LLS initial conditions; 2) the quasar completely dominates over the
UVB, so incorrectly attenuating the UVB changes $t_{\rm eq}=\Gamma_{\rm tot}^{-1}$ negligibly; 
3) given that 1) and 2) will be even more true closer to the quasar, the attenuation is correctly modeled, and the \ion{He}{2} fractions with and without \ion{He}{2}-LLSs are essentially
the same.  Examining Figure~\ref{fig:F_LLS}, we see that for $R <
30$\,cMpc, where the quasar radiation completely dominates over the
\ion{He}{2} ionizing background, all parameters are indistinguishable with and without self-shielding. 

At distances $R > 30$\, cMpc one reaches the regime where the
\ion{He}{2} background becomes comparable to the photoionization rate
of the quasar and LLSs become important. According to the three
criteria we used before this happens because: 1) for typical values of
quasar lifetimes that we consider $t_{\rm eq} \ge t_{\rm Q}$ and thus
$x_{\rm HeII}$ can still be out of equilibrium at $t \sim t_{\rm
  Q}$. Consequently, setting the initial \ion{He}{2} fraction
correctly based on the attenuated background radiation field
$\Gamma_{\rm HeII}^{\rm bkg}$ is important. If LLS attenuation is
ignored, the initial $x_{\rm HeII}$ are too low and are underestimated
at the end of each time-step; 2) because at these radii $\Gamma_{\rm
  HeII}^{\rm bkg} \sim \Gamma_{\rm QSO}$ the equilibration time
becomes strongly dependent on $\Gamma_{\rm HeII}^{\rm bkg}$ and it
will be too short if the UVB is not attenuated properly, thus the rate
of the gas evolution will be incorrect; 3) the result is the lack of
attenuation in the overdense pixels, which leads to the too quick
evolution of a proximity zone. One can see this effect in
Figure~\ref{fig:F_LLS} at $R > 30$ \,cMpc. The photoionization rate in
the no self-shielding case (green) is clearly overestimated in the
\ion{He}{2}-LLSs, where $x_{\rm HeII}$ is lower than in the
self-shielding case (black).

Finally, at distances $R > 60 $ \,cMpc, the total photoionization rate
$\Gamma_{\rm tot}$ is totally dominated by the UV background and the
quasar radiation $\Gamma_{\rm QSO}$ has a very weak effect on the IGM.
But at this point the treatment of \ion{He}{2}-LLSs is unimportant for
predicting the transmitted flux because essentially all of the
transmission in the \ion{He}{2} Ly$\alpha$ forest comes from the
underdense regions and the \ion{He}{2}-LLSs are always line black.
	
Ultimately, \ion{He}{2}-LLSs do not change the results dramatically neither in the inner part of the proximity zone, nor at larger distances ($R > 30$\ cMpc) where they have the biggest effect on the photoionization rate and thus \ion{He}{2} fraction. The reason is that in both models the optical depth $\tau \left( <R \right)$ becomes larger than unity at $R \sim 30 $\ cMpc and the resulting attenuation ${\rm e}^{-\tau}$ is effectively the same with or without self-shielding. Therefore, despite some differences at large distances, inside the proximity zone of the quasar, where the ionizations are dominated by the quasar, the effect of self-shielding in LLSs on the ionizing background radiation field is negligible.

In addition, high column density \ion{H}{1} absorption line
  systems ($N_{\rm HI} \gtrsim 10^{-17}{\rm cm^{-2}}$, i.e. \ion{H}{1} LLSs) might also
  impact our radiative transfer calculations. Although these dense
  clouds are only rarely encountered in the ambient IGM, they 
  are preferentially clustered around quasars \citep{Hennawi2006,Hennawi2007,Prochaska2013a,Prochaska2015}, and the photoevaporation
  of such systems may slow down the propagation of the ionization fronts
  and, consequently, change the structure of \ion{He}{2} transmission profiles. In
  order to investigate the impact of these systems on our results,
  we artificially increased the overdensity of pixels at $R \lesssim
  150$~ckpc radius from the location of the quasar halo by a factor of
  5. We then performed our radiative transfer calculations again and found
  no significant differences in the resulting \ion{He}{2}
  transmission. However, we note that in order to fully explore the
  effect of high column density \ion{H}{1} systems one should do a
  coupled radiative transfer calculation of \ion{H}{1}, \ion{He}{1}
  and \ion{He}{2}.

\section{Appendix C: He II ionizing radiation from recombinations}
\label{ap:recrad} 

The recombinations to the ground state of \ion{He}{2} produce
additional \ion{He}{2} ionizing radiation, that can affect the
evolution of the \ion{He}{2} fraction and subsequently change the
transmission profile in the \ion{He}{2} proximity zone. In reality
calculation of this radiation is a $3$D problem that can not be simply
addressed with our $1$D radiative transfer algorithm. In addition,
introducing this radiation in our radiative transfer algorithm would
effectively make each cell also a source of ionizing radiation and,
therefore, would complicate the calculations. In what follows,
we estimate the importance of recombination radiation using a simple analytical
toy model and argue that it can be safely neglected. 

Consider a spherically symmetric region around the quasar of size equal to the radius of the ionization front $R_{\rm IF}$ (see eqn.~\ref{eqn:RIF}). For the purpose of this calculations let us assume that \ion{He}{2} ionizing photons are produced only from the recombinations of helium ions and free electrons inside this region. We assume further that the distribution of the gas density and temperature ($T = 10^4$~K) are homogeneous inside the region with an average \ion{He}{2} fraction $x_{\rm HeII} = 0.02$, and that all recombination radiation is emitted only at a single frequency $\nu_{\rm th}$ corresponding to the \ion{He}{2} ionization threshold of $4$~Ryd. A remote observer tracks the evolution of the radiation field inside the region only along a $1$D line-of-sight towards the center of the region. 

Using the equation of radiative transfer we can estimate the monochromatic intensity $I_{\nu}$ of the recombination radiation at any location $R$ along the line-of-sight coming from the location $R'$
inside the spherical region as
\begin{equation}\label{eq:C1}
I_{\nu}=\int_{0}^{s\left( R^{\prime}-R\right)} j_{\nu}e^{- \tau_{\rm s}} {\rm d}s
\end{equation}
where $\tau_{\rm s} = \int_{0}^{s\left( R^{\prime}-R\right)}N_{\rm HeII}\sigma_{\nu}{\rm d}s$ is the optical depth in the IGM along the direction $s\left( R^{\prime} - R\right)$ connecting points $R^{\prime}$ and $R$. This distance $s\left( R^{\prime} - R\right)$ can be expressed through the viewing angle $\theta$ ($0<\theta<\pi$) between the radial direction to $R'$ and the line-of-sight as

\begin{equation}\label{eq:C2}
s\left( R^{\prime} - R\right)^2 = R^2+R'^2-2R R'\cos(\theta)
\end{equation} 
We choose to align our $z$-axis with the direction of {\it R} and hence $\theta$ is the usual polar angle in spherical coordinates. The free-bound emissivity $j_{\nu}$ is given by 
\begin{equation}\label{eq:C3}
j_{\nu} =\frac{h^4\nu^3}{(2\pi m_{\rm e}kT)^{3/2}c^2}e^{\frac{-h(\nu-\nu_{\rm th})}{k_{\rm B}T}} n_{\rm e} n_{\rm HeIII} \sigma_{\nu}\left(\nu \right)
\end{equation}
where $n_{\rm e}$ and $n_{\rm HeIII}$ are the densities of electrons and ions, respectively. 

Combining eqn.~(\ref{eq:C1}) and eqn.~(\ref{eq:C3}) the intensity of the recombination radiation can be written as

\begin{figure}[!t]
\centering
 \includegraphics[width=1.\linewidth]{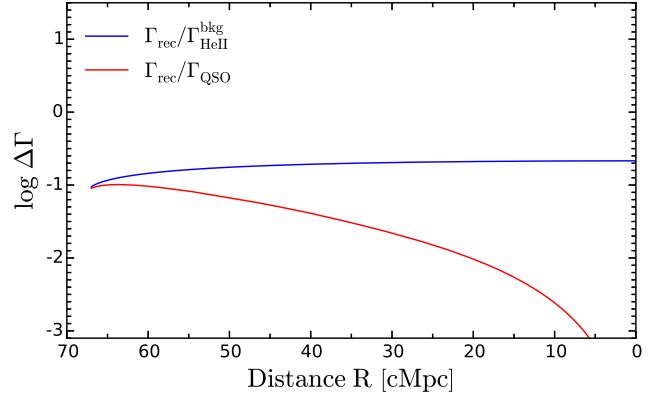}
  \caption{Comparison between \ion{He}{2} recombination radiation photoionization rate $\Gamma_{\rm rec}$ and quasar photoionization rate $\Gamma_{\rm QSO}$ (\emph{red curve}) and \ion{He}{2} ionizing background $\Gamma_{\rm HeII}^{\rm bkg} = 10^{-14.9}{\rm s^{-1}}$ (\emph{blue curve}) for mean \ion{He}{2} fraction $\langle x_{\rm HeII}\rangle = 0.02$.}
 \label{fig:F_REC}
\end{figure}

\begin{equation}\label{eq:C4}
I_{\nu}\left( \theta \right) = \int_{0}^{s\left( R^{\prime}-R\right)}\frac{h^4\nu^3 \sigma_{\nu}\left( \nu \right) n_{\rm e} n_{\rm HeIII}}{(2\pi m_{\rm e}k_{\rm B}T)^{3/2}c^2} e^{\frac{-h(\nu-\nu_{\rm th})}{k_{\rm B}T}} e^{-\tau_{s}} {\rm d}s
\end{equation}

Integrating over the solid angle $\Omega$ we obtain the mean azimuthally symmetric intensity of the recombination radiation $4\pi J_{\nu}$ at location $R$ along the line-of-sight
\begin{equation}\label{eq:C5}
4\pi J_{\nu} = \int I_{\nu} {\rm d}\Omega = 2\pi\int_{0}^{\pi}I_{\nu}\left( \theta \right)\sin{\theta}{\rm d}\theta 
\end{equation}

Lastly, given eqn.~(\ref{eq:C5}) for the mean intensity $4 \pi J_{\nu}$ we calculate the 
\ion{He}{2} photoionization rate due to recombinations $\Gamma_{\rm rec}$ as
  
\begin{equation}\label{eq:C6}
\Gamma_{\rm rec} = \int_{\nu_{\rm th}}^{2\nu_{\rm th}} \frac{4\pi J_{\nu}}{h_{\rm P}\nu}\sigma_{\nu}\left( \nu \right) {\rm d}\nu
\end{equation}
Recall, that we approximated the recombination radiation as being propagated with a single frequency corresponding to $4$~Ryd. However, this will make it impossible to calculate the frequency-averaged photoionization rate and we, therefore, assume a narrow finite frequency interval $\nu_{\rm th} \leq \nu \leq 2\nu_{\rm th}$ in order to calculate the integral in eqn.~(\ref{eq:C6}).

We compare $\Gamma_{\rm rec}$ to the values of quasar photoionization rate
$\Gamma_{\rm QSO}$ and \ion{He}{2} ionizing background $\Gamma_{\rm 
HeII}^{\rm bkg} = 10^{-14.9}{\rm s^{-1}}$ taken from our fiducial
model with quasar lifetime $t_{\rm Q} = 10$~Myr and quasar photon
production rate $Q_{\rm 4Ry} = 10^{56.1}{\rm s^{-1}}$.
The location of the ionization front is this model is approximately
$R_{\rm IF} \simeq 67$~cMpc, we thus use this value as a boarder of
the region where the recombination radiation is calculated. The
results are shown in Figure~\ref{fig:F_REC}. The blue curve shows the
ratio of the \ion{He}{2} photoionization rates due to recombinations
$\Gamma_{\rm rec}$ and the \ion{He}{2} background $\Gamma_{\rm
  HeII}^{\rm bkg}$. This ratio stays nearly constant throughout the
region we consider, with $\Gamma_{\rm rec}$ being $\simeq 10-30\%$ of
the $\Gamma_{\rm HeII}^{\rm bkg}$. Not only do recombinations
constitute a minor correction to the background, but because
$\Gamma_{\rm rec}$ does not vary significantly with distance $R$, it
could also be effectively absorbed into the background and thought of as a
small correction to it. At the same time, the red curve shows the ratio
between $\Gamma_{\rm rec}$ and quasar photoionization rate
$\Gamma_{\rm QSO}$. It is clear that the quasar photoionization rate
dominants at all distances $R$ from the quasar, making the
contribution of recombination radiation to the total radiation field
in the \ion{He}{2} proximity region essentially negligible.

Although our analtyical estimate of $\Gamma_{\rm rec}$ is admittedly rather crude,
according to Figure~\ref{fig:F_REC} including recombination radiation in our calculation will not significantly change the results of our $1$D radiative transfer.

\section{Appendix D: Generating \ion{He}{2} Skewers}
\label{ap:skewers} 

We calculate \ion{He}{2} spectra along each of the skewers drawn from the SPH simulations following the procedure described in \citet{Theuns1998}. We use densities and velocities of the SPH particles at each pixel, combined with the \ion{He}{2} fraction and temperature resulting from our radiative transfer calculation to compute the observed flux
$F\left( v \right) = e^{-\tau_{\rm HeII} \left( v \right)}$, with the \ion{He}{2} optical depth $\tau_{\rm HeII} \left( v \right)$ of the pixel given by
\begin{equation}
\begin{split}
\tau_{\rm HeII} \left( v \right) = \frac{\sigma_\alpha c}{\sqrt{\pi V}} n_{\rm He} x_{\rm HeII} \left( 1 + \delta \right) \frac{\Delta R}{1 + z} \\
\times {\rm exp} \left( - \left[\frac{v - v_H}{V}\right]^2 \right)
\end{split}
\end{equation}
where $V^2 = 2 k_{\rm B} T \slash m_{\rm He}$ is the Doppler width of helium with mass $m_{\rm He}$, $c$ is the speed of light, $T$ is the temperature inside the cell, $n_{\rm He} x_{\rm HeII}$ is the density of \ion{He}{2}, $\Delta R$ is the pixel width, $\sigma_{\rm HeII}$ is \ion{He}{2} Ly$\alpha$ cross-section, and the term ${v_{\rm H}}$ describes the combination of Hubble and peculiar velocities.

We similarly compute the \ion{H}{1} Ly$\alpha$ optical
depth. Hydrogen gas is also considered optically thin to the ionizing
radiation and analogously to helium, we introduce an \ion{H}{1}
ionizing background photoionization rate $\Gamma_{\rm HI}^{\rm bkg}$
to mimic the influence of other sources in the Universe, with a value
of $\Gamma_{\rm HI}^{\rm bkg} = 1 \times 10^{-12} {\rm s^{-1}}$ chosen to match the mean transmitted flux at $z \simeq 3.1$ in agreement with recent measurements by \citet{Becker2013}.

\end{appendix}

\end{document}